\documentclass{article}
\usepackage{epsf}
\usepackage{amsmath}
\usepackage{amssymb}
\usepackage{epubtk}

\DeclareMathOperator{\real}{Re}
\DeclareMathOperator{\imaginary}{Im}
\DeclareMathOperator{\diver}{div}
\DeclareMathOperator{\tr}{tr}

\newcounter{definition}
\newenvironment{definition}
  {\refstepcounter{definition}
   \vspace{1 em}
   \noindent{\bf Definition~\thedefinition:}
   \begin{em}}
  {\end{em}}

\newcommand{\pullback}[1]{\hbox{\lower0.5ex\hbox{${}_{\leftarrow}$}}\kern-1.9ex{#1}}
\newcommand{\pullbacklong}[1]{\hbox{\lower0.85ex\hbox{${}_{\longleftarrow}$}}\kern-3.0ex{#1}}
\newcommand{\pullbackllong}[1]{\hbox{\lower0.85ex\hbox{${}_{\longleftarrow\!\!-\!\!-\!\!-\!\!-}$}}\kern-6.4ex{#1}}

\begin{document}

\title{Isolated and Dynamical Horizons and Their Applications}

\author{\epubtkAuthorData{Abhay Ashtekar}
        {Institute for Gravitational Physics and Geometry \\
         Pennsylvania State University \\
         University Park, PA 16801, U.S.A. \\
         and \\
         Kavli Institute of Theoretical Physics \\
         University of California \\
         Santa Barbara, CA 93106-4030, U.S.A. \\
         and \\
         Max-Planck-Institut f\"ur Gravitationsphysik,
         Albert-Einstein-Institut \\
         Am M\"uhlenberg 1, 14476 Golm, Germany \\
         and \\
         Erwin-Schr\"odinger-Institut \\
         Boltzmanngasse 9, 1090 Vienna, Austria}
        {ashtekar@gravity.psu.edu}
        {http://cgpg.gravity.psu.edu/people/Ashtekar/}
        \and
        \epubtkAuthorData{Badri Krishnan}
        {Max-Planck-Institut f\"ur Gravitationsphysik,
         Albert-Einstein-Institut \\
         Am M\"uhlenberg 1, 14476 Golm, Germany \\
         and \\
         Erwin-Schr\"odinger-Institut \\
         Boltzmanngasse 9, 1090 Vienna, Austria}
        {badri.krishnan@aei.mpg.de}
        {}}

\date{}
\maketitle


\begin{abstract}
  Over the past three decades, black holes have played an important
  role in quantum gravity, mathematical physics, numerical relativity
  and gravitational wave phenomenology. However, conceptual settings
  and mathematical models used to discuss them have varied
  considerably from one area to another. Over the last five years a
  new, quasi-local framework was introduced to analyze diverse facets
  of black holes in a unified manner. In this framework, evolving
  black holes are modelled by \emph{dynamical horizons} and black
  holes in equilibrium by \emph{isolated horizons}. We review basic
  properties of these horizons and summarize applications to
  mathematical physics, numerical relativity, and quantum gravity. This
  paradigm has led to significant generalizations of several results
  in black hole physics. Specifically, it has introduced a more
  physical setting for black hole thermodynamics and for black hole
  entropy calculations in quantum gravity, suggested a
  phenomenological model for hairy black holes, provided novel
  techniques to extract physics from numerical simulations, and led to
  new laws governing the dynamics of black holes in exact general
  relativity.
\end{abstract}

\epubtkKeywords{black holes, generalized thermodynamics, numerical
  relativity, quantum gravity}

\newpage


\section{Introduction}
\label{s1}

Research inspired by black holes has dominated several areas of
gravitational physics since the early seventies. The mathematical
theory turned out to be extraordinarily rich and full of
surprises. Laws of black hole mechanics brought out deep,
unsuspected connections between classical general relativity,
quantum physics, and statistical mechanics~\cite{bch, jdb1, jdb2,
jdb3}. In particular, they provided a concrete challenge to
quantum gravity which became a driving force for progress in that
area. On the classical front, black hole uniqueness
theorems~\cite{pcreview, mh} took the community by surprise. The
subsequent analysis of the detailed properties of Kerr--Newman
solutions~\cite{bc} and perturbations thereof~\cite{chandra}
constituted a large fraction of research in mathematical general relativity in
the seventies and eighties. Just as the community had come to
terms with the uniqueness results, it was surprised yet again by
the discovery of hairy black holes~\cite{bm, pb}. Research in this
area continues to be an active branch of mathematical
physics~\cite{vgreview}. The situation has been similar in numerical
relativity. Since its inception, much of the research in this area
has been driven by problems related to black holes, particularly
their formation through gravitational collapse~\cite{collapsereview},
the associated critical phenomenon~\cite{mc, cgreview}, and the
dynamics leading to their coalescence (see,
e.g., \cite{grand1, closelimit2, closelimit1, grazing2, grazing1, lhreview}).
Finally, black holes now play a major role in relativistic
astrophysics, providing mechanisms to fuel the most powerful
engines in the cosmos. They are also among the most promising
sources of gravitational waves, leading to new avenues to confront
theory with experiments~\cite{sourcesreview}.

Thus there has been truly remarkable progress on many different
fronts. Yet, as the subject matured, it became apparent that the
basic theoretical framework has certain undesirable features from
both conceptual and practical viewpoints. Nagging questions have
persisted, suggesting the need of a new paradigm.

\begin{description}
\item[Dynamical situations]~\\
  For fully dynamical black
holes, apart from the `topological censorship' results which
restrict the horizon topology~\cite{swh2, fsw}, there has
essentially been only one major result in \emph{exact} general
relativity. This is the celebrated area theorem proved by Hawking
in the early seventies~\cite{swh, he}: If matter satisfies the null
energy condition, the area of the black hole event horizon can
never decrease. This theorem has been extremely influential
because of its similarity with the second law of thermodynamics.
However, it is a qualitative result; it does not provide an
explicit formula for the amount by which the area increases in
physical processes. Now, for a black hole of mass $M$, angular
momentum $J$, area $a$, surface gravity $\kappa$, and angular velocity
$\Omega$, the first law of black hole mechanics,
\begin{equation}
  \label{1law1}
  \delta M = \frac{\kappa}{8\pi G}\delta a + \Omega \, \delta J,
\end{equation}
does relate the change in the horizon area to that in the energy
and angular momentum, as the black hole makes a transition from
one equilibrium state to a \emph{nearby} one~\cite{bch, rwreview}.
This suggests that there may well be a fully dynamical version of
Equation~(\ref{1law1}) which relates the change in the black hole area
to the energy and angular momentum it absorbs in fully dynamical
processes in which the black hole makes a transition from a given
state to one which is \emph{far removed}. Indeed, without such a
formula, one would have no quantitative control on how black holes
grow in exact general relativity. Note however that the event
horizons can form and grow even in a flat region of space-time
(see Figure~\ref{fig:vaidya} and Section~\ref{s2.2.2}
for illustrations). During this phase, the area grows in spite of
the fact that there is no flux of energy or angular momentum
across the event horizon. Hence, in the standard framework where
the surface of the black hole is represented by an event horizon,
it is impossible to obtain the desired formula. Is there then a
more appropriate notion that can replace event horizons?
\end{description}

\begin{description}
\item[Equilibrium situations]~\\
  The zeroth and first laws of
black hole mechanics refer to equilibrium situations and small
departures therefrom. Therefore, in this context, it is natural to
focus on isolated black holes. It was customary to represent them
by \emph{stationary} solutions of field equations, i.e,
solutions which admit a time-translational Killing vector field
\emph{everywhere}, not just in a small neighborhood of the black
hole. While this simple idealization was natural as a starting
point, it is overly restrictive. Physically, it should be
sufficient to impose boundary conditions at the horizon which
ensure \emph{only that the black hole itself is isolated}. That
is, it should suffice to demand only that the intrinsic geometry
of the horizon be time independent, whereas the geometry outside
may be dynamical and admit gravitational and other radiation.
Indeed, we adopt a similar viewpoint in ordinary thermodynamics;
while studying systems such as a classical gas in a box, one
usually assumes that only the system under consideration is in
equilibrium, not the whole world. In realistic situations, one is
typically interested in the final stages of collapse where the
black hole has formed and `settled down' or in situations in which
an already formed black hole is isolated for the duration of the
experiment (see Figure~\ref{exam}). In such cases, there is likely
to be gravitational radiation and non-stationary matter far away
from the black hole. Thus, from a physical perspective, a
framework which demands global stationarity is too restrictive.

\epubtkImage{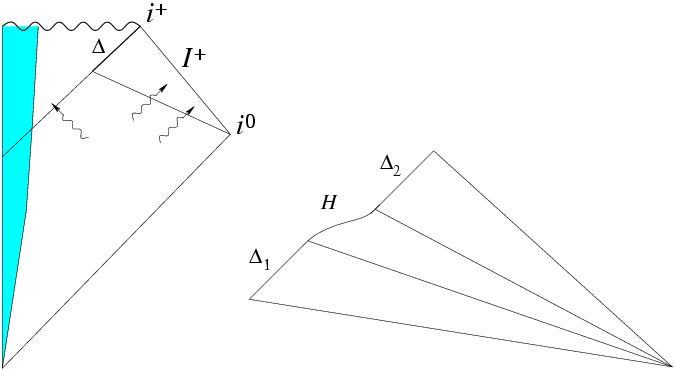}
{\begin{figure}[hptb]
   \def\epsfsize#1#2{0.5#1}
   \centerline{\epsfbox{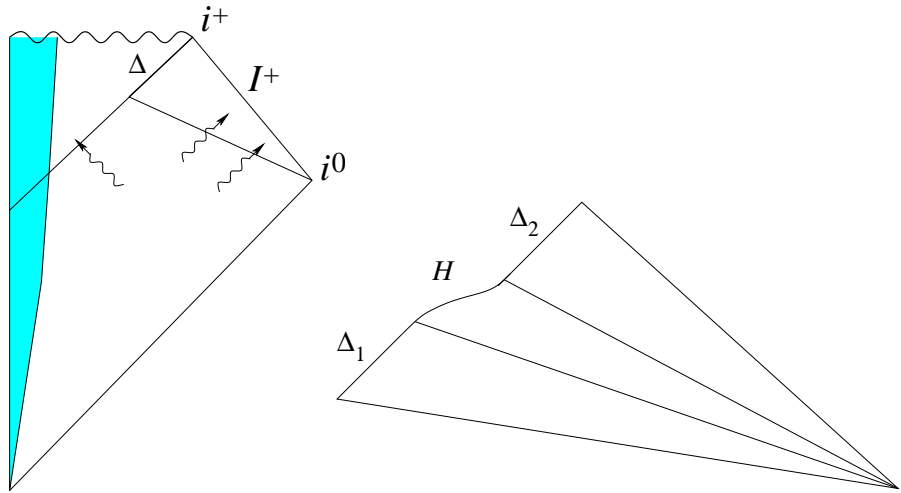}}
   \caption{\it Left panel: A typical gravitational collapse. The
     portion $\Delta$ of the event horizon at late times is
     isolated. Physically, one would expect the first law to apply to
     $\Delta$ even though the entire space-time is not stationary
     because of the presence of gravitational radiation in the
     exterior region. Right panel: Space-time diagram of a black hole which
     is initially in equilibrium, absorbs a finite amount of radiation,
     and again settles down to equilibrium. Portions $\Delta_1$ and
     $\Delta_2$ of the horizon are isolated. One would expect the
     first law to hold on both portions although the space-time is not
     stationary.}
   \label{exam}
 \end{figure}
}

Even if one were to ignore these conceptual considerations and
focus just on results, the framework has certain unsatisfactory
features. Consider the central result, the first law of
Equation~(\ref{1law1}). Here, the angular momentum $J$ and the mass
$M$ are defined at infinity while the angular velocity $\Omega$ and
surface gravity $\kappa$ are defined at the horizon. Because one has
to go back and forth between the horizon and infinity, the
physical meaning of the first law is not transparent\epubtkFootnote{The
  situation is even more puzzling in the Einstein--Yang--Mills theory
  where the right side of Equation~(\ref{1law1}) acquires an
  additional term, $V\delta Q$. In treatments based on stationary
  space-times, not only the Yang--Mills charge $Q$, but also the
  potential $V$ (the analog of $\Omega$ and $\kappa$, is evaluated
  at infinity~\cite{sw}.}.
For instance, there may be matter rings around the black hole
which contribute to the angular momentum and mass at infinity. Why
is this contribution relevant to the first law of black hole
mechanics? Shouldn't only the angular momentum and mass of the
black hole feature in the first law? Thus, one is led to ask: Is
there a more suitable paradigm which can replace frameworks based
on event horizons in stationary space-times?
\end{description}

\begin{description}
\item[Entropy calculations]~\\
  The first and the second laws
suggest that one should assign to a black hole an entropy which is
proportional to its area. This poses a concrete challenge to
candidate theories of quantum gravity: Account for this entropy
from fundamental, statistical mechanical considerations. String
theory has had a remarkable success in meeting this challenge in
detail for a subclass of extremal, stationary black holes whose
charge equals mass (the so-called BPS states)~\cite{stringreview}.
However, for realistic black holes the charge to mass ratio is
less than $10^{-18}$. It has not been possible to extend the
detailed calculation to realistic cases where charge is negligible
and matter rings may distort the black hole horizon. From a
mathematical physics perspective, the entropy calculation should
also encompass hairy black holes whose equilibrium states cannot
be characterized just by specifying the mass, angular momentum and
charges at infinity, as well as non-minimal gravitational
couplings, in presence of which the entropy is no longer a
function just of the horizon area. One may therefore ask if other
avenues are available. A natural strategy is to consider the
sector of general relativity containing an isolated black hole and
carry out its quantization systematically. A pre-requisite for
such a program is the availability of a manageable action
principle and/or Hamiltonian framework. Unfortunately, however, if
one attempts to construct these within the classical frameworks
traditionally used to describe black holes, one runs into two
difficulties. First, because the event horizon is such a global
notion, no action principle is known for the sector of general
relativity containing geometries which admit an event horizon as
an internal boundary. Second, if one restricts oneself to globally
stationary solutions, the phase space has only a finite number of
true degrees of freedom and is thus `too small' to adequately
incorporate all quantum fluctuations. Thus, again, we are led to
ask: Is there a more satisfactory framework which can serve as the
point of departure for a non-perturbative quantization to address
this problem?
\end{description}

\begin{description}
\item[Global nature of event horizons]~\\
  The future event
horizon is defined as the future boundary of the causal past of
future null infinity. While this definition neatly encodes the
idea that an outside observer can not `look into' a black hole,
it is too global for many applications. First, since it refers to
null infinity, it can not be used in spatially compact
space-times. Surely, one should be able to analyze black hole
dynamics also in these space-times. More importantly, the notion
is teleological; it lets us speak of a black hole \emph{only after
we have constructed the entire space-time}. Thus, for example, an
event horizon may well be developing in the room you are now
sitting \emph{in anticipation} of a gravitational collapse that
may occur in this region of our galaxy a million years from now.
When astrophysicists say that they have discovered a black hole in
the center of our galaxy, they are referring to something much
more concrete and quasi-local than an event horizon. Is there a
satisfactory notion that captures what they are referring to?

\epubtkImage{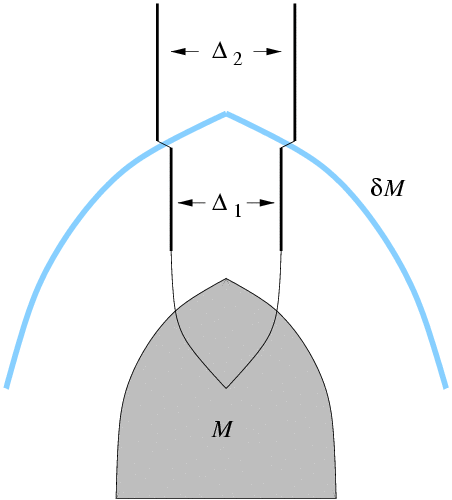}
{\begin{figure}[hptb]
   \def\epsfsize#1#2{0.6#1}
   \centerline{\epsfbox{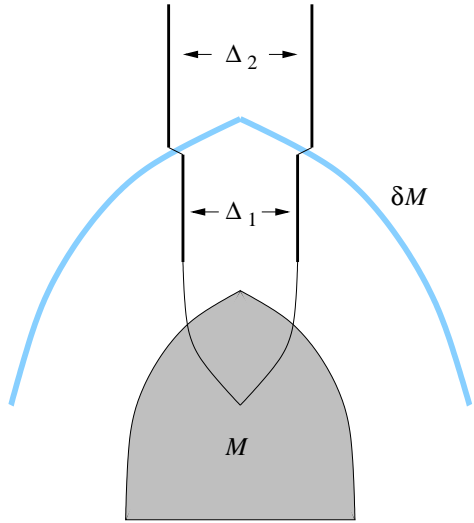}}
   \caption{\it A spherical star of mass $M$ undergoes collapse. Much
     later, a spherical shell of mass $\delta{M}$ falls into the
     resulting black hole. While $\Delta_1$ and $\Delta_2$ are both
     isolated horizons, only $\Delta_2$ is part of the event horizon.}
   \label{shell}
 \end{figure}
}

The teleological nature of event horizons is also an obstruction
to extending black hole mechanics in certain physical situations.
Consider for example, Figure~\ref{shell} in which a spherical star
of mass $M$ undergoes a gravitational collapse. The singularity is
hidden inside the null surface $\Delta_1$ at $r=2M$ which is foliated
by a family of marginally trapped surfaces and would be a part of
the event horizon if nothing further happens. Suppose instead,
after a million years, a thin spherical shell of mass $\delta M$
collapses. Then $\Delta_1$ would not be a part of the event horizon
which would actually lie slightly outside $\Delta_1$ and coincide
with the surface $r= 2(M+\delta M)$ in the distant future. On
physical grounds, it seems unreasonable to exclude $\Delta_1$ a
priori from thermodynamical considerations. Surely one should be
able to establish the standard laws of mechanics not only for the
equilibrium portion of the event horizon but also for $\Delta_1$.

Next, let us consider numerical simulations of binary black holes.
Here the main task is to \emph{construct} the space-time
containing evolving black holes. Thus, one needs to identify
initial data containing black holes without the knowledge of the
entire space-time and evolve them step by step. The notion of a
event horizon is clearly inadequate for this. One uses instead the
notion of apparent horizons (see Section~\ref{s2.2}). One may then
ask: Can we use apparent horizons instead of event horizons in
other contexts as well? Unfortunately, it has not been possible to
derive the laws of black hole mechanics using apparent horizons.
Furthermore, as discussed in section 2, while apparent horizons
are `local in time' they are still global notions, tied too
rigidly to the choice of a space-like 3-surface to be directly
useful in all contexts. Is there a truly quasi-local notion which
can be useful in all these contexts?
\end{description}

\begin{description}
\item[Disparate paradigms]~\\
  In different communities within
gravitational physics, the intended meaning of the term `black
hole' varies quite considerably. Thus, in a string theory seminar,
the term `fundamental black holes' without further qualification
generally refers to the BPS states referred to above -- \emph{a
sub-class of stationary, extremal black holes}. In a mathematical
physics talk on black holes, the fundamental objects of interest
are stationary solutions to, say, the Einstein--Higgs--Yang--Mills
equations \emph{for which the uniqueness theorem fails}. The focus
is on the ramifications of `hair', which are completely ignored in
string theory. In a numerical relativity lecture, both these
classes of objects are considered to be so exotic that they are
excluded from discussion without comment. The focus is primarily
on the \emph{dynamics of apparent horizons} in general relativity.
In astrophysically interesting situations, the distortion of black
holes by external matter rings, magnetic fields and other black
holes is often non-negligible~\cite{dis2, dis3, dis4}. While these
illustrative notions
seem so different, clearly there is a common conceptual core. Laws
of black hole mechanics and the statistical mechanical derivation
of entropy should go through for \emph{all} black holes in
equilibrium. Laws dictating the dynamics of apparent horizons
should \emph{predict} that the final equilibrium states are those
represented by the \emph{stable} stationary solutions of the
theory. Is there a paradigm that can serve as an unified framework
to establish such results in all these disparate situations?
\end{description}

These considerations led to the development of a new, quasi-local
paradigm to describe black holes. This framework was inspired by
certain seminal ideas introduced by Hayward~\cite{sh, sh2, sh3, sh4} in
the mid-nineties and has been systematically developed over the past 
five years or so. Evolving black holes are modelled by
\emph{dynamical horizons} while those in equilibrium are modelled
by \emph{isolated horizons}. Both notions are quasi-local. In
contrast to event horizons, neither notion requires the knowledge
of space-time as a whole or refers to asymptotic flatness.
Furthermore, they are \emph{space-time notions}. Therefore, in
contrast to apparent horizons, they are not tied to the choice of
a partial Cauchy slice. This framework provides a new perspective
encompassing all areas in which black holes feature: quantum
gravity, mathematical physics, numerical relativity, and
gravitational wave phenomenology. Thus, it brings out the
underlying unity of the subject. More importantly, it has overcome
some of the limitations of the older frameworks and also led to
new results of direct physical interest.

The purpose of this article is to review these developments. The
subject is still evolving. Many of the key issues are still open
and new results are likely to emerge in the coming years.
Nonetheless, as the Editors pointed out, there is now a core of
results of general interest and, thanks to the innovative style of
\emph{Living Reviews}, we will be able to incorporate new results
through periodic updates.

Applications of the quasi-local framework can be summarized as
follows:

\begin{description}
\item[Black hole mechanics]~\\
  Isolated horizons extract from
the notion of Killing horizons, just those conditions which ensure
that the horizon geometry is time independent; there may be matter
and radiation even nearby~\cite{pc}. Yet, it has been possible to
extend the zeroth and first laws of black hole mechanics to
isolated horizons~\cite{afk, cns, abl2}. Furthermore, this derivation
brings out a conceptually important fact about the first law.
Recall that, in presence of internal boundaries, time evolution
need not be Hamiltonian (i.e., need not preserve the symplectic
structure). If the inner boundary is an isolated horizon, a
necessary and sufficient condition for evolution to be Hamiltonian
turns out to be precisely the first law! Finally, while the first
law has the same form as before (Equation~(\ref{1law1})), \emph{all
quantities which enter the statement of the law now refer to the
horizon itself}. This is the case even when non-Abelian gauge
fields are included.

Dynamical horizons allow for the horizon geometry to be time
dependent. This framework has led to a \emph{quantitative}
relation between the growth of the horizon area and the flux of
energy and angular momentum across it~\cite{ak1, ak2}. The
processes can be in the non-linear regime of exact general
relativity, without any approximations. Thus, the second law is
generalized and the generalization also represents an
\emph{integral version} of the first law~(\ref{1law1}), applicable
also when the black hole makes a transition from one state to
another, which may be far removed.
\end{description}

\begin{description}
\item[Quantum gravity]~\\
  The entropy problem refers to
equilibrium situations. The isolated horizon framework provides an
action principle and a Hamiltonian theory which serves as a
stepping stone to non-perturbative quantization. Using the quantum
geometry framework, a detailed theory of the quantum horizon
geometry has been developed. The horizon states are then counted
to show that the statistical mechanical black hole entropy is
indeed proportional to the area~\cite{abck, abk, dl, km3, aev}. This
derivation is applicable to ordinary, astrophysical black holes
which may be distorted and far from extremality. It also
encompasses cosmological horizons to which thermodynamical
considerations are known to apply~\cite{gh}. Finally, the arena
for this derivation is the curved black hole geometry, rather than
a system in flat space-time which has the same number of states as
the black hole~\cite{sv, ms}. Therefore, this approach has a
greater potential for analyzing physical processes associated with
the black hole.

The dynamical horizon framework has raised some intriguing
questions about the relation between black hole mechanics and
thermodynamics in fully dynamical situations~\cite{bf}. In
particular, they provide seeds for further investigations of the
notion of entropy in non-equilibrium situations.
\end{description}

\begin{description}
\item[Mathematical physics]~\\
  The isolated horizon framework
has led to a phenomenological model to understand properties of
hairy black holes~\cite{acs, ac}. In this model, the hairy black
hole can be regarded as a bound state of an ordinary black hole
and a soliton. A large number of facts about hairy black holes had
accumulated through semi-analytical and numerical studies. Their
qualitative features are explained by the model.

The dynamical horizon framework also provides the groundwork for a
new approach to Penrose inequalities which relate the area of
cross-sections of the event horizon $A_\mathrm{e}$ on a Cauchy surface with
the ADM mass $M_\mathrm{ADM}$ at infinity~\cite{rp}: $\sqrt{A_\mathrm{e}/16\pi}
\leq M_\mathrm{ADM}$. Relatively recently, the conjecture has been proved
in time symmetric situations. The basic monotonicity formula of
the dynamical horizon framework could provide a new avenue to
extend the current proofs to non-time-symmetric situations. It may
also lead to a stronger version of the conjecture where the ADM
mass is replaced by the Bondi mass~\cite{ak2}.
\end{description}

\begin{description}
\item[Numerical relativity]~\\
  The framework has provided a
number of tools to extract physics from numerical simulations in
the near-horizon, strong field regime. First, there exist
expressions for mass and angular momentum of dynamical and
isolated horizons which enable one to monitor dynamical processes
occurring in the simulations~\cite{ak2} and extract properties of
the final equilibrium state~\cite{abl2, dkss}. These quantities can
be calculated knowing only the horizon geometry and do not
pre-suppose that the equilibrium state is a Kerr horizon.
The computational resources required in these calculations are
comparable to those employed by simulations using cruder
techniques, but the results are now invariant and interpretation is
free from ambiguities. Recent work~\cite{whiskey} has shown that
these methods are also numerically more accurate and robust than
older ones.

Surprisingly, there are simple \emph{local} criteria to decide
whether the geometry of an isolated horizon is that of the Kerr
horizon~\cite{lp}. These criteria have already been implemented in
numerical simulations. The isolated horizon framework also
provides invariant, practical criteria to compare near-horizon
geometries of \emph{different} simulations~\cite{ihprl} and leads
to a new approach to the problem of extracting wave-forms in a
gauge invariant fashion. Finally, the framework provides natural
boundary conditions for the initial value problem for black holes
in quasi-equilibrium~\cite{gc1, jgm, djk}, and to interpret certain
initial data sets~\cite{bkthesis}. Many of these ideas have
already been implemented in some binary black hole codes~\cite{dkss,
  whiskey, bks} and the process is continuing.
\end{description}

\begin{description}
\item[Gravitational wave phenomenology]~\\
  The isolated horizon
framework has led to a notion of horizon multipole moments~\cite{aepv}. They provide a diffeomorphism invariant
characterization of the isolated horizon geometry. They are
distinct from the Hansen multipoles in stationary space-times~\cite{rh} normally used in the analysis of equations of motion
because they depend only on the isolated horizon geometry and do
not require global stationarity. They represent \emph{source
multipoles} rather than Hansen's \emph{field multipoles}. In Kerr
space-time, while the mass and angular momenta agree in the two
regimes, quadrupole moments do not; the difference becomes
significant when $a \sim M$, i.e., in the fully relativistic
regime. In much of the literature on equations of motion of black
holes, the distinction is glossed over largely because only field
multipoles have been available in the literature. However, in
applications to equations of motion, it is the source multipoles
that are more relevant, whence the isolated horizon
multipoles are likely to play a significant role.

The dynamical horizon framework enables one to calculate mass and
angular momentum of the black hole as it evolves. In particular,
one can now ask if the black hole can be first formed violating
the Kerr bound $a\le M$ but then eventually settle down in the
Kerr regime. Preliminary considerations fail to rule out this
possibility, although the issue is still open~\cite{ak2}. The
issue can be explored both numerically and analytically. The
possibility that the bound can indeed be violated initially has
interesting astrophysical implications~\cite{sf}.
\end{description}

In this review, we will outline the basic ideas underlying
dynamical and isolated horizon frameworks and summarize their
applications listed above. The material is organized as follows.
In Section~\ref{s2} we recall the basic definitions, motivate the
assumptions and summarize their implications. In Section~\ref{s3}
we discuss the area increase theorem for dynamical horizons and
show how it naturally leads to an expression for the flux of
gravitational energy crossing dynamical horizons. Section~\ref{s4}
is devoted to the laws of black hole mechanics. We outline the main
ideas using both isolated and dynamical horizons. In the next three 
sections we review applications. Section~\ref{s5} summarizes
applications to numerical relativity, Section~\ref{s6} to black
holes with hair, and Section~\ref{s7} to the quantum entropy
calculation. Section~\ref{s8} discusses open issues and directions
for future work. \emph{Having read Section~\ref{s2},
  Sections~\ref{s3}, \ref{s4}, \ref{s5}, \ref{s6}, and~\ref{s7} are
  fairly self contained and the three applications can be read
  independently of each other.}

All manifolds will be assumed to be $C^{k+1}$ (with $k\ge 3$) and
orientable, the space-time metric will be $C^k$, and matter fields
$C^{k-2}$. For simplicity we will restrict ourselves to
4-dimensional space-time manifolds $\mathcal{M}$ (although most of the
classical results on isolated horizons have been extended to
3-dimensions space-times~\cite{adw}, as well as higher dimensional
ones~\cite{lp2}). The space-time metric $g_{ab}$ has signature
$(-,+,+,+)$ and its derivative operator will be denoted by
$\nabla$. The Riemann tensor is defined by $R_{abc}{}^d W_d := 2
\nabla_{[a} \nabla_{b]}W_c$, the Ricci tensor by $R_{ab} :=
R_{acb}{}^c$, and the scalar curvature by $R := g^{ab} R_{ab}$. We
will assume the field equations
\begin{equation}
  \label{fe}
  R_{ab}- \frac{1}{2}R\, g_{ab} + \Lambda g_{ab} = 8\pi G T_{ab}.
\end{equation}
(With these conventions, de Sitter space-time has positive
cosmological constant $\Lambda$.) We assume that $T_{ab}$
satisfies the dominant energy condition (although, as the reader
can easily tell, several of the results will hold under weaker
restrictions.) Cauchy (and partial Cauchy) surfaces will be denoted
by $M$, isolated horizons by $\Delta$, and dynamical horizons by $H$.

\newpage


\section{Basic Notions}
\label{s2}

This section is divided into two parts. The first introduces isolated
horizons, and the second dynamical horizons.


\subsection{Isolated horizons}
\label{s2.1}

In this part, we provide the basic definitions and discuss
geometrical properties of non-expanding, weakly isolated, and
isolated horizons which describe black holes which are in
equilibrium in an increasingly stronger sense.

These horizons model black holes which are themselves in
equilibrium, but in possibly dynamical space-times~\cite{abf1,
abf2, afk, abl1}. For early references with similar ideas,
see~\cite{newman, hajicek}. A useful example is provided by the late
stage of a gravitational collapse shown in Figure~\ref{exam}. In
such physical situations, one expects the back-scattered radiation
falling into the black hole to become negligible at late times so
that the `end portion' of the event horizon (labelled by $\Delta$
in the figure) can be regarded as isolated to an excellent
approximation. This expectation is borne out in numerical
simulations where the backscattering effects typically become
smaller than the numerical errors rather quickly after the
formation of the black hole (see, e.g., \cite{whiskey, bks}).


\subsubsection{Definitions}
\label{s2.1.1}

The key idea is to extract from the notion of a Killing horizon
the minimal conditions which are necessary to define physical
quantities such as the mass and angular momentum of the black hole
and to establish the zeroth and the first laws of black hole
mechanics. Like Killing horizons, isolated horizons are null,
3-dimensional sub-manifolds of space-time. Let us therefore begin
by recalling some essential features of such sub-manifolds, which
we will denote by $\Delta$. The intrinsic metric $q_{ab}$ on
$\Delta$ has signature (0,+,+), and is simply the pull-back of the
space-time metric to $\Delta$, $q_{ab}=\pullbacklong{g_{ab}}$, where an
underarrow indicates the pullback to $\Delta$. Since
$q_{ab}$ is degenerate, it does not have an inverse in the standard
sense. However, it does admit an inverse in a weaker sense:
$q^{ab}$ will be said to be an inverse of $q_{ab}$ if it satisfies
$q_{am}q_{bn} q^{mn} = q_{ab}$. As one would expect, the inverse
is not unique: We can always add to $q^{ab}$ a term of the type
$\ell^{(a} V^{b)}$, where $\ell^a$ is a null normal to $\Delta$
and $V^b$ any vector field tangential to $\Delta$. All our
constructions will be insensitive to this ambiguity. Given a null
normal $\ell^a$ to $\Delta$, the expansion $\Theta_{{(\ell)}}$ is
defined as
\begin{equation}
  \Theta_{(\ell)} := q^{ab}\nabla_a\ell_b.
\end{equation}
(Throughout this review, we will assume that $\ell^a$ is future
directed.) We can now state the first definition:

\begin{definition}
  \label{def_1}
  A sub-manifold $\Delta$ of a space-time $(\mathcal{M},g_{ab})$ is
  said to be a \emph{non-expanding horizon} (NEH) if
  
  \begin{enumerate}
  \item $\Delta$ is topologically $S^2\times \mathbb{R}$ and null;
    \label{cond_1_1}
  \item any null normal $\ell^a$ of $\Delta$ has vanishing
    expansion, $\Theta_{(\ell)}=0$; and
    \label{cond_1_2}
  \item all equations of motion hold at $\Delta$ and the
    stress energy tensor $T_{ab}$ is such that $-T_b^a\ell^b$ is
    future-causal for any future directed null normal $\ell^a$.
    \label{cond_1_3}
  \end{enumerate}
\end{definition}

The motivation behind this definition can be summarized as
follows. Condition~\ref{cond_1_1} is imposed for definiteness; while most
geometric results are insensitive to topology, the $S^2\times
\mathbb{R}$ case is physically the most relevant
one. Condition~\ref{cond_1_3} is
satisfied by all classical matter fields of direct physical
interest. The key condition in the above definition is
Condition~\ref{cond_1_2} which
is equivalent to requiring that every cross-section of $\Delta$ be
\emph{marginally trapped}. (Note incidentally that if $\Theta_{(\ell)}$
vanishes for one null normal $\ell^a$ to $\Delta$, it vanishes for
all.) Condition~\ref{cond_1_2} is equivalent to requiring that the
\emph{infinitesimal area element} is Lie dragged by the null
normal $\ell^a$. In particular, then, Condition~\ref{cond_1_2} implies that
the horizon area is `constant in time'. We will denote the area of
any cross section of $\Delta$ by $a_\Delta$ and define the horizon
radius as $R_\Delta:=\sqrt{a_\Delta/4\pi}$.

Because of the Raychaudhuri equation, Condition~\ref{cond_1_2} also
implies
\begin{equation}
  R_{ab}\ell^a\ell^b + \sigma_{ab}\sigma^{ab} = 0,
\end{equation}
where $\sigma^{ab}$ is the shear of $\ell_a$, defined by
$\sigma_{ab} := \pullback{\nabla_{(a} \ell_{b)}} - \frac{1}{2}
\Theta_{{(\ell)}} q_{ab}$, where the underarrow denotes `pull-back to
$\Delta$'. Now the energy condition~\ref{cond_1_3} implies that
$R_{ab}\ell^a \ell^b$ is non-negative, whence we conclude that
each of the two terms in the last equation vanishes. This in turn
implies that $\pullbackllong{T_{ab}}\ell^b =0$ \emph{and}
$\pullbackllong{\nabla_{(a}\ell_{b)}}=0$ on $\Delta$. The first of these
equations constrains the matter fields on $\Delta$ in an
interesting way, while the second is equivalent to $\mathcal{L}_\ell
q_{ab} =0$ on $\Delta$. Thus, the intrinsic metric on an NEH is
`time-independent'; this is the sense in which an NEH is in
equilibrium.

The zeroth and first laws of black hole mechanics require an
additional structure, which is provided by the concept of a weakly
isolated horizon. To arrive at this concept, let us first
introduce a derivative operator $\mathcal{D}$ on $\Delta$. Because $q_{ab}$
is degenerate, there is an infinite number of (torsion-free)
derivative operators which are compatible with it. However, on an
NEH, the property $\pullbackllong{\nabla_{(a} \ell_{b)}} =0$ implies that
the space-time connection $\nabla$ induces a unique (torsion-free)
derivative operator $\mathcal{D}$ on $\Delta$ which is compatible with
$q_{ab}$~\cite{afk, bkthesis}. Weakly isolated horizons are
characterized by the
property that, in addition to the metric $q_{ab}$, the connection
component $\mathcal{D}_a\ell^b$  is also `time independent'.

Two null normals $\ell^a$ and $\tilde{\ell}^a$ to an NEH $\Delta$
are said to belong to the same equivalence class $[\ell ]$ if
$\tilde\ell^a = c\ell^a$ for some positive \emph{constant} $c$.
Then, weakly isolated horizons are defined as follows:

\begin{definition}
  \label{def_2}
  The pair $(\Delta, [\ell])$ is said to constitute a \emph{weakly
  isolated horizon} (WIH) provided $\Delta$ is an NEH and each null
  normal $\ell^a$ in $[\ell]$ satisfies
  \begin{equation}
    \label{wihcond}
    (\mathcal{L}_\ell \mathcal{D}_a -
    \mathcal{D}_a\mathcal{L}_\ell )\ell^b = 0.
  \end{equation}
\end{definition}

It is easy to verify that every NEH admits null normals
satisfying Equation~(\ref{wihcond}), i.e., can be made a WIH with a suitable
choice of $[\ell]$. However the required equivalence class is not
unique, whence an NEH admits distinct WIH structures~\cite{abl1}.

Compared to conditions required of a Killing horizon, conditions
in this definition are very weak. Nonetheless, it turns out that
they are strong enough to capture the notion of a black hole in
equilibrium in applications ranging from black hole mechanics to
numerical relativity. (In fact, many of the basic notions such as
the mass and angular momentum are well-defined already on NEHs
although intermediate steps in derivations use a WIH structure.)
This is quite surprising at first because the laws of black hole
mechanics were traditionally proved for \emph{globally stationary
black holes}~\cite{rwreview}, and the definitions of mass and
angular momentum of a black hole first used in numerical
relativity implicitly assumed that the near horizon geometry is
\emph{isometric to Kerr}~\cite{earlynr1}.

Although the notion of a WIH is sufficient for most applications,
from a geometric viewpoint, a stronger notion of isolation is more
natural: The \emph{full} connection $\mathcal{D}$ should be
time-independent. This leads to the notion of an isolated horizon.

\begin{definition}
  \label{def_3}
  A WIH $(\Delta, [\ell])$ is said to constitute an \emph{isolated
  horizon} (IH) if
  \begin{equation}
    \label{ihcond}
    (\mathcal{L}_\ell \mathcal{D}_a -
    \mathcal{D}_a\mathcal{L}_\ell ) V^b = 0
  \end{equation}
  for arbitrary vector fields $V^a$ tangential to $\Delta$. \\
\end{definition}

While an NEH can always be given a WIH structure simply by making a
suitable choice of the null normal, not every WIH admits an IH
structure. Thus, the passage from a WIH to an IH is a genuine
restriction~\cite{abl1}. However, even for this stronger notion of
isolation,
conditions in the definition are \emph{local to $\Delta$}.
Furthermore, the definition only uses quantities \emph{intrinsic}
to $\Delta$; there are no restrictions on components of any fields
transverse to $\Delta$. (Even the full 4-metric $g_{ab}$ need not
be time independent on the horizon.) Robinson--Trautman solutions
provide explicit examples of isolated horizons which do not admit
a stationary Killing field even in an arbitrarily small
neighborhood of the horizon~\cite{pc}. In this sense, the
conditions in this definition are also rather weak. One expects
them to be met to an excellent degree of approximation in a wide
variety of situations representing late stages of gravitational
collapse and black hole mergers\epubtkFootnote{However,
  Condition~(\ref{ihcond}) may be too strong in some problems, e.g.,
  in the construction of quasi-equilibrium initial data sets, where
  the notion of WIH is more useful~\cite{jgm} (see
  Section~\ref{s5.2}).}.


\subsubsection{Examples}
\label{s2.1.2}

The class of space-times admitting NEHs, WIHs, and IHs is quite
rich. First, it is trivial to verify that any Killing horizon
which is topologically $S^2\times\mathbb{R}$ is also an isolated
horizon. This in particular implies that the event horizons of all
globally stationary black holes, such as the Kerr--Newman solutions
(including a possible cosmological constant), are isolated
horizons. (For more exotic examples, see~\cite{plj}.)

\epubtkImage{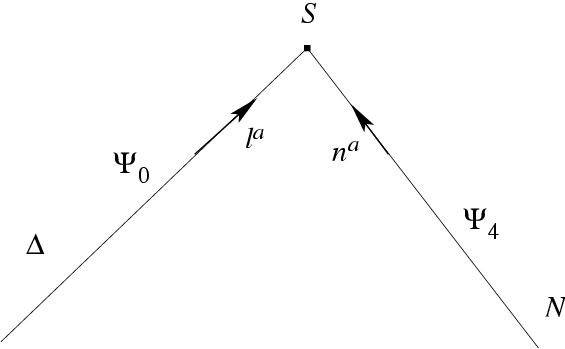}
{\begin{figure}[hptb]
   \def\epsfsize#1#2{0.6#1}
   \centerline{\epsfbox{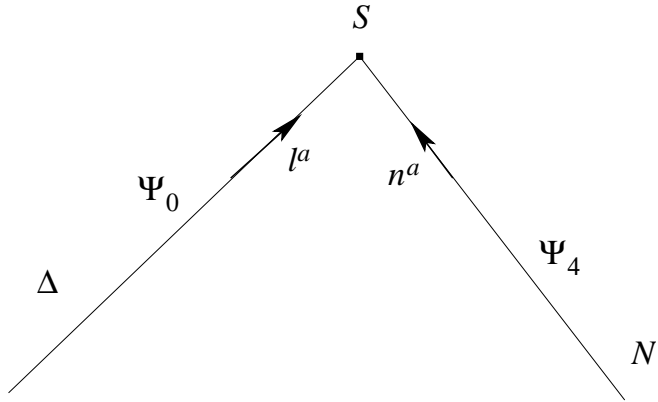}}
   \caption{\it Set-up of the general characteristic initial value
     formulation. The Weyl tensor component $\Psi_0$ on the null
     surface $\Delta$ is part of the free data which vanishes if
     $\Delta$ is an IH.}
   \label{fig:friedrich}
 \end{figure}
}

But there exist other non-trivial examples as well. These arise
because the notion is \emph{quasi-local}, referring only to fields
defined intrinsically on the horizon. First, let us consider the
sub-family of Kastor--Traschen solutions~\cite{kt, kt1} which are
asymptotically de Sitter and admit event horizons. They are
interpreted as containing multiple charged, dynamical black holes
in presence of a positive cosmological constant. Since these
solutions do not appear to admit any stationary Killing fields, no
Killing horizons are known to exist. Nonetheless, the event
horizons of individual black holes are WIHs. However, to our
knowledge, no one has checked if they are IHs. A more striking
example is provided by a sub-family of Robinson--Trautman
solutions, analyzed by Chrusciel~\cite{pc}. These space-times
admit IHs whose intrinsic geometry is isomorphic to that of the
Schwarzschild isolated horizons but in which there is radiation
\emph{arbitrarily close} to $\Delta$.

More generally, using the characteristic initial value
formulation~\cite{hf, ar}, Lewandowski~\cite{jl} has constructed an
infinite dimensional set of local examples. Here, one considers two null
surfaces $\Delta$ and $\mathcal{N}$ intersecting in a 2-sphere
$S$ (see Figure~\ref{fig:friedrich}). One can freely specify
certain data on these two surfaces which then determines a
solution to the vacuum Einstein equations in a neighborhood of $S$
bounded by $\Delta$ and $\mathcal{N}$, in which $\Delta$ is an
isolated horizon.


\subsubsection{Geometrical properties}
\label{s2.1.3}

\begin{description}
\item[Rescaling freedom in {\boldmath $\ell^a$}]~\\
  As we remarked in
Section~\ref{s2.1.1}, there is a functional rescaling freedom in
the choice of a null normal on an NEH and, while the choice of null
normals is restricted by the weakly isolated horizon
condition~(\ref{wihcond}), considerable freedom still remains. That
is, a given NEH $\Delta$ admits an infinite number of WIH structures
$(\Delta, [\ell ])$~\cite{abl1}.

On IHs, by contrast, the situation is dramatically different.
Given an IH $(\Delta, [\ell])$, generically the
Condition~(\ref{ihcond}) in Definition~\ref{def_3} can not be
satisfied by a \emph{distinct} equivalence class of null normals $[\ell^\prime
]$. Thus on a generic IH, the only freedom in the choice of the
null normal is that of a rescaling by a positive
constant~\cite{abl1}. This freedom mimics the properties of a Killing
horizon since one can also rescale the Killing vector by an
arbitrary constant. \emph{The triplet $(q_{ab}, \mathcal{D}_a, [\ell^a])$
is said to constitute
the geometry of the isolated horizon}.
\end{description}

\begin{description}
\item[Surface gravity]~\\
  Let us begin by defining a
1-form $\omega_a$ which will be used repeatedly. First note that,
by Definition~\ref{def_1}, $\ell^a$ is expansion free and shear free. It is
automatically twist free since it is a normal to a smooth
hypersurface. This means that the contraction of $\nabla_a\ell_b$
with any two vectors tangent to $\Delta$ is identically zero,
whence there must exist a 1-form $\omega_a$ on $\Delta$ such that
for any $V^a$ tangent to $\Delta$,
\begin{equation}
  \label{eq:omegadef}
  V^a\nabla_a\ell^b = V^a\omega_a\ell^b.
\end{equation}
Note that the WIH condition~(\ref{wihcond}) requires simply that
$\omega_a$ be time independent, ${\mathcal L}_\ell \omega_a =0$.
Given $\omega_a$, the \emph{surface gravity} $\kappa_{(\ell)}$
associated with a null normal $\ell^a$ is defined as
\begin{equation}
  \label{eq:surfacegravity}
  \kappa_{(\ell)} := \ell^a\omega_a.
\end{equation}
Thus, $\kappa_{(\ell)}$ is simply the acceleration of $\ell^a$. Note
that the surface gravity is not an intrinsic property of a WIH
$(\Delta, [\ell])$. Rather, it is a property of a null normal to
$\Delta$: $\kappa_{(c\ell)} = c \kappa_{(\ell)}$. An isolated
horizon with $\kappa_{{(\ell)}} =0$ is said to be an \emph{extremal}
isolated horizon. Note that while the value of surface gravity
refers to a specific null normal, whether a given WIH is extremal
or not is insensitive to the permissible rescaling of the normal.
\end{description}

\begin{description}
\item[Curvature tensors on {\boldmath $\Delta$}]~\\
  Consider any
(space-time) null tetrad $(\ell^a, n^a, m^a, \bar{m}^a)$ on
$\Delta$ such that $\ell^a$ is a null normal to $\Delta$. Then, it
follows from Definition~\ref{def_1} that two of the Newman--Penrose
Weyl components vanish on $\Delta$: $\Psi_0 := C_{abcd} \ell^am^b\ell^c
m^d =0$ and $\Psi_1 := C_{abcd}\ell^am^b\ell^cn^d=0$. This in turn
implies that $\Psi_2 := C_{abcd} \ell^a m^b \bar{m}^c n^d$ is
gauge invariant (i.e., does not depend on the specific choice of
the null tetrad satisfying the condition stated above.) The
imaginary part of $\Psi_2$ is related to the curl of $\omega_a$,
\begin{equation}
  \label{domega}
  d\omega = 2 \left(\mathrm{Im}\Psi_2\right) \epsilon,
\end{equation}
where $\epsilon_{ab}$ is the natural area 2-form on $\Delta$.
Horizons on which $\imaginary \Psi_2$ vanishes are said to be
\emph{non-rotating}: On these horizons all angular momentum
multipoles vanish~\cite{aepv}. Therefore, $\mathrm{Im} \Psi_2$ is
sometimes referred to as the \emph{rotational scalar} and
$\omega_a$ as the \emph{rotation 1-form} of the horizon.

Next, let us consider the Ricci-tensor components. On any NEH
$\Delta$ we have: $\Phi_{00} := \frac{1}{2} R_{ab} \ell^a \ell^b
:=0$, $\Phi_{01} := \frac{1}{2}R_{ab}\ell^a m^b =0$. In the
Einstein--Maxwell theory, one further has: On $\Delta$, $\Phi_{02}
:= \frac{1}{2} R_{ab}m^a m^b =0$ and $\Phi_{20} := \frac{1}{2}
R_{ab} \bar{m}^a\bar{m}^b =0$.
\end{description}

\begin{description}
\item[Free data on an isolated horizon]~\\
  Given the geometry $(q_{ab},\mathcal{D}, [\ell])$ of an IH, it is
  natural to ask for the minimum amount of information, i.e., the free
  data, required to construct it. This question has been answered in
  detail (also for WIHs)~\cite{abl1}. For simplicity, here we will
  summarize the results only for the non-extremal case. (For the
  extremal case, see~\cite{abl1, lp3}.)

Let $S$ be a spherical cross section of $\Delta$. The degenerate
metric $q_{ab}$ naturally projects to a Riemannian metric
$\tilde{q}_{ab}$ on $S$, and similarly the 1-form $\omega_a$ of
Equation~(\ref{eq:omegadef}) projects to a 1-form
$\tilde{\omega}_a$ on $S$. If the vacuum equations hold on
$\Delta$, then given $(\tilde{q}_{ab},\tilde{\omega}_a)$ on $S$,
there is, up to diffeomorphisms, a unique \emph{non-extremal}
isolated horizon geometry $(q_{ab},\mathcal{D}, [\ell])$ such that
$\tilde{q}_{ab}$ is the projection of $q_{ab}$, $\tilde{\omega}_a$
is the projection of the $\omega_a$ constructed from $\mathcal{D}$, and
$\kappa_{{(\ell)}} =\omega_a\ell^a\neq 0$. (If the vacuum equations do
not hold, the additional data required is the projection on $S$ of
the space-time Ricci tensor.)

The underlying reason behind this result can be sketched as
follows. First, since $q_{ab}$ is degenerate along $\ell^a$, its
non-trivial part is just its projection $\tilde{q}_{ab}$. Second,
$\tilde{q}_{ab}$ fixes the connection on $S$; it is only the
quantity $S_{ab}:=\mathcal{D}_an_b$ that is not constrained by
$\tilde{q}_{ab}$, where $n_a$ is a 1-form on $\Delta$ orthogonal
to $S$, normalized so that $\ell^an_a=-1$. It is easy to show that
$S_{ab}$ is symmetric and the contraction of one of its indices
with $\ell^a$ gives $\omega_a$: $\ell^aS_{ab}=\omega_a$.
Furthermore, it turns out that if $\omega_a\ell^a\neq 0$, the
field equations completely determine the angular part of $S_{ab}$
in terms of $\tilde{\omega}_a$ and $\tilde{q}_{ab}$. Finally,
recall that the surface gravity is not fixed on $\Delta$ because
of the rescaling freedom in $\ell^a$; thus the $\ell$-component of
$\omega_a$ is not part of the free data. Putting all these facts
together, we see that the pair $(\tilde{q}_{ab},\tilde{\omega}_a)$
enables us to reconstruct the isolated horizon geometry uniquely
up to diffeomorphisms.
\end{description}

\begin{description}
\item[Rest frame of a non-expanding horizon]~\\
  As at null infinity, a preferred foliation of $\Delta$ can be thought of
as providing a `rest frame' for an isolated horizon. On the
Schwarzschild horizon, for example, the 2-spheres on which the
Eddington--Finkelstein advanced time coordinate is constant -- which
are also integral manifolds of the rotational Killing
fields -- provide such a rest frame. For the Kerr metric, this
foliation generalizes naturally. The question is whether a general
prescription exists to select such a preferred foliation.

On any non-extremal NEH, the 1-form $\omega_a$ can be used to
construct preferred foliations of $\Delta$. Let us first examine
the simpler, non-rotating case in which $ \imaginary \Psi_2 = 0$.
Then Equation~(\ref{domega}) implies that $\omega_a$ is \emph{curl-free}
and therefore hypersurface orthogonal. The 2-surfaces orthogonal
to $\omega_a$ must be topologically $S^2$ because, on any
non-extremal horizon, $\ell^a\omega_a\neq 0$. Thus, in the
non-rotating case, every isolated horizon comes equipped with a
preferred family of cross-sections which defines the rest frame~\cite{afk}. Note that the projection $\tilde\omega_a$ of
$\omega_a$ on any leaf of this foliation vanishes identically.

The rotating case is a little more complicated since $\omega_a$ is
then no longer curl-free. Now the idea is to exploit the fact that
the \emph{divergence} of the projection $\tilde\omega_a$ of
$\omega_a$ on a cross-section is sensitive to the choice of the
cross-section, and to select a preferred family of cross-sections by
imposing a suitable condition on this divergence~\cite{abl1}. A
mathematically natural choice is to ask that this divergence
vanish.
However, (in the case when the angular momentum is
non-zero) this condition does not pick out the $v = \mathrm{const.}$
cuts of the Kerr horizon where $v$ is the (Carter generalization
of the) Eddington--Finkelstein coordinate. Pawlowski has provided
another condition that also selects a preferred foliation and
reduces to the $v = \mathrm{const.}$ cuts of the Kerr horizon:
\begin{equation}
  \diver \tilde{\omega} = - \tilde\Delta
  \ln\left|\Psi_2\right|^{1/3}\!\!\!\!\!\!\!\!,
\end{equation}
where $\tilde\Delta$ is the Laplacian of $\tilde{q}_{ab}$.
On isolated horizons on which $\left|\Psi_2\right|$ is nowhere zero
-- a condition satisfied if the horizon geometry is `near' that of
the Kerr isolated horizon -- this selects a preferred foliation
and hence a rest frame. This construction is potentially useful to
numerical relativity.
\end{description}

\begin{description}
\item[Symmetries of an isolated horizon]~\\
  By definition, a symmetry of an IH $(\Delta, [\ell])$ is a diffeomorphism of
$\Delta$ which preserves the geometry
$(q_{ab},\mathcal{D},[\ell])$. (On a WIH, the symmetry has to
  preserve $(q_{ab}, \omega_a, [\ell])$. There are again three
  universality classes of symmetry groups as on an IH.)
Let us denote the symmetry group by $G_\Delta$. First note that
diffeomorphisms generated by the null normals in $[\ell^a]$ are
symmetries; this is already built into the very definition of an
isolated horizon. The other possible symmetries are related to
the cross-sections of $\Delta$. Since we have assumed the
cross-sections to be topologically spherical and since a metric on
a sphere can have either exactly three, one or zero Killing
vectors, it follows that $G_\Delta$ can be of only three
types~\cite{abl2}:

\begin{itemize}
\item Type I: The pair $(q_{ab}, \mathcal{D}_a)$ is spherically symmetric;
  $G_\Delta$ is four dimensional.
\item Type II: The pair $(q_{ab}, \mathcal{D}_a)$ is axisymmetric;
  $G_\Delta$ is two dimensional.
\item Type III: Diffeomorphisms generated by $\ell^a$ are the
  only symmetries of the pair $(q_{ab}, \mathcal{D}_a)$; $G_\Delta$ is one
  dimensional.
\end{itemize}

In the asymptotically flat context, boundary conditions select a
\emph{universal} symmetry group at spatial infinity, e.g., the
Poincar\'e group, because the space-time metric approaches a fixed
Minkowskian one. The situation is completely different in the
strong field region near a black hole. Because the geometry at the
horizon can vary from one space-time to another, the symmetry
group is not universal. However, the above result shows that the
symmetry group can be one of only three universality classes.
\end{description}


\subsection{Dynamical horizons}
\label{s2.2}

This section is divided into three parts. In the first, we
discuss basic definitions, in the second we introduce an explicit
example, and in the third we analyze the issue of uniqueness of
dynamical horizons and their role in numerical relativity.


\subsubsection{Definitions}
\label{s2.2.1}

To explain the evolution of ideas and provide points of
comparison, we will introduce the notion of dynamical horizons
following a chronological order. Readers who are not familiar with
causal structures can go directly to Definition~\ref{def_5} of dynamical
horizons (for which a more direct motivation can be found in~\cite{ak2}).

As discussed in Section~\ref{s1}, while the notion of an event
horizon has proved to be very convenient in mathematical
relativity, it is too global and teleological to be directly
useful in a number of physical contexts ranging from quantum
gravity to numerical relativity to astrophysics. This limitation
was recognized early on (see, e.g., \cite{he}, page 319) and
alternate notions were introduced to capture the intuitive idea of
a black hole in a quasi-local manner. In particular, to make the
concept `local in time', Hawking~\cite{swh, he} introduced the
notions of a trapped region and an apparent horizon, both of which
are associated to a space-like 3-surface $M$ representing `an
instant of time'. Let us begin by recalling these ideas.

Hawking's  \emph{outer trapped surface $S$} is a compact,
space-like 2-dimensional sub-manifold in $(\mathcal{M}, g_{ab})$ such
that the expansion $\Theta_{(\ell)}$ of the outgoing null normal
$\ell^a$ to $S$ is non-positive. Hawking then defined the
\emph{trapped region $\mathcal{T}(M)$ in a surface $M$} as the set
of all points in $M$ through which there passes an outer-trapped
surface, lying entirely in $M$. Finally, Hawking's \emph{apparent
horizon} $\partial \mathcal{T}(M)$ is the boundary of a connected
component of $\mathcal{T}(M)$. The idea then was to regard each
apparent horizon as the instantaneous surface of a black hole. One
can calculate the expansion $\Theta_{(\ell)}$ of $S$ knowing only
the intrinsic 3-metric $q_{ab}$ and the extrinsic curvature
$K_{ab}$ of $M$. Hence, to find outer trapped surfaces and
apparent horizons on $M$, one does not need to evolve $(q_{ab},
K_{ab})$ away from $M$ even locally. In this sense the notion is
local to $M$. However, this locality is achieved at the price of
restricting $S$ to lie in $M$. If we wiggle $M$ even slightly, new
outer trapped surfaces can appear and older ones may disappear. In
this sense, the notion is still very global. Initially, it was
hoped that the laws of black hole mechanics can be extended to
these apparent horizons. However, this has not been possible
because the notion is so sensitive to the choice of $M$.

To improve on this situation, in the early nineties Hayward proposed
a novel modification of this framework~\cite{sh}. The main idea is
to free these notions from the complicated dependence on $M$. He
began with Penrose's notion of a trapped surface. A \emph{trapped
surface} $S$ a la Penrose is a compact, space-like 2-dimensional
sub-manifold of space-time on which $\Theta_{(\ell)}
\Theta_{(n)}>0$, where $\ell^a$ and $n^a$ are the two null normals
to $S$. We will focus on \emph{future} trapped surfaces on which
both expansions are negative. Hayward then defined a
\emph{space-time} trapped region. A \emph{trapped region}
$\mathcal{T}$ a la Hayward is a subset of space-time through each
point of which there passes a trapped surface. Finally, Hayward's
\emph{trapping boundary} $\partial\mathcal{T}$ is a connected
component of the boundary of an inextendible trapped region. Under
certain assumptions (which appear to be natural intuitively but
technically are quite strong), he was able to show that the trapping
boundary is foliated by \emph{marginally trapped surfaces} (MTSs),
i.e., compact, space-like 2-dimensional sub-manifolds on which the
expansion of one of the null normals, say $\ell^a$, vanishes and that
of the other, say $n^a$, is everywhere non-positive. Furthermore,
$\mathcal{L}_{n} \Theta_{(\ell)}$ is also everywhere of one
sign. These general considerations led him to define a quasi-local
analog of future event horizons as follows:

\begin{definition}
  \label{def_4}
  A \emph{future, outer, trapping horizon} (FOTH) is a smooth
  3-dimensional sub-manifold $\underbar{\it H}$ of space-time, foliated by
  closed 2-manifolds $\underbar{\it S}$, such that
  \begin{enumerate}
  \item the expansion of one future directed null normal to the
    foliation, say $\ell^a$, vanishes, $\Theta_{(\ell)} =0$;
    \label{cond_4_1}
  \item the expansion of the other future directed null normal $n^a$
    is negative, $\Theta_{(n)}<0$; and
    \label{cond_4_2}
  \item the directional derivative of $\Theta_{(\ell)}$ along $n^a$ is
    negative, $\mathcal{L}_{n}\, \Theta_{(\ell)} <0$.
    \label{cond_4_3}
  \end{enumerate}
\end{definition}

In this definition, Condition~\ref{cond_4_2} captures the idea that
$\underbar{\it H}$ is a future horizon (i.e., of black hole rather than
white hole type), and Condition~\ref{cond_4_3} encodes the idea that it is
`outer' since infinitesimal motions along the `inward' normal
$n^a$ makes the 2-surface trapped. (Condition~\ref{cond_4_3} also serves to
distinguish black hole type horizons from certain cosmological
ones~\cite{sh} which are not ruled out by
Condition~\ref{cond_4_2}). Using the
Raychaudhuri equation, it is easy to show that $\underbar{\it H}$ is either
space-like or null, being null if and only if the shear
$\sigma_{ab}$ of $\ell^a$ as well as the matter flux
$T_{ab}\ell^a\ell^b$ across $H$ vanishes. Thus, when $\underbar{\it H}$ is
null, it is a non-expanding horizon introduced in
Section~\ref{s2.1}. Intuitively, $\underbar{\it H}$ is space-like in the
dynamical region where gravitational radiation and matter fields are
pouring into it and is null when it has reached equilibrium.

In truly dynamical situations, then, $\underbar{\it H}$ is expected to be
space-like. Furthermore, it turns out that most of the key results
of physical interest~\cite{ak1, ak2}, such as the area increase law
and generalization of black hole mechanics, do not require the
condition on the sign of  $\mathcal{L}_n \Theta_{(\ell)}$. It is
therefore convenient to introduce a simpler and at the same time
`tighter' notion, that of a dynamical horizon, which is better
suited to analyze how black holes grow in exact general
relativity~\cite{ak1, ak2}:

\begin{definition}
  \label{def_5}
  A smooth, three-dimensional, space-like sub-manifold (possibly with
  boundary) $H$ of space-time is said to be a \emph{dynamical horizon}
  (DH) if it can be foliated by a family of closed 2-manifolds such
  that
  \begin{enumerate}
  \item on each leaf $S$ the expansion $\Theta_{(\ell)}$ of one null
    normal $\ell^a$ vanishes; and
  \item the expansion $\Theta_{(n)}$ of the other null normal $n^a$ is
    negative.
  \end{enumerate}
\end{definition}

\noindent
Note first that, like FOTHs, dynamical horizons are
`space-time notions', defined quasi-locally. They are not defined
relative to a space-like surface as was the case with Hawking's
apparent horizons nor do they make any reference to infinity as is
the case with event horizons. In particular, they are well-defined
also in the spatially compact context. Being quasi-local, they are
not teleological. Next, let us spell out the relation between
FOTHs and DHs. A \emph{space-like} FOTH is a DH on which the
additional condition $\mathcal{L}_{n}\, \Theta_{(\ell)} <0$ holds.
Similarly, a DH \emph{satisfying} $\mathcal{L}_{n}\, \Theta_{(\ell)}
<0$ is a space-like FOTH. Thus, while neither definition implies
the other, the two are closely related. The advantage of
Definition~\ref{def_5} is that it refers only to the intrinsic structure of
$H$, without any conditions on the evolution of fields in
directions transverse to $H$. Therefore, it is easier to verify in
numerical simulations. More importantly, as we will see, this
feature makes it natural to analyze the structure of $H$ using
only the constraint (or initial value) equations on it. This
analysis will lead to a wealth of information on black hole
dynamics. Reciprocally, Definition~\ref{def_4} has the advantage that, since
it permits $\underbar{\it H}$ to be space-like \emph{or} null, it is better
suited to analyze the transition to equilibrium~\cite{ak2}.

A DH which is also a FOTH will be referred to as a
\emph{space-like future outer horizon} (SFOTH). To fully capture
the physical notion of a \emph{dynamical} black hole, one should
require both sets of conditions, i.e., restrict oneself to SFOTHs.
For, stationary black holes admit FOTHS and there exist
space-times~\cite{jmms} which admit dynamical horizons but no
trapped surfaces; neither can be regarded as containing a
dynamical black hole. However, it is important to keep track of
precisely which assumptions are needed to establish specific
results. Most of the results reported in this review require only
those conditions which are satisfied on DHs. This fact may well
play a role in conceptual issues that arise while generalizing
black hole thermodynamics to non-equilibrium
situations\epubtkFootnote{Indeed, the situation is similar for black holes
  in equilibrium. While it is physically reasonable to restrict
  oneself to IHs, most results require only the WIH boundary
  conditions. The distinction can be important in certain
  applications, e.g., in finding boundary conditions on the
  quasi-equilibrium initial data at inner horizons.}.


\subsubsection{Examples}
\label{s2.2.2}

Let us begin with the simplest examples of space-times admitting
DHs (and SFOTHs). These are provided by the spherically symmetric
solution to Einstein's equations with a null fluid as source, the
Vaidya metric~\cite{pcv, kuroda, lw}. (Further details and the
inclusion of a cosmological constant are discussed  in~\cite{ak2}.)
Just as the Schwarzschild--Kruskal solution provides a great deal of
intuition for general static black holes, the Vaidya
metric furnishes some of the much needed intuition in the
dynamical regime by bringing out the key differences between the
static and dynamical situations. However, one should bear in mind
that both Schwarzschild and Vaidya black holes are the simplest
examples and certain aspects of geometry can be much more
complicated in more general situations. The 4-metric of the Vaidya
space-time is given by
\begin{equation}
  g_{ab} = -\left(1-\frac{2GM(v)}{r}\right)\nabla_av\nabla_bv +
  2\nabla_{(a}v\nabla_{b)}r + r^2\left(\nabla_a\theta\nabla_b\theta +
  \sin^2\theta\nabla_a \phi\nabla_b\phi\right),
\end{equation}
where $M(v)$ is any smooth, non-decreasing function of $v$. Thus,
$(v, r,\theta,\phi)$ are the ingoing Eddington--Finkelstein
coordinates. This is a solution of Einstein's equations, the
stress-energy tensor $T_{ab}$ being given by
\begin{equation}
  T_{ab} = \frac{\dot{M}(v)}{4\pi r^2}\nabla_av\nabla_bv,
\end{equation}
where $\dot{M}=dM/dv$. Clearly, $T_{ab}$ satisfies the dominant
energy condition if $\dot{M} \geq 0$, and vanishes if and only if
$\dot{M} =0$. Of special interest to us are the cases illustrated in
Figure~\ref{fig:vaidya}: $M(v)$ is non-zero until a certain finite retarded
time, say $v=0$, and then grows monotonically, either reaching an
asymptotic value $M_0$ as $v$ tends to infinity (panel a), or,
reaching this value at a finite retarded time, say $v=v_0$, and
then remaining constant (panel b). In either case, the
space-time region $v \le 0$ is \emph{flat}.

Let us focus our attention on the metric 2-spheres, which are all
given by $v =\mathrm{const.} $ and $ r= \mathrm{const.}$. It is easy to verify
that the expansion of the outgoing null normal $\ell^a$ vanishes if
and only if ($v =\mathrm{const.}$ and) $r = 2GM(v)$. Thus, these are
the \emph{only spherically symmetric marginally trapped surfaces}
MTSs. On each of them, the expansion $\Theta_{(n)}$ of the ingoing
normal $n^a$ is negative. By inspection, the 3-metric on the world
tube $r= 2GM(v)$ of these MTSs has signature $(+,+,+)$ when
$\dot{M}(v)$ is non-zero and $(0,+,+)$ if $\dot{M}(v)$ is zero. Hence,
in the left panel of Figure~\ref{fig:vaidya} the surface $r=2GM(v)$ is
the DH $H$. In the right panel of Figure~\ref{fig:vaidya}
the portion of this surface $v\le v_0$ is the DH $H$, while the
portion $v\ge v_0$ is a non-expanding horizon. (The general issue
of transition of a DH to equilibrium is briefly discussed in
Section~\ref{s5}.) Finally, note that at these MTSs,
$\mathcal{L}_{{n}}\theta_{(\ell)} = -2/r^2 < 0$. Hence in both cases, the
DH is an SFOTH. Furthermore, in the case depicted in the right panel
of Figure~\ref{fig:vaidya} the \emph{entire} surface $r= 2GM(v)$ is a
FOTH, part of which is dynamical and part null.

\epubtkImage{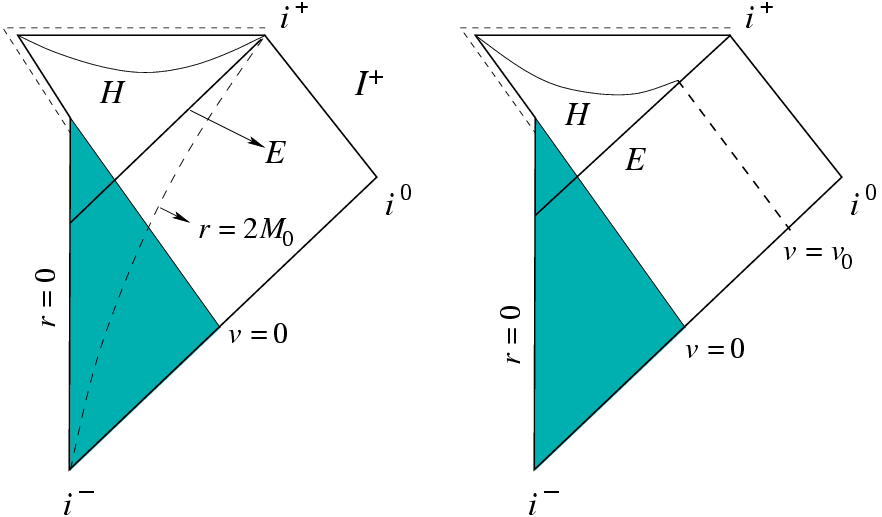}
{\begin{figure}[hptb]
   \def\epsfsize#1#2{0.75#1}
   \centerline{\epsfbox{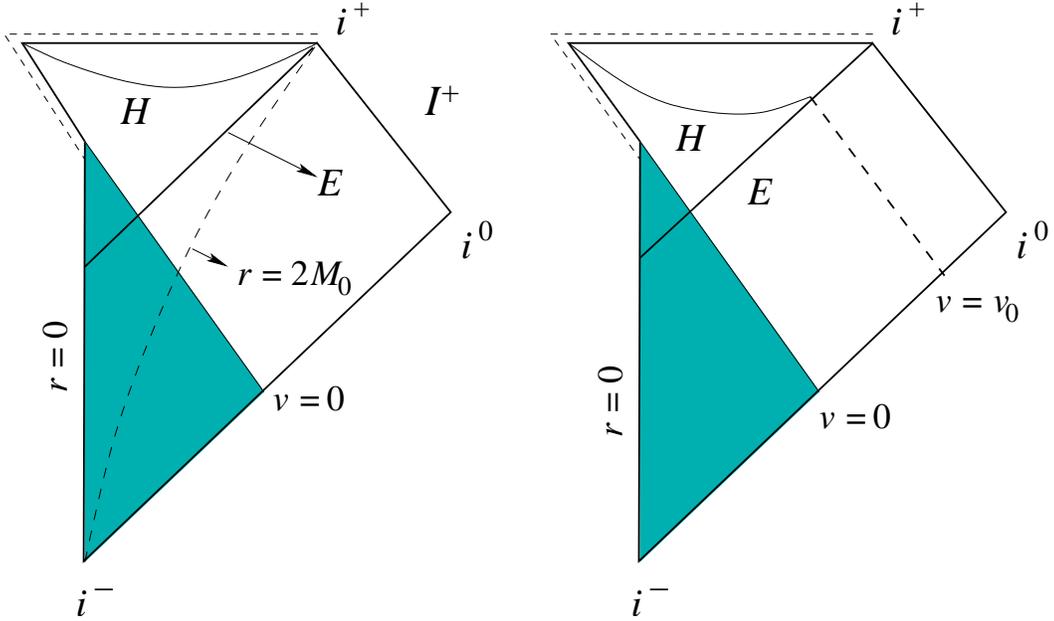}}
   \caption{\it Penrose diagrams of Schwarzschild--Vaidya metrics for
     which the mass function $M(v)$ vanishes for $v\le
     0$~\cite{kuroda}. The space-time metric is flat in the past of
     $v=0$ (i.e., in the shaded region). In the left panel, as $v$ tends to
     infinity, $\dot{M}$ vanishes and $M$ tends to a constant value
     $M_0$. The space-like dynamical horizon $H$, the null event
     horizon $E$, and the time-like surface $r = 2M_0$ (represented by
     the dashed line) all meet tangentially at $i^+$. In the right panel,
     for $v\ge v_0$ we have $\dot{M} =0$. Space-time in the future of $v=v_0$
     is isometric with a portion of the Schwarzschild space-time. The
     dynamical horizon $H$ and the event horizon $E$ meet tangentially
     at $v=v_0$. In both figures, the event horizon originates in the
     shaded flat region, while the dynamical horizon exists only in the
     curved region.}
   \label{fig:vaidya}
 \end{figure}
}

This simple example also illustrates some interesting features
which are absent in the stationary situations. First, by making
explicit choices of $M(v)$, one can plot the event horizon using, say, 
Mathematica~\cite{mathematica} and show that they originate in the
flat space-time region $v<0$, \emph{in anticipation of the null fluid
  that is going to fall in} after $v=0$. The dynamical horizon, on the
other hand, originates in the curved region of space-time, where the
metric is time-\emph{dependent}, and steadily expands until it
reaches equilibrium. Finally, as Figures~\ref{fig:vaidya}
illustrate, the dynamical and event horizons can be well
separated. Recall that in the equilibrium situation depicted by
the Schwarzschild space-time, a spherically symmetric trapped
surface passes through every point in the interior of the event
horizon. In the dynamical situation depicted by the Vaidya
space-time, they all lie in the interior of the DH. However, in
both cases, the event horizon is the boundary of $J^-(\mathcal{I}^+)$.
Thus, the numerous roles played by the event horizon in
equilibrium situations get split in dynamical contexts, some taken
up by the DH.

What is the situation in a more general gravitational collapse? As
indicated in the beginning of this section, the geometric
structure can be much more subtle. Consider 3-manifolds
$\mathcal\tau$ which are foliated by marginally trapped compact
2-surfaces $S$. We denote by $\ell^a$ the normal whose expansion
vanishes. If the expansion of the other null normal $n^a$ is
negative, $\mathcal\tau$ will be called a \emph{marginally trapped
tube} (MTT). If the tube $\mathcal\tau$ is space-like, it is a
dynamical horizon. If it is time-like, it will be called
\emph{time-like membrane}. Since future directed causal curves can
traverse time-like membranes in either direction, they are not
good candidates to represent surfaces of black holes; therefore
they are not referred to as horizons.

In Vaidya metrics, there is precisely one MTT to which all three
rotational Killing fields are tangential and this is the DH $H$.
In the Oppenheimer--Volkoff dust collapse, however, the situation
is just the opposite; the unique MTT on which each MTS $S$ is
spherical is time-like~\cite{cvb, ibd}. Thus we have a time-like
membrane rather than a dynamical horizon. However, in this case
the metric does not satisfy the smoothness conditions spelled out
at the end of Section~\ref{s1} and the global time-like character
of $\mathcal{\tau}$ is an artifact of the lack of this smoothness.
In the general perfect fluid spherical collapse, if the solution
is smooth, one can show analytically that the spherical MTT is
space-like at sufficiently late times, i.e., in a neighborhood of
its intersection with the event horizon~\cite{gv2}. For the
spherical scalar field collapse, numerical simulations show that,
as in the Vaidya solutions, the spherical MTT is space-like
everywhere~\cite{gv2}. Finally, the geometry of the numerically
evolved MTTs has been examined in two types of non-spherical
situations: the axi-symmetric collapse of a neutron star to a Kerr
black hole and in the head-on collision of two non-rotating black
holes~\cite{bks}. In both cases, in the initial phase the MTT is
neither space-like nor time-like all the way around its
cross-sections $S$. However, it quickly becomes space-like and has
a long space-like portion which approaches the event horizon. This
portion is then a dynamical horizon. There are no hard results on
what would happen in general, physically interesting situations.
The current expectation is that the MTT of a numerically evolved
black hole space-time which asymptotically approaches the event
horizon will become space-like rather soon after its formation.
Therefore most of the ongoing detailed work focuses on this
portion, although basic analytical results are available also on
how the time-like membranes evolve (see Appendix A of~\cite{ak2}).


\subsubsection{Uniqueness}
\label{s2.2.3}

Even in the simplest, Vaidya example discussed above, our explicit
calculations were restricted to spherically symmetric marginally
trapped surfaces. Indeed, already in the case of the Schwarzschild
space-time, very little is known analytically about
non-spherically symmetric marginally trapped surfaces. It is then
natural to ask if the Vaidya metric admits other, non-spherical
dynamical horizons which also asymptote to the non-expanding one.
Indeed, even if we restrict ourselves to the 3-manifold $r =
2GM(v)$, can we find another foliation by non-spherical,
marginally trapped surfaces which endows it with another dynamical
horizon structure? These considerations illustrate that in general
there are two uniqueness issues that must be addressed.

First, in a general space-time $(\mathcal{M}, g_{ab})$, can a space-like
3-manifold $H$ be foliated by two distinct families of marginally
trapped surfaces, each endowing it with the structure of a
dynamical horizon? Using the maximum principle, one can show that
this is not possible~\cite{gg}. Thus, if $H$ admits a dynamical
horizon structure, it is unique.

Second, we can ask the following question: How many DHs can a
space-time admit? Since a space-time may contain several distinct
black holes, there may well be several distinct DHs. The relevant
question is if distinct DHs can exist within each connected
component of the (space-time) trapped region. On this issue there
are several technically different uniqueness results~\cite{ag}. It
is simplest to summarize them in terms of SFOTHs. First, if two
non-intersecting SFOTHs $H$ and $H^\prime$ become tangential to
the same non-expanding horizon at a finite time (see
the right panel in Figure~\ref{fig:vaidya}), then they coincide (or
one is contained
in the other). Physically, a more interesting possibility, associated
with the late stages of collapse or mergers, is that $H$ and $H'$
become asymptotic to the event horizon. Again, they must coincide
in this case. At present, one can not rule out the existence of
more than one SFOTHs which asymptote to the event horizon if they
intersect each other repeatedly. However, even if this were to
occur, the two horizon geometries would be non-trivially
constrained. In particular, none of the marginally trapped
surfaces on $H$ can lie entirely to the past of $H'$.

A better control on uniqueness is perhaps the most important open
issue in the basic framework for dynamical horizons and there is
ongoing work to improve the existing results. Note however that all
results of Sections~\ref{s3} and~\ref{s5}, including the area increase
law and the generalization of black hole mechanics, apply to all DHs
(including the `transient ones' which may not asymptote to the event
horizon). This makes the framework much more useful in practice.

The existing results also provide some new insights for numerical
relativity~\cite{ag}. First, suppose that a MTT $\mathcal{\tau}$
is generated by a foliation of a region of space-time by partial
Cauchy surfaces $M_t$ such that each MTS $S_t$ is the outermost
MTS in $M_t$. Then $\mathcal{\tau}$ can not be a time-like
membrane. Note however that this does not imply that
$\mathcal{\tau}$ is necessarily a dynamical horizon because
$\mathcal{\tau}$ may be partially time-like and partially
space-like on each of its marginally trapped surfaces $S$. The
requirement that $\mathcal{\tau}$ be space-like
-- i.e., be a dynamical horizon -- would restrict the choice of
the foliation $M_t$ of space-time and reduce the unruly freedom in
the choice of gauge conditions that numerical simulations
currently face. A second result of interest to numerical
relativity is the following. Let a space-time $(\mathcal{M}, g_{ab})$
admit a DH $H$ which asymptotes to the event horizon. Let $M_0$ be
\emph{any} partial Cauchy surface in $(\mathcal{M}, g_{ab})$ which
intersects $H$ in one of the marginally trapped surfaces, say
$S_0$. Then, $S_0$ is the outermost marginally trapped surface
-- i.e., apparent horizon in the numerical relativity
terminology -- on $M_0$.

\newpage


\section{Area Increase Law}
\label{s3}

As mentioned in the introduction, the dynamical horizon framework
has led to a monotonicity formula governing the growth of black
holes. In this section, we summarize this result. Our discussion
is divided into three parts. The first spells out the strategy,
the second presents a brief derivation of the basic formula, and
the third is devoted to interpretational issues.


\subsection{Preliminaries}
\label{s3.1}

The first law of black hole mechanics~(\ref{1law1}) tells us
how the area of the black hole increases when it makes a
transition from an initial equilibrium state to a nearby
equilibrium state. The question we want to address is: Can one
obtain an integral generalization to incorporate fully dynamical
situations? Attractive as this possibility seems, one immediately
encounters a serious conceptual and technical problem. For, the
generalization requires, in particular, a precise notion of the
flux of gravitational energy across the horizon. Already at null
infinity, the expression of the gravitational energy flux is
subtle: One needs the framework developed by Bondi, Sachs, Newman,
Penrose, and others to introduce a viable, gauge invariant
expression of this flux~\cite{bondiflux, as, wz}. In the strong
field regime, there is no satisfactory generalization of this
framework and, beyond perturbation theory, no viable, gauge
invariant notion of the flux of gravitational energy across a
general surface.

Yet, there are at least two general considerations that suggest
that something special may happen on DHs. Consider a stellar
collapse leading to the formation of a black hole. At the end of
the process, one has a black hole and, from general physical
considerations, one expects that the energy in the final black
hole should equal the total matter plus gravitational energy that
fell across the horizon. Thus, at least the total integrated flux
across the horizon should be well defined. Indeed, it should equal
the depletion of the energy in the asymptotic region, i.e., the
difference between the ADM energy and the energy radiated across
future null infinity. The second consideration involves the
Penrose inequality~\cite{rp} introduced in Section~\ref{s1}.
Heuristically, the inequality leads us to think of the radius of a
marginally trapped surface as a measure of the mass in its
interior, whence one is led to conclude that the change in the
area is due to influx of energy. Since a DH is foliated by
marginally trapped surfaces, it is tempting to hope that something
special may happen, enabling one to define the flux of energy and
angular momentum across it. This hope is borne out.

In the discussion of DHs (Sections~\ref{s3} and~\ref{s4.2}) we
will use the following conventions (see Figure~\ref{dhfig}). The
DH is denoted by $H$ and marginally trapped surfaces that foliate
it are referred to as \emph{cross-sections}. The unit, time-like
normal to $H$ is denoted by $\widehat{\tau}^a$ with
$g_{ab}\widehat{\tau}^a\widehat{\tau}^b = -1$. The
intrinsic metric and the extrinsic curvature of $H$ are denoted by
$q_{ab}:= g_{ab} + \widehat{\tau}_a\widehat{\tau}_b$ and
$K_{ab}:={q_a}^c{q_b}^d\nabla_c\widehat{\tau}_d$, respectively. $D$ is the
derivative operator on $H$ compatible with $q_{ab}$, $\mathcal{R}_{ab}$ its
Ricci tensor, and $\mathcal{R}$ its scalar curvature. The unit space-like
vector orthogonal to $S$ and tangent to $H$ is denoted by
$\widehat{r}^a$. Quantities intrinsic to $S$ are generally written
with a tilde. Thus, the two-metric on $S$ is $\widetilde{q}_{ab}$ and the
extrinsic curvature of $S\subset H$ is
$\widetilde{K}_{ab}:=\widetilde{q}_a^{\,\,\,\,c}\widetilde{q}_b^{\,\,\,\,d}
D_c\widehat{r}_d$; the derivative operator on $(S, \widetilde{q}_{ab})$ is
$\widetilde{D}$ and its Ricci tensor is $\widetilde{\mathcal{R}}_{ab}$. Finally, we
fix the rescaling freedom in the choice of null normals to
cross-sections via $\ell^a:=\widehat{\tau}^a+\widehat{r}^a$ and
$n^a:=\widehat{\tau}^a-\widehat{r}^a$ (so that $\ell^a n_a = -2$). To keep
the discussion reasonably focused, we will not consider gauge
fields with non-zero charges on the horizon. Inclusion of these
fields is not difficult but introduces a number of subtleties and
complications which are irrelevant for numerical relativity and
astrophysics.

\epubtkImage{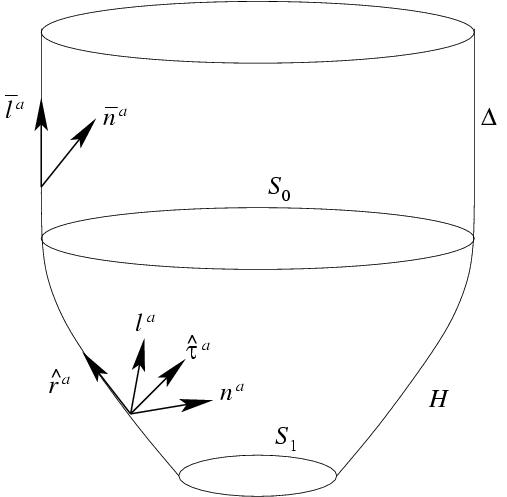}
{\begin{figure}[hptb]
   \def\epsfsize#1#2{0.6#1}
   \centerline{\epsfbox{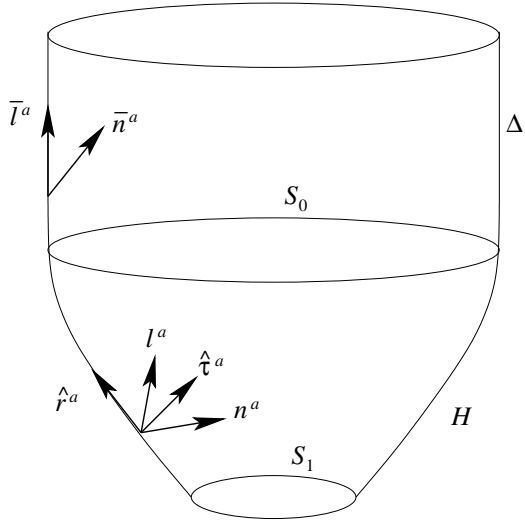}}
   \caption{\it $H$ is a dynamical horizon, foliated by marginally
     trapped surfaces $S$. $\widehat{\tau}^a$ is the unit time-like normal to $H$
     and $\widehat{r}^a$ the unit space-like normal within $H$ to the
     foliations. Although $H$ is space-like, motions along $\widehat{r}^a$
     can be regarded as `time evolution with respect to observers at
     infinity'. In this respect, one can think of $H$ as a hyperboloid
     in Minkowski space and $S$ as the intersection of the hyperboloid
     with space-like planes. In the figure, $H$ joins on to a weakly
     isolated horizon $\Delta$ with null normal $\bar{\ell}^a$ at a
     cross-section $S_0$.}
   \label{dhfig}
 \end{figure}
}


\subsection{Area increase law}
\label{s3.2}

The qualitative result that the area $a_S$ of cross-sections $S$
increases monotonically on $H$ follows immediately from the
definition,
\begin{equation}
  \widetilde{K} = \tilde{q}^{ab} D_a \widehat{r}_b =
  \frac{1}{2} \tilde{q}^{ab} \nabla_a (\ell_b - n_b) =
  - \frac{1}{2} \Theta_{(n)} > 0,
\end{equation}
since $\Theta_{(\ell)} =0$ and $\Theta_{(n)} <0$. Hence $a_S$ increases
monotonically in the direction of $\widehat{r}^a$. The non-trivial
task is to obtain a quantitative formula for the \emph{amount} of
area increase.

To obtain this formula, one simply uses the scalar and vector
constraints satisfied by the Cauchy data $(q_{ab},K_{ab})$ on $H$:
\begin{eqnarray}
  H_S &:=& \mathcal{R} + K^2 - K^{ab}K_{ab} =
  16\pi G \bar{T}_{ab}\,\widehat{\tau}^a\widehat{\tau}^b,
  \label{hamconstr} \\
  H_V^a &:=& D_b\left(K^{ab} - Kq^{ab}\right) =
  8\pi G \bar{T}^{bc}\,\widehat{\tau}_c{q^a}_b,
  \label{momconstr}
\end{eqnarray}%
where
\begin{equation}
  \label{T}
  \bar{T}_{ab} = T_{ab} - \frac{1}{8\pi G} \Lambda g_{ab},
\end{equation}
and $T_{ab}$ is the matter stress-energy tensor. The strategy is
entirely straightforward: One fixes two cross-sections $S_1$ and
$S_2$ of $H$, multiplies $H_S$ and $H_V^a$ with appropriate lapse
and shift fields and integrates the result on a portion $\Delta H
\subset H$ which is bounded by $S_1$ and $S_2$. Somewhat
surprisingly, if the cosmological constant is non-negative, the
resulting area balance law also provides strong constraints on the
topology of cross sections $S$.

Specification of lapse $N$ and shift $N^a$ is equivalent to the
specification of a vector field $\xi^a = N\widehat{\tau}^a + N^a$ with
respect to which energy-flux across $H$ is defined. The definition
of a DH provides a preferred direction field, that along $\ell^a$.
Hence it is natural set $\xi^a = N\ell^a \equiv N\widehat{\tau}^a + N\widehat{r}^a$. We
will begin with this choice and defer the possibility of choosing
more general vector fields until Section~\ref{s4.2}.

The object of interest now is the flux of energy associated with
$\xi^a = N \ell^a$ across $\Delta H$. We denote the flux of
\emph{matter} energy across $\Delta H$ by
$\mathcal{F}^{(\xi)}_\mathrm{matter}$:
\begin{equation}
  \mathcal{F}^{(\xi)}_\mathrm{matter} :=
  \int_{\Delta H} \!\!\! T_{ab}\widehat{\tau}^a\xi^b \, d^3V.
\end{equation}
By taking the appropriate combination of Equations~(\ref{hamconstr})
and~(\ref{momconstr}) we obtain
\begin{eqnarray}
  \label{flux1}
  \mathcal{F}^{(\xi)}_\mathrm{matter} &=&
  \frac{1}{16\pi G} \int_{\Delta H} \!\!\! N
  \left(H_S + 2\widehat{r}_a H_V^a \right)\, d^3V
  \nonumber \\
  &=&\frac{1}{16\pi G} \int_{\Delta H} \!\!\!
  N\left(\mathcal{R} + K^2 - K^{ab}K_{ab} +
  2\widehat{r}_a D_b (K^{ab}- Kq^{ab}) \right) \, d^3V.
\end{eqnarray}%
Since $H$ is foliated by compact 2-manifolds $S$, one can perform
a 2\,+\,1 decomposition of various quantities on $H$. In particular,
one first uses the Gauss--Codazzi equation to express $\mathcal{R}$ in terms
of $\widetilde{\mathcal{R}}$, $\widetilde{K}_{ab}$, and a total
divergence. Then, one uses the identity
\begin{equation}
  \label{eq:expansion0}
  \widetilde{q}^{ab}(K_{ab} + \widetilde{K}_{ab}) =
  \Theta_{(\ell)} = 0
\end{equation}
to simplify the expression. Finally one sets
\begin{equation}
  \sigma_{ab} = \widetilde{q}^{ac} \widetilde{q}^{bd}\,
  \nabla_a \ell_b,
  \qquad
  \zeta^a = {\widetilde{q}}^{ab}\widehat{r}^{c}\nabla_c\ell_b.
\end{equation}
(Note that $\sigma_{ab}$ is just the shear tensor since the
expansion of $\ell^a$ vanishes.) Then, Equation~(\ref{flux1}) reduces to
\begin{equation}
  \label{flux2}
  \int_{\Delta H} \!\!\! N \widetilde{\mathcal{R}}\,d^3V =
  16\pi G \int_{\Delta H} \!\!\! \bar{T}_{ab}
  \widehat{\tau}^a\xi^b\,d^3V +
  \int_{\Delta H} \!\!\! N \left( |\sigma|^2 + 2|\zeta|^2\right)\,d^3V.
\end{equation}

To simplify this expression further, we now make a specific choice
of the lapse $N$. We denote by $R$ the area-radius function; thus $R$
is constant on each $S$ and satisfies $a_S = 4\pi R^2$. Since we
already know that area increases monotonically, $R$ is a good
coordinate on $H$, and using it the 3-volume $d^3V$ on $H$ can be
decomposed as $d^3V = |\partial R|^{-1}dR d^2V$, where $\partial$
denotes the gradient on $H$. Therefore calculations simplify if we
choose
\begin{equation}
  \label{NR}
  N = |\partial R|\equiv N_R.
\end{equation}
We will set $N_R \ell^a = \xi^a_{(R)}$. Then, the integral on the
left side of Equation~(\ref{flux2}) becomes
\begin{equation}
  \int_{\Delta H} \!\!\! N_R \widetilde{\mathcal{R}}\,d^3V =
  \int_{R_1}^{R_2} \!\!\! dR \oint \widetilde{\mathcal{R}}\, d^2V =
  \mathcal{I} (R_2 - R_1),
\end{equation}
where $R_1$ and $R_2$ are the (geometrical) radii of $S_1$ and
$S_2$, and $\mathcal{I}$ is the Gauss--Bonnet topological invariant
of the cross-sections $S$. Substituting back in Equation~(\ref{flux2})
one obtains
\begin{equation}
  \label{ab1}
  \mathcal{I}\, (R_2 - R_1) = 16\pi G \int_{\Delta H} \!
  \left( T_{ab} - \frac{\Lambda}{8\pi G} g_{ab} \right)
  \widehat{\tau}^a\xi_{(R)}^b\,d^3V + 
  \int_{\Delta H} \!\!\! N_R
  \left( |\sigma|^2 + 2|\zeta|^2\right)\,d^3V.
\end{equation}
This is the general expression relating the change in area to
fluxes across $\Delta H$. Let us consider its ramifications in the
three cases, $\Lambda$ being positive, zero, or negative:

\begin{itemize}
\item If $\Lambda >0$, the right side is positive definite whence the
  Gauss--Bonnet invariant $\mathcal{I}$ is positive definite, and the
  topology of the cross-sections $S$ of the DH is necessarily that of
  $S^2$.
\item If $\Lambda =0$, then $S$ is either spherical or toroidal. The
  toroidal case is exceptional: If it occurs, the matter and the
  gravitational energy flux across $H$ vanishes (see
  Section~\ref{s3.3}), the metric $\widetilde{q}_{ab}$ is flat,
  $\mathcal{L}_n \Theta_{(\ell)} = 0$ (so $H$ can not be a FOTH), and
  $\mathcal{L}_\ell \Theta_{(\ell)} =0$. In view of these highly
  restrictive conditions, toroidal DHs appear to be unrelated to the
  toroidal topology of cross-sections of the event horizon discussed
  by Shapiro, Teukolsky, Winicour, and others~\cite{sah, st, jw}. In
  the generic spherical case, the area balance law~(\ref{ab1})
  becomes
  \begin{equation}
    \label{ab3}
    \frac{1}{2G} (R_2 - R_1) = \int_{\Delta H} \!\!\! T_{ab}
    \widehat{\tau}^a\xi_{(R)}^b\,d^3V + \frac{1}{16\pi G}\int_{\Delta H}
    \!\!\! N_R\left( |\sigma|^2 + 2|\zeta|^2\right)\,d^3V.
  \end{equation}
\item If $\Lambda <0$, there is no control on the sign of the right
  hand side of Equation~(\ref{ab1}). Hence, a priori any topology is
  permissible. Stationary solutions with quite general topologies are
  known for black holes which are asymptotically locally anti-de
  Sitter. Event horizons of these solutions are the potential
  asymptotic states of these DHs in the distant future.
\end{itemize}

\noindent
\emph{For simplicity, the remainder of our discussion of DHs will
be focused on the zero cosmological constant case with 2-sphere
topology.}


\subsection{Energy flux due to gravitational waves}
\label{s3.3}

Let us interpret the various terms appearing in the area balance
law~(\ref{ab3}).

The left side of this equation provides us with the change in the
horizon radius caused by the dynamical process under
consideration. Since the expansion $\Theta_{(\ell)}$ vanishes,
this is also the change in the \emph{Hawking mass} as one moves
from the cross section $S_1$ to $S_2$. The first integral on the
right side of this equation is the flux $\mathcal{F}^{(R)}_\mathrm{matter}$ of
matter energy associated with the vector field $\xi_{(R)}^a$. The
second term is purely geometrical and accompanies the term
representing the matter energy flux. Hence it is interpreted as
\emph{the flux $\mathcal{F}^{(R)}_\mathrm{grav}$ of $\xi_{(R)}^a$-energy
carried by the gravitational radiation}:
\begin{equation}
  \label{gravflux1}
  \mathcal{F}^{(R)}_\mathrm{grav} := \frac{1}{16\pi G}
  \int_{\Delta H} \!\!\! N_R\left( |\sigma|^2 + 2|\zeta|^2\right)\,d^3V.
\end{equation}

A priori, it is surprising that there should exist a meaningful
expression for the gravitational energy flux in the strong field
regime where gravitational waves can no longer be envisaged as
ripples on a flat space-time. Therefore, it is important to
subject this interpretation to viability criteria analogous to the
`standard' tests one uses to demonstrate the viability of the
Bondi flux formula at null infinity. It is known that it passes
most of these tests. However, to our knowledge, the status is
still partially open on one of these criteria. The situation can
be summarized as follows:

\begin{description}
\item[Gauge invariance]~\\
  Since one did not have to
introduce any structure, such as coordinates or tetrads, which is
auxiliary to the problem, the expression is obviously gauge
invariant. This is to be contrasted with definitions involving
pseudo-tensors or background fields.
\end{description}

\begin{description}
\item[Positivity]~\\
  The energy flux~(\ref{gravflux1}) is
manifestly non-negative. In the case of the Bondi flux, positivity
played a key role in the early development of the gravitational
radiation theory. It was perhaps the most convincing evidence
that gravitational waves are not coordinate artifacts but carry
\emph{physical} energy. It is quite surprising that a simple,
manifestly non-negative expression can exist in the strong field
regime of DHs. One can of course apply our general strategy to any
space-like 3-surface $\bar{H}$, foliated by 2-spheres. However, if
$\bar{H}$ is not a DH, the sign of the geometric terms in the
integral over $\Delta\bar{H}$ can not be controlled, not even when
$\bar{H}$ lies in the black hole region and is foliated by trapped
(rather than marginally trapped) surfaces $\bar{S}$. Thus, the
positivity of $\mathcal{F}^{(R)}_\mathrm{grav} $ is a rather subtle property,
not shared by 3-surfaces which are foliated by non-trapped
surfaces, nor those which are foliated by trapped surfaces; one
needs a foliation \emph{precisely by marginally trapped surfaces}.
The property is delicately matched to the definition of DHs~\cite{ak2}.
\end{description}

\begin{description}
\item[Locality]~\\
  All fields used in Equation~(\ref{gravflux1}) are
defined by the \emph{local} geometrical structures on
cross-sections of $H$. This is a non-trivial property, shared also
by the Bondi-flux formula. However, it is not shared in other
contexts. For example, the proof of the positive energy theorem by
Witten~\cite{ew} provides a positive definite energy density on
Cauchy surfaces. But since it is obtained by solving an elliptic
equation with appropriate boundary conditions at infinity, this
energy density is a highly non-local function of geometry.
Locality of $\mathcal{F}^{(R)}_\mathrm{grav}$ enables one to associate it
with the energy of gravitational waves instantaneously falling
across any cross section $S$.
\end{description}

\begin{description}
\item[Vanishing in spherical symmetry]~\\
  The fourth
criterion is that the flux should vanish in presence of spherical
symmetry. Suppose $H$ is spherically symmetric. Then one can show
that each cross-section of $S$ must be spherically symmetric. Now,
since the only spherically symmetric vector field and trace-free,
second rank tensor field on a 2-sphere are the zero fields,
$\sigma_{ab}=0$ and $\zeta^a=0$.
\end{description}

\begin{description}
\item[Balance law]~\\
  The Bondi--Sachs energy flux also has
the important property that there is a \emph{locally} defined
notion of the Bondi energy $E(C)$ associated with any 2-sphere
cross-section $C$ of future null infinity, and the difference
$E(C_1) - E(C_2)$ equals the Bondi--Sachs flux through the portion
of null infinity bounded by $C_2$ and $C_1$. Does the
expression~(\ref{gravflux1}) share this property? The answer is in the
affirmative: As noted in the beginning of this section, the
integrated flux is precisely the difference between the
\emph{locally defined} Hawking mass associated with the
cross-section. In Section~\ref{s5} we will extend these
considerations to include angular momentum.
\end{description}

Taken together, the properties discussed above provide a strong
support in favor of the interpretation of Equation~(\ref{gravflux1}) as the
$\xi_{(R)}$-energy flux carried by gravitational waves into the
portion $\Delta H$ of the DH. Nonetheless, it is important to
continue to think of new criteria and make sure that
Equation~(\ref{gravflux1}) passes these tests. For instance, in
physically reasonable, stationary, vacuum solutions to Einstein's equations,
one would expect that the flux should vanish. However, on DHs the
area must increase. Thus, one is led to conjecture that these
space-times do not admit DHs. While special cases of this
conjecture have been proved, a general proof is still lacking.
Situation is similar for non-spherical DHs in spherically
symmetric space-times.

We will conclude this section with two remarks:

\begin{itemize}
\item The presence of the shear term $|\sigma|^2$ in the integrand of
  the flux formula~(\ref{gravflux1}) seems natural from one's
  expectations based on perturbation theory at the event horizon of
  the Kerr family~\cite{hh, chandra}. But the term $|\zeta|^2$ is new
  and can arise only because $H$ is space-like rather than null: On a
  null surface, the analogous term vanishes identically. To bring out
  this point, one can consider a more general case and allow the
  cross-sections $S$ to lie on a horizon which is partially null and
  partially space-like. Then, using a 2\,+\,2 formulation~\cite{sh}
  one can show that flux on the null portion is given \emph{entirely}
  by the term $|\sigma|^2$~\cite{ahk}. However, on the space-like
  portion, the term $|\zeta|^2$ does not vanish in general. Indeed, on
  a DH, it \emph{cannot} vanish in presence of rotation: The angular
  momentum is given by the integral of $\zeta_a \varphi^a$, where
  $\varphi^a$ is the rotational symmetry.
\item The flux refers to a \emph{specific} vector field
  $\xi^a_{(R)}$ and measures the change in the Hawking mass associated
  with the cross-sections. However, this is not a good measure of the
  mass in presence of angular momentum (see, e.g., \cite{whiskey} for
  numerical simulations). Generalization of the balance law to include
  angular momentum is discussed in Section~\ref{s4.2}.
\end{itemize}

\newpage


\section{Black Hole Mechanics}
\label{s4}

As mentioned in the introduction, the discovery of the laws of
black hole mechanics has led to fundamental insights in both
classical and quantum gravity. In this section we discuss how the
standard framework tied to stationary space-times can be extended
using WIHs and DHs.


\subsection{Mechanics of weakly isolated horizons}
\label{s4.1}

The isolated horizon framework has not only extended black hole
mechanics, but it has also led to a deeper insight into the
`origin' of the laws of black hole mechanics. In this section we
will summarize these developments using WIHs. Along the way we
shall also obtain formulas for the mass and angular momentum of a
WIH. For simplicity, in the main part of the discussion, we will
restrict ourselves to type II (i.e., axi-symmetric) WIHs on which
all matter fields vanish. Generalizations including various types
of matter field can be found in~\cite{acdilaton, afk, abl2, cns, cns2, cs}.


\subsubsection{The zeroth law}
\label{s4.1.1}

The zeroth law of thermodynamics says that the temperature of a
system in thermodynamic equilibrium is constant. Its counterpart
for black hole mechanics says that surface gravity of a weakly
isolated horizon is constant. This result is non-trivial because
the horizon geometry is \emph{not} assumed to be spherically
symmetric; the result holds even when the horizon itself is highly
distorted so long as it is in equilibrium. It is established as
follows.

Recall from Section~\ref{s2.1.3} that the notion of surface
gravity is tied to the choice of a null normal $\ell^a$ of the
isolated horizon: $\kappa_{(\ell)} := \ell^a\omega_a$. Now, using
Equation~(\ref{wihcond}) in Definition~\ref{def_2} (of WIHs), we obtain:
\begin{equation}
  \mathcal{L}_\ell\omega_a = 0.
\end{equation}
Next, recall from Equation~(\ref{domega}) that the curl of $\omega_a$ is
related to the imaginary part of $\Psi_2$:
\begin{equation}
  \label{eq:twoeps}
  d\omega = 2\left(\mathrm{Im}\Psi_2\right) \epsilon
\end{equation}
where ${\epsilon}$ is the natural area 2-form on $\Delta$
satisfying $\mathcal{L}_\ell{\epsilon} =0$ and $\ell\cdot{\epsilon}=0$.
Hence we conclude $\ell\cdot d\omega = 0$ which in turn
implies that $\kappa_{(\ell)}$ is constant on the horizon:
\begin{equation}
  0 = \mathcal{L}_\ell\omega = d(\ell\cdot\omega) =
  d \kappa_{(\ell)}.
\end{equation}
This completes the proof of the zeroth law. As the argument shows,
given an NEH, the main condition~(\ref{wihcond}) in the definition
of a WIH is equivalent to constancy of surface gravity. Note that
no restriction has been imposed on $\Psi_2$ which determines the
mass and angular momentum multipoles~\cite{aepv}: as emphasized
above, the zeroth law holds even if the WIH is highly distorted
and rapidly rotating.

If electromagnetic fields are included, one can also show that the
electric potential is constant on the horizon~\cite{afk}. Finally,
there is an interesting interplay between the zeroth law the
action principle. Let us restrict ourselves to space-times which
admit a non expanding horizon as inner boundary. Then the standard
Palatini action principle is not well defined because the
variation produces a non-vanishing surface term at the horizon.
The necessary and sufficient condition for this surface term to
vanish is precisely that the gravitational (and the
electromagnetic) zeroth laws hold~\cite{afk}. Consequently, the
standard action principle \emph{is} well-defined if inner
boundaries are WIHs.


\subsubsection{Phase space, symplectic structure, and angular momentum}
\label{s4.1.2}

In field theories, conserved quantities such as energy and angular
momentum can be universally defined via a Hamiltonian framework:
they are the numerical values of Hamiltonians generating canonical
transformations corresponding to time translation and rotation
symmetries. In absence of inner boundaries, it is this procedure
that first led to the notion of the ADM energy and angular
momentum at spatial infinity~\cite{adm}. At null infinity, it can
also be used to define fluxes of Bondi energy and angular momentum
across regions of ${\mathcal{I}}^+$~\cite{as}, and values of these
quantities associated with any cross-section of ${\mathcal{I}}^+$~\cite{abr, wz}.

This procedure can be extended to allow inner boundaries which are
WIHs. The first ingredient required for a Hamiltonian framework
is, of course, a phase space. The appropriate phase space now
consists of fields living in a region of space-time outside the
black hole, satisfying suitable boundary conditions at infinity
\emph{and} horizon. Let $\mathcal{M}$ be the region of space-time that we
are interested in. The boundary of $\mathcal{M}$ consists of four
components: the time-like cylinder $\tau$ at spatial infinity, two
space-like surfaces $M_1$ and $M_2$ which are the future and past
boundaries of $\mathcal{M}$, and an inner boundary $\Delta$ which is to be
the WIH (see Figure~\ref{fig:phasespace}). At infinity, all
fields are assumed to satisfy the fall-off conditions needed to
ensure asymptotic flatness. To ensure that $\Delta$ is a type II
horizon, one fixes a rotational vector field $\varphi^a$ on $\Delta$
and requires that physical fields on $\mathcal{M}$ are such that the
induced geometry on $\Delta$ is that of a type II horizon with
$\varphi^a$ as the rotational symmetry.

\epubtkImage{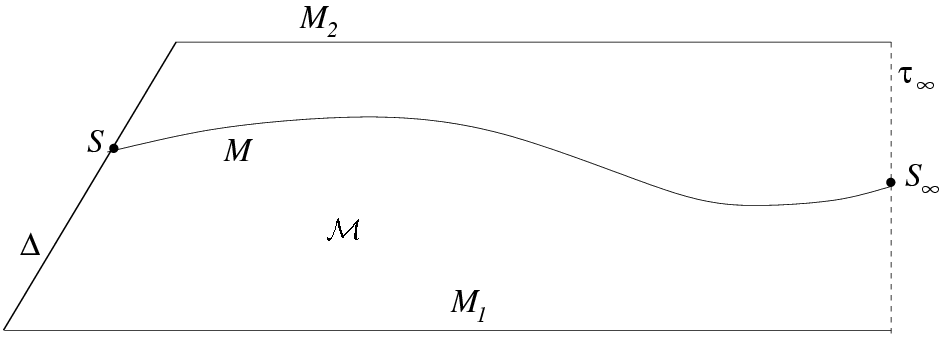}
{\begin{figure}[hptb]
   \def\epsfsize#1#2{0.6#1}
   \centerline{\epsfbox{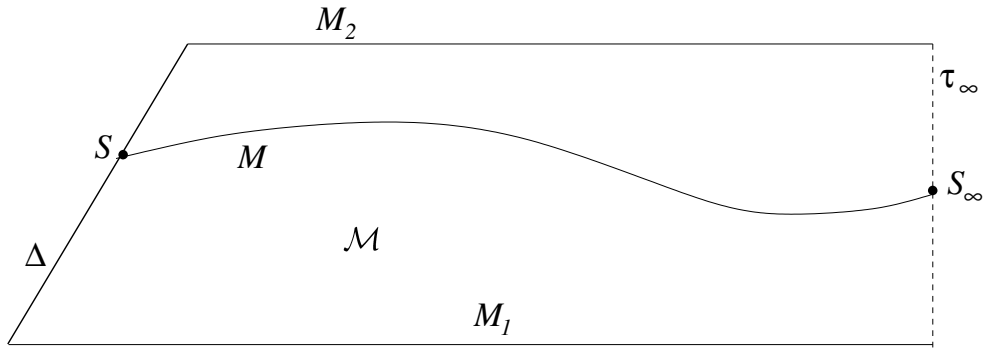}}
   \caption{\it The region of space-time $\mathcal{M}$ under consideration has
     an internal boundary $\Delta$ and is bounded by two Cauchy
     surfaces $M_1$ and $M_2$ and the time-like cylinder $\tau_\infty$
     at infinity. $M$ is a Cauchy surface in $\mathcal{M}$ whose intersection
     with $\Delta$ is a spherical cross-section $S$ and the
     intersection with $\tau_\infty$ is $S_\infty$, the sphere at
     infinity.}
   \label{fig:phasespace}
 \end{figure}
}

Two Hamiltonian frameworks are available. The first uses a
\emph{covariant phase space} which consists of the solutions to
field equations which satisfy the required boundary
conditions~\cite{afk, abl2}. Here the calculations are simplest if one
uses a first order formalism for gravity, so that the basic gravitational
variables are orthonormal tetrads and Lorentz connections. The
second uses a \emph{canonical phase space} consisting of initial
data on a Cauchy slice $M$ of $\mathcal{M}$~\cite{booth}. In the
gravitational sector, this description is based on the standard
ADM variables. Since the conceptual structure underlying the main
calculation and the final results are the same, the details of the
formalism are not important. For definiteness, in the main
discussion, we will use the covariant phase space and indicate the
technical modifications needed in the canonical picture at the
end.

The phase space $\mathbf{\Gamma}$ is naturally endowed with a
(pre-)symplectic structure $\mathbf{\Omega}$ -- a closed 2-form
(whose degenerate directions correspond to infinitesimal gauge
motions). Given any two vector fields (i.e., infinitesimal
variations) $\delta_1$ and $\delta_2$ on $\mathbf{\Gamma}$, the
action $\mathbf{\Omega}(\delta_1, \delta_2)$ of the symplectic
2-form on them provides a function on $\mathbf{\Gamma}$. A vector
field $X$ on $\mathbf{\Gamma}$ is said to be a Hamiltonian vector
field (i.e., to generate an infinitesimal canonical
transformation) if and only if ${\mathcal L}_X \mathbf{\Omega} =0$.
Since the phase space is topologically trivial, it follows that
this condition holds if and only if there is a function $H$ on
$\mathbf{\Gamma}$ such that $\mathbf{\Omega}(\delta, X) = \delta H$ for
all vector fields $\delta$. The function $H$ is called a
Hamiltonian and $X$ its Hamiltonian vector field; alternatively,
$H$ is said to generate the infinitesimal canonical transformation
$X$.

Since we are interested in energy and angular momentum, the
infinitesimal canonical transformations $X$ will correspond to
time translations and rotations. As in any generally covariant
theory, when the constraints are satisfied, values of Hamiltonians
generating such diffeomorphisms can be expressed purely as surface
terms. In the present case, the relevant surfaces are the sphere
at infinity and the spherical section $S=M\cap \Delta$ of the
horizon. Thus the numerical values of Hamiltonians now consist of
two terms: a term at infinity and a term at the horizon. The
terms at infinity reproduce the ADM formulas for energy and
angular momentum. The terms at the horizon define the energy and
angular momentum of the WIH.

Let us begin with angular momentum (see~\cite{abl2} for details).
Consider a vector field $\phi^a$ on $M$ which satisfies the
following boundary conditions: (i) At infinity, $\phi^a$ coincides
with a fixed rotational symmetry of the fiducial flat metric; and,
(ii) on $\Delta$, it coincides with the vector field $\varphi^a$. Lie
derivatives of physical fields along $\phi^a$ define a vector
field $X_{(\phi)}$ on $\mathbf{\Gamma}$. The question is whether this
is an infinitesimal canonical transformation, i.e., a generator of
the phase space symmetry. As indicated above, this is the case if
and only if there exists a phase space function $J^{(\phi)}$
satisfying:
\begin{equation}
  \delta J^{(\phi)} = \Omega(\delta,  X_{(\phi)}).
\end{equation}
for all variations $\delta$. If such a phase space function
$J^{(\phi)}$ exists, it can be interpreted as the Hamiltonian
generating rotations.

Now, a direct calculation~\cite{abl1} shows that, in absence of
gauge fields on $\Delta$, one has:
\begin{equation}
  \label{eq:hphi}
  \Omega(\delta, X_{(\phi)}) = -\frac{1}{8\pi G}\,\delta 
  \oint_S \left[\left(\varphi^a\omega_a\right) \;{}^2\!\epsilon \right] -
  \delta J_\mathrm{ADM}^{(\phi)} =: \delta J^{(\phi)}.
\end{equation}
As expected, the expression for $\delta J^{(\phi)}$ consists of
two terms: a term at the horizon and a term at infinity. The term
at infinity is the variation of the familiar ADM angular momentum
$J_\mathrm{ADM}^{(\phi)}$ associated with $\phi^a$. The surface integral
at the horizon is interpreted as the variation of the horizon
angular momentum $J_\Delta$. Since variations $\delta$ are
arbitrary, one can recover, up to additive constants,
$J^{(\phi)}_\mathrm{ADM}$ and $J_\Delta$ from their variations, and these
constants can be eliminated by requiring that both of these
angular momenta should vanish in static axi-symmetric space-times.
One then obtains:
\begin{equation}
  \label{eq:ihangmom}
  J_{\Delta} :=
  -\frac{1}{8\pi}\oint_{S} \! (\omega_a\varphi^a) \; {}^2\!\epsilon =
  -\frac{1}{4\pi}\oint_S \! f \imaginary[\Psi_2] \; {}^2\!\epsilon
\end{equation}
where the function $f$ on $\Delta$ is related to $\varphi^a$ by
$\partial_af= \epsilon_{ba}\varphi^b$. In the last step we have
used Equation~(\ref{eq:twoeps}) and performed an integration by
parts. Equation~(\ref{eq:ihangmom}) is the expression of the horizon
angular momentum. Note that all fields that enter this expression
are \emph{local} to the horizon and $\phi^a$ is \emph{not}
required to be a Killing field of the space-time metric even in a
neighborhood of the horizon. Therefore, $J_{\Delta}$ can be
calculated knowing only the horizon geometry of a type II horizon.

We conclude our discussion of angular momentum with some comments:

\begin{description}
\item[The Hamiltonian {\boldmath $J^{(\phi)}$}]~\\
  It follows from Equation~(\ref{eq:hphi}) and~(\ref{eq:ihangmom})
  that the total Hamiltonian generating the rotation along $\phi^a$ is
  the difference between the ADM and the horizon angular momenta
  (apart from a sign which is an artifact of conventions). Thus, it
  can be interpreted as the angular momentum of physical fields in the
  space-time region $\mathcal{M}$ \emph{outside} the black hole.
\end{description}

\begin{description}
\item[Relation to the Komar integral]~\\
  If $\phi^a$ happens to be a space-time Killing field in a
  neighborhood of $\Delta$, then $J_\Delta$ agrees with the Komar
  integral of $\phi^a$~\cite{abl2}. If $\phi^a$ is a global,
  space-time Killing field, then both $J^{(\phi)}_\mathrm{ADM}$ as
  well as $J_\Delta$ agree with the Komar integral, whence the total
  Hamiltonian $J^{(\phi)}$ vanishes identically. Since the fields in
  the space-time region $\mathcal{M}$ are all axi-symmetric in this
  case, this is just what one would expect from the definition of
  $J^{(\phi)}$.
\end{description}

\begin{description}
\item[{\boldmath $J_\Delta$} for general axial fields {\boldmath $\phi^a$}]~\\
  If the vector field $\phi^a$ is tangential to cross-sections of
  $\Delta$, $J^{(\phi)}$ continues to the generator of the canonical
  transformation corresponding to rotations along $\phi^a$, even if
  its restriction ${\tilde\varphi}^a$ to $\Delta$ does not agree with
  the axial symmetry $\varphi^a$ of horizon geometries of our phase
  space fields. However, there is an infinity of such vector fields
  ${\tilde\varphi}^a$ and there is no physical reason to identify the
  surface term $J_{\Delta}$ arising from any one of them with the
  horizon angular momentum.
\end{description}

\begin{description}
\item[Inclusion of gauge fields]~\\
  If non-trivial gauge fields are present at the horizon,
  Equation~(\ref{eq:ihangmom}) is incomplete. The horizon angular
  momentum $J_\Delta$ is still an integral over $S$; however it now
  contains an additional term involving the Maxwell field. Thus
  $J_\Delta$ contains not only the `bare' angular momentum but also a
  contribution from its electromagnetic hair (see~\cite{abl2} for
  details).
\end{description}

\begin{description}
\item[Canonical phase space]~\\
  The conceptual part of the above discussion does not change if one
  uses the canonical phase space~\cite{booth} in place of the
  covariant. However, now the generator of the canonical
  transformation corresponding to rotations has a volume term in
  addition to the two surface terms discussed above. However, on the
  constraint surface the volume term vanishes and the numerical value
  of the Hamiltonian reduces to the two surface terms discussed above.
\end{description}


\subsubsection{Energy, mass, and the first law}
\label{s4.1.3}

To obtain an expression of the horizon energy, one has to find the
Hamiltonian on $\mathbf{\Gamma}$ generating diffeomorphisms along a
time translation symmetry $t^a$ on $\mathcal{M}$. To qualify as a
symmetry, at infinity $t^a$ must approach a fixed time translation
of the fiducial flat metric. At the horizon, $t^a$ must be an
infinitesimal symmetry of the type II horizon geometry. Thus, the
restriction of $t^a$ to $\Delta$ should be a linear combination of
a null normal $\ell^a$ and the axial symmetry vector $\varphi^a$,
\begin{equation}
  t^a = B_{(\ell, t)}\ell^a - \Omega_{(t)}\varphi^a,
\end{equation}
where $B_{(\ell, t)}$ and the angular velocity $\Omega_{(t)}$ are
constants on $\Delta$.

However there is subtlety: Unlike in the angular momentum
calculation where $\phi^a$ is required to approach a \emph{fixed}
rotational vector $\varphi^a$ on $\Delta$, the restriction of $t^a$
to $\Delta$ can not be a fixed vector field. For physical reasons,
the constants $B_{(\ell, t)}$ and $\Omega_{(t)}$ should be allowed to
vary from one space-time to another; they are to be
\emph{functions} on phase space. For instance, physically one
expects $\Omega_{(t)}$ to vanish on the Schwarzschild horizon but
not on a generic Kerr horizon. In the terminology of numerical
relativity, unlike $\varphi^a$, the time translation $t^a$ must be a
\emph{live} vector field. As we shall see shortly, this generality
is essential also for mathematical reasons: without it, evolution
along $t^a$ will not be Hamiltonian!

At first sight, it may seem surprising that there exist choices of
evolution vector fields $t^a$ for which no Hamiltonian exists. But
in fact this phenomenon can also happen in the derivations of the
ADM
energy for asymptotically flat space-times in the absence of any
black holes. Standard treatments usually consider only those $t^a$
that asymptote to the \emph{same} unit time translation at
infinity for \emph{all} space-times included in the phase space.
However, if we drop this requirement and choose a \emph{live}
$t^a$ which approaches different asymptotic time-translations for
different space-times, then in general there exists no Hamiltonian
which generates diffeomorphisms along such a $t^a$. Thus, the
requirement that the evolution be Hamiltonian restricts
permissible $t^a$. This restriction can be traced back to the fact
that there is a fixed fiducial flat metric at infinity. At the
horizon, the situation is the opposite: The geometry is not fixed and
this forces one to adapt $t^a$ to the space-time under consideration,
i.e., to make it live.

Apart from this important caveat, the calculation of the
Hamiltonian is very similar to that for angular momentum. First,
one evaluates the 1-form $Y_{(t)}$ on $\mathbf{\Gamma}$ whose action
on any tangent vector field $\delta$ is given by
\begin{equation}
  Y_{(t)}(\delta) := \Omega(\delta, X_{(t})),
\end{equation}
where $X_{(t)}$ is the vector field on $\mathbf{\Gamma}$ induced by
diffeomorphisms along $t^a$. Once again, $Y_{(t)}(\delta)$ will
consist of a surface term at infinity and a surface term at the
horizon. A direct calculation yields
\begin{equation}
  \label{eq:xtdelta}
  Y_{(t)}(\delta) = -\frac{\kappa_{(t)}}{8\pi G}
  \delta a_\Delta - \Omega_{(t)}\delta J_\Delta +
  \delta E^{(t)}_\mathrm{ADM},
\end{equation}
where $\kappa_{(t)}:=B_{(\ell, t)}\ell^a\omega_a$ is the surface
gravity associated with the restriction of $t^a$ to $\Delta$,
$a_\Delta$ is the area of $\Delta$, and $E^{(t)}_{\mathrm{ADM}}$ is the ADM
energy associated with $t^a$. The first two terms in the right hand side of
this equation are associated with the horizon, while the
$E^t_{\mathrm{ADM}}$ term is associated with an integral at infinity.
Since the term at infinity gives the ADM energy, it is natural to
hope that terms at the horizon will give the horizon energy.
However, at this point, we see an important difference from the
angular momentum calculation. Recall that the right hand side of
Equation~(\ref{eq:hphi}) is an exact variation which means that
$J^{(\phi)}$ is well defined. However, the right hand side of
Equation~(\ref{eq:xtdelta}) is not guaranteed to be an exact
variation; in other words, $X_{(t)}$ need not be a \emph{Hamiltonian} vector
field in phase space. It is Hamiltonian if and only if there is a
phase space function $E^{(t)}_\Delta$ -- the would be energy of
the WIH -- satisfying
\begin{equation}
  \label{eq:etdelta}
  \delta E^{(t)}_\Delta = \frac{\kappa_{(t)}}{8\pi G}
  \delta a_\Delta + \Omega_{(t)}\delta J_\Delta.
\end{equation}
In particular, this condition implies that, of the infinite number
of coordinates in phase space, $E^t_\Delta$, $\kappa_{(t)}$, and
$\Omega_{(t)}$ can depend only on two: $a_\Delta$ and $J_\Delta$.

Let us analyze Equation~(\ref{eq:etdelta}). Clearly, a necessary condition
for existence of $E^{(t)}_\Delta$ is just the integrability
requirement
\begin{equation}
  \label{eq:integrability}
  \frac{\partial\kappa_{(t)}}{\partial J_\Delta} =
  8\pi G \frac{\partial \Omega_{(t)}}{\partial a_\Delta}.
\end{equation}
Since $\kappa_{(t)}$ and $\Omega_{(t)}$ are determined by $t^a$,
Equation~(\ref{eq:integrability}) is a constraint on the restriction
to the horizon of the time evolution vector field $t^a$. A vector field
$t^a$ for which $E^{(t)}_\Delta$ exists is called a
\emph{permissible} time evolution vector field. Since
Equation~(\ref{eq:etdelta}) is precisely the first law of black hole
mechanics, $t^a$ is permissible if and only if the first law
holds. Thus \emph{the first law is the necessary and sufficient
condition that the evolution generated by $t^a$ is Hamiltonian!}

There are infinitely many permissible vector fields $t^a$. To
construct them, one can start with a suitably regular function
$\kappa_0$ of $a_\Delta$ and  $J_\Delta$, find $B_{(t)}$ so that
$\kappa_{(t)} = \kappa_0$, solve Equation~(\ref{eq:integrability}) to
obtain $\Omega_{(t)}$, and find a permissible $t^a$ with $t^a =
B_{(\ell, t)}\ell^a - \Omega_{(t)}\varphi^a$ on $\Delta$~\cite{abl2}. Each
permissible
$t^a$ defines a horizon energy $E^{(t)}_\Delta$ and provides a
first law~(\ref{eq:etdelta}). A question naturally arises: Can one
select a preferred $t^a_0$ or, alternatively, a canonical function
$\kappa_0(a_\Delta, J_\Delta)$? Now, thanks to the no-hair
theorems, we know that for each choice of $(a_\Delta, J_\Delta)$,
there is precisely one stationary black hole in vacuum general
relativity: the Kerr solution. So, it is natural to set $\kappa_0
(a_\Delta, J_\Delta) = \kappa_\mathrm{Kerr} (a_\Delta, J_\Delta)$,
or, more explicitly,
\begin{equation}
  \label{eq:kerrkappa}
  \kappa_0(a_\Delta, J_\Delta) =
  \frac{R_\Delta^4-4J_\Delta^2}{2R_\Delta^3\sqrt{R_\Delta^4+4J_\Delta^2}},
\end{equation}
where $R_\Delta$ is the area radius of the horizon, $R_\Delta
=(a_\Delta/4\pi)^{1/2}$. Via Equation~(\ref{eq:integrability}), this
choice then leads to
\begin{equation}
  \label{eq:kerromega}
  \Omega_{(t)} = \Omega_\mathrm{Kerr}(a_\Delta, J_\Delta) =
  \frac{\sqrt{R^4_\Delta + 4 G J_\Delta^2}}{2GR_\Delta}.
\end{equation}
The associated horizon energy is then:
\begin{equation}
  \label{eq:ihmass}
  E^{(t_0)}_\Delta = \frac{1}{2GR_\Delta}\sqrt{R_\Delta^4 + 4G^2J_\Delta^2}.
\end{equation}
This canonical horizon energy is called the \emph{horizon mass}:
\begin{equation}
  M_\Delta := E^{(t_0)}_\Delta
\end{equation}
Note that, its dependence on the horizon area and angular momentum
is the same as that in the Kerr space-time. Although the final
expression is so simple, it is important to keep in mind that this
is not just a postulate. Rather, this result is \emph{derived}
using a systematic Hamiltonian framework, following the same
overall procedure that leads to the definition of the ADM
4-momentum at spatial infinity. Finally, note that the quantities
which enter the first law refer just to physical fields on the
horizon; one does not have to go back and forth between the
horizon and infinity.

We will conclude with three remarks:

\begin{description}
\item[Relation to the ADM and Bondi energy]~\\
  Under certain physically reasonable assumptions on the behavior of
fields near future time-like infinity $i^+$, one can argue that,
if the WIH extends all the way to $i^+$, then the difference
$M_\mathrm{ADM} - M_\Delta$ equals the energy radiated across future
null-infinity~\cite{abf2}. Thus, as one would expect physically,
$M_\Delta$ is the \emph{mass that is left over after
all the gravitational radiation has left the system}.
\end{description}

\begin{description}
\item[Horizon angular momentum and mass]~\\
  To obtain a
well-defined action principle and Hamiltonian framework, it is
essential to work with WIHs. However, the final
expressions~(\ref{eq:ihangmom}) and~(\ref{eq:ihmass}) of the horizon
angular momentum and mass do not refer to the preferred null normals
$[\ell]$ used in the transition from an NEH to a WIH. Therefore,
the expressions can be used on any NEH. This fact is useful in the
analysis of transition to equilibrium (Section~\ref{s4.3}) and
numerical relativity (Section~\ref{s5.1}).
\end{description}

\begin{description}
\item[Generalizations of the first law]~\\
  The derivation
of the first law given here can be extended to allow the presence
of matter fields at the horizon~\cite{afk, abl2}. If gauge fields
are present, the expression of the angular momentum has an extra
term and the first law~(\ref{eq:etdelta}) also acquires the
familiar extra term `$\Phi \delta Q$', representing work done on
the horizon in increasing its charge. Again, all quantities are
defined locally on the horizon. The situation is similar in lower~\cite{adw} and higher~\cite{klp} space-time dimensions. However, a
key difference arises in the definition of the horizon mass. Since
the uniqueness theorems for stationary black holes fail to extend
beyond the Einstein--Maxwell theory in four space-time dimensions,
it is no longer possible to assign a \emph{canonical} mass to the
horizon. However, as we will see in Section~\ref{s6}, the
ambiguity in the notion of the horizon mass can in fact be
exploited to obtain new insights into the properties of black
holes and solitons in these more general theories.
\end{description}


\subsection{Mechanics of dynamical horizons}
\label{s4.2}

The variations $\delta$ in the first law~(\ref{eq:etdelta})
represent infinitesimal changes in equilibrium states of horizon
geometries. In the derivation of Section~\ref{s4.1}, these
variations relate nearby but distinct space-times in each of which
the horizon is in equilibrium. Therefore Equation~(\ref{eq:etdelta}) is
interpreted as the first law in a \emph{passive} form. Physically,
it is perhaps the \emph{active form} of the first law that is of
more direct interest where a \emph{physical process}, such as the
one depicted in the right panel of Figure~\ref{exam} causes a
transition from one equilibrium
state to a nearby one. Such a law can be established in the
dynamical horizon framework. In fact, one can consider fully
non-equilibrium situations, allowing physical processes in a given
space-time in which there is a \emph{finite} -- rather than an
infinitesimal -- change in the state of the horizon. This leads to
an \emph{integral} version of the first law.

Our summary of the mechanics of DHs is divided in to three parts.
In the first, we begin with some preliminaries on angular
momentum. In the second, we extend the area balance law~(\ref{ab3})
by allowing more general lapse and shift functions, which leads to
the integral version of the first law. In the third, we introduce
the notion of horizon mass.


\subsubsection{Angular momentum balance}
\label{s4.2.1}

As one might expect, the angular momentum balance law results from
the momentum constraint~(\ref{momconstr}) on the DH $H$. Fix
\emph{any} vector field $\varphi^a$ on $H$ which is tangential to all
the cross-sections $S$ of $H$, contract  both sides of
Equation~(\ref{momconstr}) with $\varphi^a$, and integrate the
resulting equation over the region $\Delta H$ to obtain\epubtkFootnote{Note
  that we could replace $\bar{T}_{ab}$ with $T_{ab}$ because
  $g_{ab}\widehat{\tau}^a\varphi^b =0$. Thus the cosmological constant
  plays no role in this section.}
\begin{equation}
  \frac{1}{8\pi G}\oint_{S_2} \!\!\! K_{ab}
  \varphi^a\widehat{r}^b \, d^2V - \frac{1}{8\pi G}
  \oint_{S_1} \!\!\! K_{ab}\varphi^a\widehat{r}^b \, d^2V
  = \int_{\Delta H} \!\! \left( \! T_{ab}\widehat{\tau}^a\varphi^b +
  \frac{1}{16\pi G} (K^{ab}- K q^{ab})\mathcal{L}_\varphi
  q_{ab} \!\right) d^3V.
  \label{balanceJ}
\end{equation}
It is natural to identify the surface integrals with the
\emph{generalized angular momentum} $J_S^{\varphi}$ associated with
cross-sections $S$ and set
\begin{equation}
  \label{jdynamic1}J_S^{\varphi} =
  -\frac{1}{8\pi G} \oint_{S} \!\!\! K_{ab} \varphi^a\widehat{r}^b \,
  d^2V \equiv \frac{1}{8\pi G}\oint_{S} \! j^\varphi d^2V,
\end{equation}
where the overall sign ensures compatibility with conventions
normally used in the asymptotically flat context, and where we have
introduced an angular momentum density $j^\varphi := - K_{ab}\varphi^a
\widehat{r}^b$ for later convenience. The term `generalized' emphasizes
the fact that the vector field $\varphi^a$ need not be an axial
Killing field even on $S$; it only has to be tangential to our
cross-sections. If $\varphi^a$ happens to be the restriction of a
space-time Killing field to $S$, then $J^\varphi_S$ agrees with the
Komar integral. If the pair $(q_{ab},K_{ab})$ is spherically symmetric
on $S$, as one would expect, the angular momenta
associated with the rotational Killing fields vanish.

Equation~(\ref{balanceJ}) is a balance law; the right side provides
expressions of fluxes of the generalized angular momentum across
$\Delta H$. The contributions due to matter and gravitational
waves are cleanly separated and given by
\begin{equation}
  \mathcal{J}^{\varphi}_\mathrm{matter} =
  -\int_{\Delta H} \!\!\! T_{ab}\widehat{\tau}^a\varphi^b\, d^3V,
  \qquad
  \mathcal{J}^{\varphi}_\mathrm{grav} = -\frac{1}{16\pi G}
 \int_{\Delta H} \!\!\! P^{ab}\mathcal{L}_\varphi q_{ab}\, d^3V,
\end{equation}
with $P^{ab} = K^{ab} - Kq^{ab}$, so that
\begin{equation}
  J_{S_2}^{\varphi} - J_{S_1}^{\varphi} =
  \mathcal{J}^{\varphi}_\mathrm{matter} +
  \mathcal{J}^{\varphi}_\mathrm{grav}.
\end{equation}
As expected, if $\varphi^a$ is a Killing vector of the three-metric
$q_{ab}$, then the gravitational angular momentum flux vanishes:
$\mathcal{J}^{\varphi}_\mathrm{grav} = 0$.

As with the area balance law, here we worked directly with the
constraint equations rather than with a Hamiltonian framework.
However, we could also have used, e.g., the standard ADM phase
space framework based on a Cauchy surface $M$ with internal
boundary $S$ and the outer boundary at infinity. If $\phi^a$ is a
vector field on $M$ which tends to $\varphi^a$ on $S$ and to an
asymptotic rotational symmetry at infinity, we can ask for the
phase space function which generates the canonical transformation
corresponding to the rotation generated by $\phi^a$. When the
constraints are satisfied, as usual the value of this generating
function is given by just surface terms. The term at infinity
provides the total angular momentum and, as in Section~\ref{s4.1.2},
it is natural to interpret the surface term at $S$
as the $\varphi$-angular momentum of $S$. This term can be expressed
in terms of the Cauchy data $(\bar{q}_{ab}, \bar{K}_{ab})$ on $M$
as
\begin{equation}
  \label{jdynamic2}
  \bar{J}_S^{\varphi} = -\frac{1}{8\pi G} \oint_{S} \!
  \bar{K}_{ab} \varphi^a\bar{r}^b \, d^2V,
\end{equation}
where $\bar{r}^a$ is the unit normal to $S$ within $M$. However,
since the right side involves the extrinsic curvature of $M$, in
general the value of the integral is sensitive to the choice of
$M$. Hence, the notion of the $\varphi$-angular momentum
associated with an arbitrary cross-section is ambiguous. This
ambiguity disappears \emph{if $\varphi^a$ is divergence-free on $S$}.
In particular, in this case, one has  $J^\varphi_S = \bar{J}^\varphi_S$.
Thus, although the balance law~(\ref{balanceJ}) holds for more
general vector fields $\varphi^a$, it is robust only
when $\varphi^a$ is divergence-free on $S$. (These considerations shed
some light on the interpretation of the field $\zeta^a$ in the area
balance law~(\ref{ab3}). For, the form of the right side of
Equation~(\ref{jdynamic1}) implies that the field $\zeta^a$ vanishes
identically on $S$ if and only if $J^\varphi_S$ vanishes for every
divergence-free $\varphi^a$ on $S$. In particular then, if the horizon
has non-zero angular momentum, the $\zeta$-contribution to the energy
flux can not vanish.)

Finally, for $J^\varphi_S$ to be interpreted as `the' angular momentum,
$\varphi^a$ has to be a symmetry. An obvious possibility is that it be a
Killing field of $q_{ab}$ on $S$. A more general scenario is
discussed in Section~\ref{s8}.


\subsubsection{Integral form of the first law}
\label{s4.2.2}

To obtain the area balance law, in Section~\ref{s3.2} we
restricted ourselves to vector fields $\xi_{(R)}^a = N_R \ell^a$,
i.e., to lapse functions $N_{R} = |\partial R|$ and shifts $N^a =
N_R \widehat{r}^a$. We were then led to a conservation law for the Hawking
mass. In the spherically symmetric context, the Hawking mass can be
taken to be the physical mass of the horizon. However, as the Kerr
space-time already illustrates, in presence of rotation this
interpretation is physically incorrect. Therefore, although
Equation~(\ref{ab3}) continues to dictate the dynamics of the Hawking
mass even in presence of rotation, a more general procedure is needed
to obtain \emph{physically interesting} conservation laws in this
case. In the case of WIHs, the first law incorporating rotations
required us to consider suitable linear combinations of $\ell^a$ and
the rotational symmetry field $\varphi^a$ on the horizon. In the same
spirit, on DHs, one has to consider more general vector fields
than $\xi^a_{(R)}$, i.e., more general choices of lapses and
shifts.

As on WIHs, one first restricts oneself to situations in which the
metric $q_{ab}$ on $H$ admits a Killing field $\varphi^a$ so that
$J^\varphi_S$ can be unambiguously interpreted as the angular
momentum associated with each $S$. In the case of a WIH, $t^a$ was
given by $t^a = c \ell^a + \Omega \varphi^a$ and the freedom was in
the choice of constants $c$ and $\Omega$. On a DH, one must allow
the corresponding coefficients to be `time-dependent'. The
simplest generalization is to choose, in place of $\xi^a_{(R)}$,
vector fields
\begin{equation}
  \label{t} t^a := N_r \ell^a + \Omega \varphi^a \equiv
  N_r \widehat{\tau}^a + (N_r \widehat{r}^a + \Omega \varphi^a),
\end{equation}
where $\Omega$ is an arbitrary function of $R$, and the lapse $N_r$
is given by $N_r = |\partial r|$ for \emph{any} function $r$ of
$R$. Note
that one is free to rescale $N_r$ and $\Omega$ by functions of $R$
so that on each cross-section (`instant of time') one has the same
rescaling freedom as on a WIH. One can consider even more general
lapse-shift pairs to allow, e.g., for differential rotation
(see~\cite{ak2}).

Using $t^a$ in place of $\xi_{(R)}^a$, one obtains the following
generalization of the area balance equation~\cite{ak2}:
\begin{eqnarray}
  && \frac{r_2-r_1}{2G} + \frac{1}{8\pi G}
  \left( \oint_{S_2} \!\! \Omega j^\varphi\,d^2V -
  \oint_{S_1} \!\! \Omega j^\varphi\,d^2V -
  \int_{\Omega_1}^{\Omega_2} \!\!\! d\Omega \oint_S \!
  j^\varphi\,d^2V \right) =
  \nonumber \\
  && \int_{\Delta H} \!\!\! T_{ab}\widehat{\tau}^a t^b \,d^3V +
  \frac{1}{16\pi G}\int_{\Delta H} \!\!\! N_r
  \left(|\sigma|^2 + 2|\zeta|^2\right)\,d^3V -
  \frac{1}{16 \pi G} \int_{\Delta H} \!\!\! \Omega
  P^{ab}\mathcal{L}_\varphi q_{ab} \, d^3V.
  \label{balance1}
\end{eqnarray}%
Note that there is one balance equation for every vector field
$t^a$ of the form~(\ref{t}); as in Section~\ref{s4.1}, we have an
infinite number of relations, now ensured by the constraint part
of Einstein's equations.

The right side of Equation~(\ref{balance1}) can be naturally interpreted as
the flux $\mathcal{F}^t_{\Delta H}$ of the `energy' $E^t$
associated with the vector field $t^a$ across $\Delta H$. Hence,
we can rewrite the equation as
\begin{equation}
  \label{flux}
  \mathcal{F}^t_{\Delta H} = \frac{r_2-r_1}{2G} +
  \frac{1}{8\pi G} \left( \oint_{S_2} \!\! \Omega j^\varphi\,d^2V -
  \oint_{S_1} \!\!\Omega j^\varphi\,d^2V -
  \int_{\Omega_1}^{\Omega_2} \!\!\! d\Omega \oint_S \! j^\varphi\,d^2V \right).
\end{equation}
If $S_1$ and $S_2$ are only infinitesimally separated, this
integral equation reduces to the differential condition
\begin{eqnarray}
  \delta E^t_S &=& \frac{1}{8\pi G}
  \left(\frac{1}{2R} \frac{dr}{dR}\right)\biggr|_S \delta a_s +
  \Omega \, \delta J_S^\varphi
  \nonumber \\
  &\equiv & \frac{\bar{\kappa}}{8\pi G} \delta a_S +
  \Omega \, \delta J_S^\varphi.
\end{eqnarray}%
Thus, the infinitesimal form of Equation~(\ref{balance1}) is a familiar
first law, provided $[({1}/{2R})({dr}/{dR})](S)$ is identified as
an \emph{effective surface gravity} on the cross-section $S$. This
identification can be motivated as follows. First, on a
spherically symmetric DH, it is natural to choose $r=R$. Then the
surface gravity reduces to $1/(2R)$, just as one would hope from
one's experience with the Schwarzschild metric and more generally
with static but possibly distorted horizons (See Appendix~A
of~\cite{abf2}). Under the change $R \mapsto r(R)$, we have
$\bar\kappa_r = (dr/dR)\, \bar\kappa_R$, which is the natural
generalization of the transformation property $\kappa_{c\ell} = c
\kappa_\ell$ of surface gravity of WIHs under the change $\ell \mapsto
c\ell$. Finally, $\bar\kappa_r$ can also be regarded as a 2-sphere
average of a geometrically defined surface gravity associated with
certain vector fields on $H$~\cite{bf, ak2}; hence the adjective
`effective'.

To summarize, Equation~(\ref{balance1}) represents an integral
generalization of the first law of mechanics of weakly isolated
horizons to dynamical situations in which the horizon is permitted
to make a transition from a given state to one far away, not just
nearby. The left side represents the flux $\mathcal{F}^t_{\Delta
H}$ of the energy associated with the vector field $t^a$,
analogous to the flux of Bondi energy across a portion of null
infinity. A natural question therefore arises: Can one integrate
this flux to obtain an energy $E^t_S$ which depends only on fields
defined \emph{locally} on the cross-section, as is possible at
null infinity? As discussed in the next section, the answer is in
the affirmative and the procedure leads to a canonical notion of
horizon mass.


\subsubsection{Horizon mass}
\label{s4.2.3}

In general relativity, the notion of energy always refers to a
vector field. On DHs, the vector field is $t^a$. Therefore, to
obtain an unambiguous notion of horizon mass, we need to make a
canonical choice of $t^a= N_r \ell^a+ \Omega \varphi^a$, i.e., of
functions $N_r$ and $\Omega$ on $H$. As we saw in
Section~\ref{s4.1.3}, on WIHs of 4-dimensional Einstein--Maxwell
theory, the pair $(a_\Delta, J_\Delta)$ suffices to pick a canonical time
translation field $t_0^a$ on $\Delta$. The associated horizon energy
$E^{t_0}_\Delta$ is then interpreted as the mass $M_\Delta$. This
suggests that the pair ($a_S, J_S^\varphi$) be similarly used to make
canonical choices $N_{r^0}$ and $\Omega^0$ on $S$. Thanks to the
black hole uniqueness theorems of the 4-dimensional
Einstein--Maxwell theory, this strategy is again viable.

Recall that the horizon surface gravity and the
horizon angular velocity in a Kerr solution can be
expressed as a function only of the horizon radius $R$ and angular
momentum $J$:
\begin{equation}
  \label{kerr}
  \kappa_\mathrm{Kerr}(R, J) :=
  \frac{R^4-4G^2J^2}{2R^3\sqrt{R^4+4G^2J^2}},
  \qquad
  \Omega_\mathrm{Kerr}(R, J) :=
  \frac{2GJ}{R\sqrt{R^4+4G^2J^2}}.
\end{equation}
Given a cross section $S$ of $H$, the idea is to consider the
unique Kerr solution in which the horizon area is given by $a_S$
and angular momentum by $J_S$, and assign to $S$ effective surface
gravity $\bar\kappa_S$ and angular velocity $\Omega_S$ through
\begin{equation}
  \bar\kappa_S := \kappa_\mathrm{Kerr}(R_S, J_S^\varphi),
  \qquad
  \Omega_S := \Omega_\mathrm{Kerr}(R_S, J_S^\varphi)
\end{equation}
Repeating this procedure on every cross-section, one obtains
functions $\bar\kappa^0(R)$ and $\Omega^0(R)$ on $H$, since
$J^\varphi$ is a function of $R$ alone. The definition of the
effective surface gravity then determines a function $r^0$ of $R$
and hence $N_{r^0}$ uniquely. Thus, using Equation~(\ref{kerr}), one can
select a canonical vector field $t_0^a$ and Equation~(\ref{flux}) then
provides a canonical balance law:
\begin{equation}
  \mathcal{F}^{t_0}_{\Delta H} = \frac{r^0_2-r^0_1}{2G} +
  \frac{1}{8\pi G} \left( \oint_{S_2} \!\! \Omega^0 j^\varphi\,d^2V -
  \oint_{S_1} \!\! \Omega^0 j^\varphi\,d^2V -
  \int_{\Omega^0_1}^{\Omega^0_2} \!\!\! d\Omega^0 \oint_S
  j^\varphi\,d^2V \right).
\end{equation}
The key question is whether this equation is integrable, i.e., if
\begin{equation}
  \mathcal{F}^{t_0}_{\Delta H} = E^{t_0}_{S_2} - E^{t_0}_{S_1}
\end{equation}
for some $E^{t_0}_S$ which depends \emph{locally} on fields
defined on $S$. The answer is in the affirmative. Furthermore, the
expression of $E^{t_0}$ is remarkably simple and is identified
with the horizon mass:
\begin{equation}
  \label{mass}
  M(R) := E^{t_0}(R) = \frac{\sqrt{R^4 + 4G^2J^2}}{2GR}.
\end{equation}
Thus, on any cross-section $S$, $M_S$ is just the mass of the Kerr
space-time which has horizon area $a_S$ and angular momentum
$J^\varphi_S$: As far as the mass is concerned, one can regard the DH as
an evolution through `a sequence of Kerr horizons'. The
non-triviality of the result lies in the fact that, although this
definition of mass is so `elementary', thanks to the balance
law~(\ref{balance1}) it obeys a Bondi-type flux formula,
\begin{equation}
  M_{S_2} - M_{S_1} = \int_{\Delta H} \!\!\!
  T_{ab}\widehat{\tau}^a t_0^b \,d^3V +
  \frac{1}{16\pi G}\int_{\Delta H} \!\!\! N_{r^0} \left(|\sigma|^2 +
  2|\zeta|^2\right)\,d^3V
  \label{balance2}
\end{equation}
for a specific vector field $t_0^a = N_{r^0}\ell^a + \Omega^0
\varphi^a$, where each term on the right has a well-defined physical
meaning. Thus, DHs admit a locally-defined notion of mass and an
associated, canonical conservation law~(\ref{balance2}). The
availability of a mass formula also provides a canonical integral
version of the first law through Equation~(\ref{balance1}):
\begin{equation}
  M_{S_2} - M_{S_1} = \frac{r^0_2-r^0_1}{2G} +
  \frac{1}{8\pi G} \left( \oint_{S_2} \!\! \Omega^0 j^\varphi\,d^2V -
  \oint_{S_1} \!\! \Omega^0 j^\varphi\,d^2V - \int_{\Omega^0_1}^{\Omega^0_2}
  \!\!\! d\Omega^0 \oint_S \! j^\varphi\,d^2V \right).
\end{equation}
The infinitesimal version of this equation yields the familiar first
law $\delta M = (\bar{\kappa}/8\pi G)\delta a + \Omega\delta J$.

On weakly isolated as well as dynamical horizons, area $a$ and
angular momentum $J$ arise as the fundamental quantities and mass
is expressed in terms of them. The fact that the horizon mass is
the same function of $a$ and $J$ in both dynamical and equilibrium
situations is extremely convenient for applications to numerical
relativity~\cite{whiskey}. Conceptually, this simplicity is a
direct consequence of the first law and the non-triviality lies in
the existence of a balance equation~(\ref{balance1}), which makes
it possible to integrate the first law.


\subsection{Passage of dynamical horizons to equilibrium}
\label{s4.3}

In physical situations, such as a gravitational collapse or black
hole mergers, one expects the dynamical horizon to approach
equilibrium at late times and become isolated. Because of back
scattering, generically the approach is only asymptotic. However,
the back scattering is generally quite weak and in simulations,
within numerical errors, equilibrium is reached at finite times.
The passage to equilibrium can be studied in detail in Vaidya
solutions discussed in Section~\ref{s2.2.2}. Moreover, in
spherically symmetric examples such as these solutions,
\emph{exact} equilibrium can be reached at a finite time (see
right panel of Figure~\ref{fig:vaidya}). The question then arises: In
these situations, do various notions introduced on dynamical horizons
go over smoothly to those introduced on WIHs? This issue has been
analyzed only in a preliminary fashion~\cite{bf, ak2}. In this section
we will summarize the known results.

First, if the dynamical horizon is a FOTH, as the flux of matter
and shear across $H$ tends to zero, $H$ becomes null and
furthermore a non-expanding horizon. By a suitable choice of null
normals, it can be made weakly isolated. Conditions under which it
would also become an isolated horizon are not well-understood.
Fortunately, however, the final expressions of angular momentum
and horizon mass refer only to that structure which is already
available on non-expanding horizons (although, as we saw in
Section~\ref{s4.1}, the underlying
Hamiltonian framework does require the horizon to be weakly
isolated~\cite{afk, abl2}). Therefore, it is meaningful to ask if
the angular momentum and mass defined on the DHs match with those
defined on the non-expanding horizons. In the case when the
approach to equilibrium is only asymptotic, it is rather
straightforward to show that the answer is in the affirmative.

In the case when the transition occurs at a finite time, the
situation is somewhat subtle. First, we now have to deal with both
regimes and the structures available in the two regimes are
entirely different. Second, since the intrinsic metric becomes
degenerate in the transition from the dynamical to isolated
regimes, limits are rather delicate. In particular, the null
vector field $\ell^a = \widehat{\tau}^a + \widehat{r}^a$ on $H$
diverges, while $n^a= \widehat{\tau}^a-\widehat{r}^a$ tends to zero at
the boundary. A priori therefore, it is not at all clear that angular
momentum and mass would join smoothly if the transition occurs at a
finite time. However, a detailed analysis shows that the two sets of
notions in fact agree.

More precisely, one has the following results. Let $\mathcal{Q} =
H\cup \Delta$ be a $C^{k+1}$ 3-manifold (with $k \ge 2$),
topologically $\mathbb{S}^2 \times \mathbb{R}$ as in the second
Penrose diagram of Figure~\ref{fig:vaidya}. Let the space-time
metric $g_{ab}$ in a neighborhood of $\mathcal{Q}$ be $C^k$. The
part $H$ of $\mathcal{Q}$ is assumed to have the structure of a DH
and the part $\Delta$ of a non-expanding horizon. Finally, the
pull-back $q_{ab}$ of $g_{ab}$ to $\mathcal{Q}$ is
assumed to admit an axial Killing field $\varphi^a$. Then we have:

\begin{itemize}
\item The angular momentum $J^\varphi$ and the mass $M$ defined in the
  two regimes agree on the boundary between $H$ and $\Delta$.
\item The vector field $t_0^a$ defined on $H$ and used in the
  definition of mass matches with a preferred vector field $t_0^a$
  used to define mass on $\Delta$.
\end{itemize}

This agreement provides an independent support in favor of the
strategy used to introduce the notion of mass in the two regimes.

\newpage


\section{Applications in Numerical Relativity}
\label{s5}

By their very nature, numerical simulations of space-times are
invariably tied to choices of coordinates, gauge conditions,
dynamical variables, etc. Therefore, it is non-trivial to extract
from them gauge invariant physics, especially in the strong
curvature regions. Traditionally, the analytical infrastructure
available for this purpose has been based on properties of the
Kerr solution and its perturbations. However, a priori it is not
clear if this intuition is reliable in the fully dynamical, strong
curvature regime. On the numerical side, a number of significant
advances have occurred in this area over the past few years. In
particular, efficient algorithms have been introduced to find
apparent horizons (see, e.g., \cite{ah1, ah2, ah3}), black hole excision 
techniques have been successfully implemented~\cite{excision2,
  excision1}, and the stability of numerical codes
has steadily improved~\cite{stability}. To take full advantage of
these ongoing improvements, one must correspondingly `upgrade' the
analytical infra-structure so that one can extract physics more
reliably and with greater accuracy.

These considerations provided stimulus for a significant body of
research at the interface of numerical relativity and the
dynamical and isolated horizon frameworks. In this section, we
will review the most important of these
developments. Section~\ref{s5.1} summarizes calculations of mass and
angular momentum of
black holes. Section~\ref{s5.2} discusses applications to problems
involving initial data. Specifically, we discuss the issue of
constructing the `quasi-equilibrium initial data' and the
calculation of the gravitational binding energy for a binary black
hole problem. Section~\ref{s5.3} describes how one can calculate
the source multipole moments for black holes, and Section~\ref{s5.4}
presents a `practical' approach for extracting gauge
invariant waveforms. Throughout this section we assume that vacuum
equations hold near horizons.


\subsection{Numerical computation of black hole mass and angular momentum}
\label{s5.1}

As we saw in Section~\ref{s4}, the mechanics of IHs and DHs
provides expressions of angular momentum and mass of the horizon.
These expressions involve geometric quantities defined
intrinsically on the IH $\Delta$ and DH $H$. Numerical
simulations, on the other hand, deal with the 3-metric
$\bar{q}_{ab}$ and extrinsic curvature $\bar{K}_{ab}$ on (partial)
Cauchy surfaces $M$. Therefore, the first task is to recast the
formulas in terms of this Cauchy data.

Simulations provide us with a foliation of space-time $\mathcal{M}$ by
partial Cauchy surfaces $M$, each of which has a marginally
trapped 2-surface $S$ as (a connected component of) its inner
boundary. The world  
tube of these 2-surfaces is a candidate for a DH or an IH. If it
is space-like, it is a DH $H$ and if it is null (or, equivalently,
if the shear $\sigma_{ab}$ of the outward null normal to $S$ is
zero) it is a WIH $\Delta$. The situation is depicted in
Figure~\ref{fig:cauchy}. It is rather simple to numerically verify if
these restrictions are met. To calculate mass and angular
momentum, one assumes that the intrinsic 2-metric on the
cross-sections $S$ admits a rotational Killing field $\varphi^a$ (see,
however, Section~\ref{s8} for weakening of this assumption). A
rather general and convenient method, based on the notion of
Killing transport, has been introduced and numerically implemented
to explicitly find this vector field $\varphi^a$~\cite{dkss}.

\epubtkImage{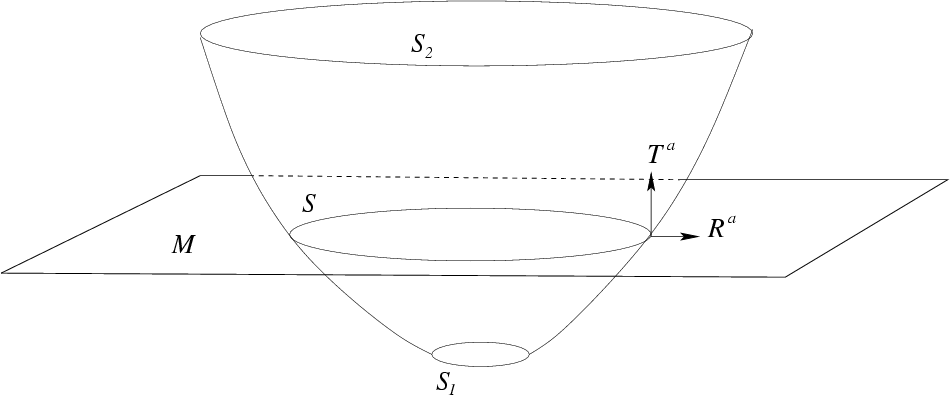}
{\begin{figure}[hptb]
   \def\epsfsize#1#2{0.35#1}
   \centerline{\epsfbox{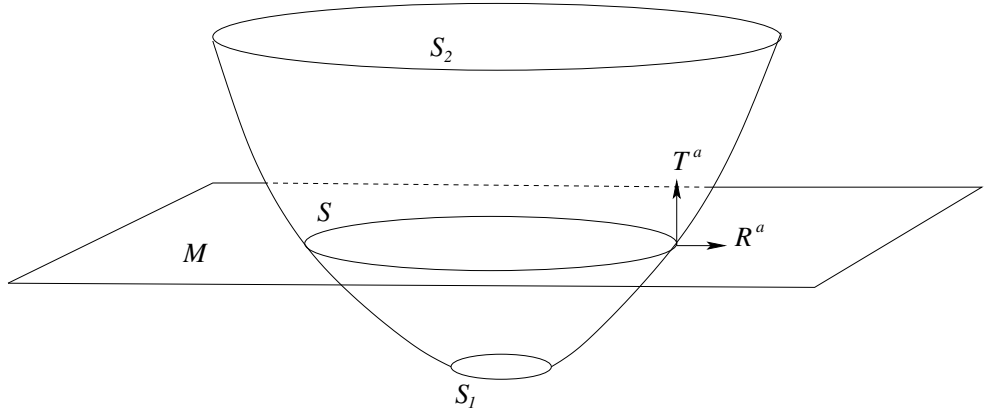}}
   \caption{\it The world tube of apparent horizons and a Cauchy
     surface $M$ intersect in a 2-sphere $S$. $T^a$ is the unit
     time-like normal to $M$ and $R^a$ is the unit space-like normal
     to $S$ within $M$.}
   \label{fig:cauchy}
 \end{figure}
}

Let us first suppose that, in a neighborhood of the cross-section
$S$ of interest, the world tube of marginally trapped surfaces
constitutes an IH. Then the task is to recast
Equation~(\ref{eq:ihangmom}) in terms of the Cauchy data
$(\bar{q}_{ab},\bar{K}_{ab})$ on $M$. This task is also
straightforward~\cite{dkss} and one arrives at\epubtkFootnote{This formula
  has a different sign from that given in~\cite{dkss} due to a
  difference in the sign convention in the definition of the extrinsic
  curvature.}:
\begin{equation}
  \label{eq:angmomK}
  J_\Delta = -\frac{1}{8\pi G}\oint_S
  \varphi^a R^b\bar{K}_{ab}\,d^2V,
\end{equation}
where $R^a$ is the unit radial normal to $S$ in $M$. This formula
is particularly convenient numerically since it involves the
integral of a single component of the extrinsic curvature.

Now consider the dynamical regime, i.e., assume that, in a small
neighborhood of $S$, the world tube of marginally trapped surfaces
is a DH $H$. The angular momentum formula~(\ref{jdynamic2}) on DHs
involves the Cauchy data on $H$. However, it is easy to
show~\cite{ak2} that it equals the expression~(\ref{eq:angmomK})
involving Cauchy data on $M$. Thus, \emph{Equation~(\ref{eq:angmomK})
  is in fact applicable in both the isolated and dynamical regimes.}
The availability of a single formula is
extremely convenient in numerical simulations. From now on, we
will drop the subscript $\Delta$ and denote the angular momentum
of $S$ simply by $J_S^{(\varphi)}$:
\begin{equation}
  \label{eq:generalJ}
  J_S^{(\varphi)} = -\frac{1}{8\pi G}
  \oint_S \varphi^a R^b\bar{K}_{ab}\,d^2V.
\end{equation}
$J_S^{(\varphi)}$ is the intrinsic angular momentum (i.e., spin) of
the black hole at the instant of `time' represented by $S$. Note that
Equation~(\ref{eq:generalJ}) differs from the standard formula for the
ADM angular momentum $J_{\mathrm{ADM}}$ only in that the integral is over
the apparent horizon instead of the sphere at infinity. Finally, we
emphasize that one only assumes that $\varphi^a$ is a rotational Killing
field of the intrinsic 2-metric on $S$; it does not have to extend to
a Killing field even to a neighborhood of $S$. In fact, the
expressions~(\ref{eq:ihangmom}) and~(\ref{jdynamic2}) on IHs and DHs
are meaningful if $\varphi^a$ is replaced by \emph{any} vector field
$\tilde\phi^a$ tangential to $S$; on $\Delta$ the expression is
conserved and on $H$ it satisfies a balance law. Furthermore, they
both equal Equation~(\ref{eq:generalJ}) if $\tilde\varphi^a$ is divergence-free;
it has to Lie-drag only the area 2-form (rather than the intrinsic
metric) on $S$. However, since $S$ admits an infinity of
divergence-free vector fields $\tilde\varphi^a$, there is no a priori
reason to interpret $J_S^{(\tilde\phi)}$ as the `angular momentum' of
the black hole. For this interpretation to be meaningful,
$\tilde\varphi^a$ must be chosen `canonically'. The most unambiguous way
to achieve this is to require that it be a Killing field of the
intrinsic geometry of $S$. However, in Section~\ref{s8} we will
introduce a candidate vector field which could be used in more general
situations which are only `near axi-symmetry'.

Having calculated $J_S$, it is easy to evaluate the mass $M_S$
using Equation~(\ref{eq:ihmass}):
\begin{equation}
  \label{eq:ihmass2} M_S =
  \frac{1}{2GR_\Delta}\sqrt{R_\Delta^4 + 4G^2J_\Delta^2}.
\end{equation}
This method for calculating $J_S$ and $M_S$ has already been used
in numerical simulations, particularly in the analysis of the
gravitational collapse of a neutron star to a black
hole~\cite{whiskey, bks}. It is found that this method is numerically
more accurate than the other commonly used methods.

\begin{description}
\item[Comparison with other methods]~\\
  The approach summarized above has the key advantage that it is
  rooted in Hamiltonian methods which can be universally used to
  define conserved quantities. In particular, it is a natural
  extension of the approach used to arrive at the ADM and Bondi--Sachs
  quantities at spatial and null infinity~\cite{adm, abr, wz}. From
  more `practical' considerations, it has three important features:

  \begin{itemize}
  \item The procedure does not presuppose that the horizon geometry is
    precisely that of the Kerr horizon.
  \item It is coordinate independent.
  \item It only requires data that is intrinsic to the apparent horizon.
  \end{itemize}

  None of the commonly used alternatives share all three of these
  features.

Before the availability of the IH and DH frameworks, standard
procedures of calculating angular momentum were based on
properties of the Kerr geometry. The motivation comes from the
common belief, based on black-hole uniqueness theorems, that a
black hole created in a violent event radiates away all its higher
multipole moments and as it settles down, its near horizon geometry
can be approximated by that of the Kerr solution. The strategy
then is to identify the geometry of $S$ with that of a suitable
member of the Kerr family and read off the corresponding angular
momentum and mass parameters.

The very considerations that lead one to this strategy also show
that it is not suitable in the dynamical regime where the horizon
may be distorted and not well-approximated by any Kerr horizon.
For horizons which have very nearly reached equilibrium, the
strategy is physically well motivated.
However, even in this case, one has to find a way to match the
horizon of the numerical simulation with that of a specific member
in the Kerr family. This is non-trivial because the coordinate
system used in the given simulation will, generically, not bear
any relation to any of the standard coordinate systems used to
describe the Kerr solution. Thus, one cannot just look at, say, a
metric component to extract mass and angular momentum.

A semi-heuristic but most-commonly used procedure is the
\emph{great circle method}. It is based on an observation of the
properties of the Kerr horizon made by Smarr~\cite{smarr} using
Kerr--Schild coordinates. Let $L_\mathrm{e}$ be the length of the equator
and $L_\mathrm{p}$ the length of a polar meridian on the Kerr horizon,
where the equator is the coordinate great circle of maximum proper
length and a polar meridian is a great circle of minimum proper
length. Define a distortion parameter $\delta$ as
$\delta = (L_\mathrm{e}-L_\mathrm{p})/L_\mathrm{e}$. The
knowledge of $\delta$, together with one other quantity such as
the area, $L_\mathrm{e}$, or $L_\mathrm{p}$, is sufficient to find the parameters
$m$ and $a$ of the Kerr geometry. However, difficulties arise when
one wishes to use these ideas to calculate $M$ and $a$ for a
general apparent horizon $S$. For, notions such as great circles,
equator or polar meridian are all highly coordinate dependent.
Indeed, if we represent the standard two-metric on the Kerr
horizon in different coordinates, the great circles in one
coordinate system will not agree with great circles in the other
system. Therefore, already for the Kerr horizon, two coordinate
systems will lead to different answers for $M$ and $a$! In certain
specific situations where one has a good intuition about the
coordinate system being used and the physical situation being
modelled, this method can be useful as a quick way of estimating
angular momentum. However, it has the conceptual drawback that it
is not derived from a well-founded, general principle and the
practical drawback that it suffers from too many ambiguities.
Therefore it is inadequate as a general method.

Problems associated with coordinate dependence can be satisfactorily
resolved on axi-sym\-metric horizons, even when the coordinate system
used in the numerical code is not adapted to the axial symmetry. The
idea is to use the orbits of the Killing vector as analogs of the
lines of latitude on a metric two-sphere. The analog of the equator
is then the orbit of the Killing vector which has maximum proper
length. This defines $L_\mathrm{e}$ in an invariant way. The north and south
poles are the points where the Killing vector vanishes, and the analog
of $L_\mathrm{p}$ is the length of a geodesic joining these two
points. (Because of axial symmetry, all geodesics joining the poles
will have the same length). This geodesic is necessarily perpendicular
to the Killing vector. Hence one just needs to find the length of a
curve joining the north and south poles which is everywhere
perpendicular to the Killing orbits. With $L_\mathrm{e}$ and $L_\mathrm{p}$ defined in
this coordinate invariant way, one can follow the same procedure as in
the great circle method to calculate the mass and angular
momentum. This procedure has been named~\cite{dkss} the
\emph{generalized} great circle method.

How does the generalized great circle method compare to that based
on IHs and DHs? Since one must find the Killing vector, the first
step is the same in the two cases. In the IH and DH method, one
is then left simply with an integration of a component of the
extrinsic curvature on the horizon. In the generalized great
circle method, by contrast, one has to determine the orbit of the
Killing vector with maximum length and also to calculate the
length of a curve joining the poles which is everywhere orthogonal
to the Killing orbits. Numerically, this requires more work and
the numerical errors are at least as large as those in the IH-DH
method. Thus, even if one ignores conceptual considerations
involving the fundamental meaning of conserved quantities, and
furthermore restricts oneself to the non-dynamical regime, the
practical simplicity of the great circle method is lost when it is
made coordinate invariant. To summarize, conceptually,
Equations~(\ref{eq:generalJ}) and~(\ref{eq:ihmass2}) provide the
fundamental definitions of angular momentum and mass, while the great circle
method provides a quick way of estimating these quantities in
suitable situations. By comparing with Equations~(\ref{eq:generalJ})
and~(\ref{eq:ihmass2}), one can calculate errors and sharpen intuition
on the reliability of the great circle method.

A completely different approach to finding the mass and angular
momentum of a black hole in a numerical solution is to use the
concept of a Killing horizon. Assume the existence of Killing
vectors in the neighborhood of the horizon so that mass and
angular momentum are defined as the appropriate Komar integrals.
This method is coordinate independent and does not assume, at
least for angular momentum, that the near horizon geometry is
isometric with the Kerr geometry. But it has two disadvantages.
First, since the Komar integral can involve derivatives of the
Killing field away from the horizon, one has to find the Killing
fields in a neighborhood of the horizon. Second, existence of such
a stationary Killing vector is a strong assumption; it will not be
satisfied in the dynamical regime. Even when the Killing field
exists, computationally it is much more expensive to find it in a
\emph{neighborhood} of the horizon rather than the horizon itself.
Finally, it is not a priori clear how the stationary Killing
vector is to be normalized if it is only known in a neighborhood
of the horizon. In a precise sense, the isolated horizon framework
extracts just the minimum amount of information from a Killing
horizon in order to carry out the Hamiltonian analysis and define
conserved quantities by by-passing these obstacles.
\end{description}


\subsection{Initial data}
\label{s5.2}

In order to accurately calculate the gravitational waveforms for,
say, the coalescence of binary black holes, one must begin with
astrophysically relevant initial data. While there has been some
progress on the construction of such data, the general problem 
is yet to be solved (see, e.g., \cite{gcreview} for a recent review).
However, there has been notable progress in the case of black
holes which are in quasi-equilibrium. Physically, this situation
arises when the black holes are sufficiently far apart for their
orbits to be quasi-periodic. In his section, we will first
summarize this development~\cite{gc1, jgm} and then review an
approach to the calculation of the binding energy in the initial
data. The discussion is based on the IH framework.


\subsubsection{Boundary conditions at the inner boundary}
\label{s5.2.1}

Consider the problem of constructing initial data
$(\bar{q}_{ab},\bar{K}_{ab})$ on $M$, representing a binary black
hole system. This problem has two distinct aspects. The first has
to do with the inner boundaries. A natural avenue to handle the
singularities is to `excise' the region contained in each black
hole (see, e.g., \cite{excision1, excision2}). However, this procedure
creates inner boundaries on $M$ and one must specify suitable
boundary conditions there. The boundary conditions should be
appropriate for the elliptic system of constraint equations, and
they should also capture the idea that the excised region
represents black holes which have certain physical parameters
\emph{and} which are in quasi-equilibrium at the `instant'
represented by $M$. The second aspect of the problem is the
choice of the free data in the bulk. To be of physical interest,
not only must the free data satisfy the appropriate boundary
conditions, but the values in the bulk must also have certain
properties. For example, we might want the black holes to move in
approximately circular orbits around each other, and require that
there be only a minimal amount of spurious gravitational radiation
in the initial data. The black holes would then remain
approximately isolated for a sufficiently long time, and orbit
around each other before finally coalescing.

While a fully satisfactory method of prescribing such initial data
is still lacking, there has been significant progress in recent
years. When the black holes are far apart and moving on
approximately circular orbits, one might expect the trajectory of
the black holes to lie along an approximate helical Killing vector~\cite{bd, friedman1, meudon1, meudon2, abbbbcop}. Using concepts
from the helical Killing vector approximation and working in the
conformal thin-sandwich decomposition of the initial data~\cite{jy}, Cook has introduced the `quasi-equilibrium' boundary
conditions which require that each of the black holes be in
instantaneous equilibrium~\cite{gc1}; see also~\cite{ycsb, cp} for
a similar approach. The relation between these quasi-equilibrium
boundary conditions and the isolated horizon formalism has also
been recently studied~\cite{jgm}.

In this section, we consider only the first aspect mentioned above,
namely the inner boundaries. Physically, the quasi-equilibrium
approximation ought to be valid for time intervals much smaller than
other dynamical time scales in the problem, and the framework assumes
only that the approximation holds infinitesimally `off $M$'. So,
in this section, the type II NEH $\Delta$ will be an
\emph{infinitesimal} world tube of apparent horizons. We assume that there
is an axial symmetry vector $\varphi^a$ on the horizon, although, as
discussed in Section~\ref{s8}, this assumption can be weakened.

Depending on the degree of isolation one wants to impose on the
individual black holes, the inner boundary may be taken to be the
cross-section of either an NEH, WIH, or an IH. The strategy is to
first start with an NEH and impose successively stronger conditions
if necessary. Using the local geometry of intersections $S$ of $M$
with $\Delta$, one can easily calculate the area radius
$R_\Delta$, the angular momentum $J_\Delta$ given by
Equation~(\ref{eq:generalJ}), the canonical surface gravity $\kappa =
\kappa_{\mathrm{Kerr}}(R_\Delta, J_\Delta)$ given by
Equation~(\ref{eq:kerrkappa}), and angular velocity
$\Omega=\Omega_{\mathrm{Kerr}}(R_\Delta,  J_\Delta)$ given by
Equation~(\ref{eq:kerromega}). (Note that while a WIH structure is
used to arrive at the expressions for $J_\Delta$, $\kappa$, and $\Omega$,
the expressions themselves are unambiguously defined also on a
NEH.) These considerations translate directly into restrictions on
the shift vector $\beta^a$ at the horizon. If, as in
Section~\ref{s4.1.3}, one requires that the restriction of the
evolution vector field $t^a$ to $\Delta$ be of the form
\begin{equation}
  t^a = B \ell_0^a - \Omega_{\mathrm{Kerr}}\varphi^a = B T^a +
  (BR^a-\Omega_{\mathrm{Kerr}}\varphi^a),
\end{equation}
where $\ell_0^a = T^a + R^a$, then the values of the lapse and shift
fields at the horizon are given by
\begin{equation}
  \label{eq:shiftbc} \alpha = B,
  \qquad
  \beta^a = \alpha R^a - \Omega_{\mathrm{Kerr}}\varphi^a.
\end{equation}
We are free to choose the lapse $\alpha$ arbitrarily on $S$,
but this fixes the shift $\beta^a$. The tangential part of $\beta^a$
can, if desired, be chosen differently depending on the asymptotic
value of $\beta^a$ at infinity which determines the angular velocity
of inertial observers at infinity.

Next, one imposes the condition that the infinitesimal world-tube of
apparent horizons is an `instantaneous' non-expanding horizon.
This requirement is equivalent to
\begin{equation}
  \label{eq:qebc}
  \mathcal{L}_t \tilde{q}_{ab} = 0,
  \qquad
  \mathcal{L}_\varphi\tilde{q}_{ab} = 0
\end{equation}
on $\Delta$, which imply that the shear and expansion of $\ell^a$
vanish identically. These can be easily translated to conditions
that the Cauchy data $(\bar{q}_{ab}, \bar{K}_{ab})$ must satisfy
at the horizon~\cite{gc1, jgm}. These boundary conditions are
equivalent to the quasi-equilibrium boundary conditions developed
by Cook~\cite{gc1}. However, the isolated horizon formalism allows
greater control on the physical parameters of the black hole. In
particular, we are not restricted just to the co-rotational or
irrotational cases. Furthermore, the use of IHs streamlines the
procedure: Rather than adding conditions one by one as needed, one
begins with a physically well-motivated notion of horizons in
equilibrium and systematically derives its consequences.

Up to this point, the considerations are general in the sense that
they are not tied to a particular method of solving the initial
value problem. However, for the quasi-equilibrium problem, it is
the \emph{conformal thin-sandwich} method~\cite{jy, gcreview} that
appears to be best suited. This approach is based on the
conformal method~\cite{lich, jy1} where we write the 3-metric as
$\bar{q}_{ab}=\psi^4\hat{q}_{ab}$. The free data consists of the
conformal 3-metric $\hat{q}_{ab}$, its time derivative
$\mathcal{L}_t\hat{q}_{ab}$, the lapse $\alpha$, and the trace of the
extrinsic curvature $\bar{K}$. Given this free data, the remaining
quantities, namely the conformal factor and the shift, are
determined by elliptic equations provided appropriate boundary
conditions are specified for them on the horizon\epubtkFootnote{Cook's
  boundary condition on the conformal factor $\psi$ (Equation~(82)
  in~\cite{gc1}) is equivalent to $\Theta_{(\ell)}=0$ which (in the
  co-rotating case, or more generally, when the 2-metric on $S$ is
  axi-symmetric) reduces to $\mathcal{L}_t \psi =0$ on $S$. The Yo et
  al.\ boundary condition on $\psi$ (Equation~(48) of~\cite{ycsb}) is
  equivalent to $\mathcal{L}_{\bar{t}} \psi =0$ on $S$, where,
  however, the evolution vector field $\bar{t}^a$ is obtained by a
  superposition of two Kerr--Schild data.}.
It turns out~\cite{jgm} that the horizon conditions~(\ref{eq:qebc})
are well-tailored for this purpose. While the
issue of existence and uniqueness of solutions using these
boundary conditions has not been proven, it is often the case that
numerical calculations are convergent and the resulting solutions are
well behaved. Thus, these conditions might therefore be sufficient
from a practical point of view.

In the above discussion, the free data consisted of
$(\hat{q}_{ab},\mathcal{L}_t\hat{q}_{ab},\alpha,\bar{K})$ and one solved
elliptic equations for $(\psi,\beta^a)$. However, it is common to
consider an enlarged initial value problem by taking
$\mathcal{L}_t\bar{K}$ as part of the free data (usually set to zero) and
solving an elliptic equation for $\alpha$. We now need to
prescribe an additional boundary condition for $\alpha$. It turns
out that this can be done by using WIHs, i.e., by bringing in
surface gravity, which did not play any role so far. From the
definition of surface gravity in Equations~(\ref{eq:omegadef},
\ref{eq:surfacegravity}), it is clear that the expression for
$\kappa$ will involve a time derivative; in particular, it turns
out to involve the time derivative of $\alpha$. It can be shown
that by choosing $\mathcal{L}_t\alpha$ on $S$ (e.g., by taking
$\mathcal{L}_t\alpha=0$) and requiring surface gravity to be constant on
$S$ and equal to $\kappa_{\mathrm{Kerr}}$, one obtains a suitable
boundary condition for $\alpha$. (The freedom to choose freely the
function $\mathcal{L}_t\alpha$ mirrors the fact that fixing surface
gravity does not uniquely fix the rescaling freedom of the null
normal.)  Note that $\kappa$ is required to be constant only on
$S$, not on $\Delta$. To ask it to be constant on $\Delta$ would
require $\mathcal{L}_t\kappa=0$, which in turn would restrict the
\emph{second} time derivative; this necessarily involves the
evolution equations, and they are not part of the initial data scheme.

One may imagine using the yet stronger notion of an IH, to
completely fix the value of the lapse at the horizon. But this
requires solving an elliptic equation on the horizon and the
relevant elliptic operator has a large kernel~\cite{abl1, gc1}.
Nonetheless, the class of initial data on which its inverse exists
is infinite dimensional so that the method may be useful in
practice. However, this condition would genuinely restrict the
permissible initial data sets. In this sense, while the degree of
isolation implied by the IH boundary condition is likely to be met
in the asymptotic future, for quasi-equilibrium \emph{initial
data} it is too strong in general. It is the WIH boundary
conditions that appear to be well-tailored for this application.

Finally, using methods introduced by Dain~\cite{sd2}, a variation
of the above procedure was recently introduced to establish the
existence and uniqueness of solutions and to ensure that the
conformal factor $\psi$ is everywhere positive~\cite{djk}. One
again imposes Equation~(\ref{eq:qebc}). However, in place of Dirichlet
boundary conditions~(\ref{eq:shiftbc}) on the shift, one now
imposes Neumann-type conditions on certain components of
$\beta^a$. This method is expected to be applicable all initial
data constructions relying on the conformal method. Furthermore,
the result might also be of practical use in numerical
constructions to ensure that the codes converge to a well behaved
solution.


\subsubsection{Binding energy}
\label{5.2.2}

For initial data representing a binary black hole system, the
quantity $E\mathrm{b} = M_\mathrm{ADM} - M_1-M_2$ is called \emph{the effective
binding energy}, where $M_\mathrm{ADM}$ is the ADM mass, and $M_{1, 2}$ are
the individual masses of the two black holes. Heuristically, even
in vacuum general relativity, one would expect $E\mathrm{b}$ to have
several components. First there is the analog of the Newtonian
potential energy and the spin-spin interaction, both of which may
be interpreted as contributing to the binding energy. But $E\mathrm{b}$
also contains contributions from kinetic energy due to momentum
and orbital angular momentum of black holes, and energy in the
gravitational radiation in the initial data. It is only when these
are negligible that $E\mathrm{b}$ is a good measure of the physical
binding energy.

The first calculation of binding energy was made by Brill and
Lindquist in such a context. They considered two non-spinning
black holes initially at rest~\cite{bl}. For large separations, they
found that, in a certain mathematical sense, the leading contribution
to binding energy comes just from the usual Newtonian gravitational
potential. More recently, Dain~\cite{sd} has extended this calculation
to the case of black holes with spin and has shown that the spin-spin
interaction energy is correctly incorporated in the same sense.

In numerical relativity, the notion of binding energy has been
used to locate sequences of quasi-circular orbits. The underlying
heuristic idea is to minimize $E\mathrm{b}$ with respect to the proper
separation between the holes, keeping the physical parameters of
the black holes fixed. The value of the separation which minimizes
$E\mathrm{b}$ provides an estimate of sizes of stable `circular' orbits~\cite{gc2, tb, ptc1}. One finds that these orbits do not exist if
the orbital angular momentum is smaller than a critical value
(which depends on other parameters) and uses this fact to
approximately locate the `inner-most stable circular orbit'
(ISCO). In another application, the binding energy has been used
to compare different initial data sets which are meant to describe
the same physical system. If the initial data sets have the same
values of the black hole masses, angular momenta, linear momenta,
orbital angular momenta, and relative separation, then any
differences in $E\mathrm{b}$ should be due only to the different radiation
content. Therefore, minimization of $E\mathrm{b}$ corresponds to
minimization of the amount of radiation in the initial data~\cite{pct2}.

In all these applications, it is important that the physical
parameters of the black holes are calculated accurately. To
illustrate the potential problems, let us return to the original
Brill--Lindquist calculation~\cite{bl}. The topology of the spatial
slice is $\mathbb{R}^3$ with two points (called `punctures')
removed. These punctures do not represent curvature singularities.
Rather, each of them represents a spatial infinity of an
asymptotically flat region which is hidden behind an apparent
horizon. This is a generalization of the familiar Einstein--Rosen
bridge in the maximally extended Schwarzschild solution. The
black hole masses $M_1$ and $M_2$ are taken to be the ADM
masses of the corresponding hidden asymptotic regions.
(Similarly, in~\cite{sd}, the angular momentum of each hole is
defined to be the ADM angular momentum at the corresponding
puncture.) Comparison between $E\mathrm{b}$ and the Newtonian binding
energy requires us to define the distance between the holes. This
is taken to be the distance between the punctures in a
\emph{fiducial flat background metric}; the physical distance
between the two punctures is infinite since they represent
asymptotic ends of the spatial 3-manifold. Therefore, the sense in
which one recovers the Newtonian binding energy as the leading
term is physically rather obscure.

Let us re-examine the procedure with an eye to extending it to a
more general context. Let us begin with the definition of masses
of individual holes which are taken to be the ADM masses in the
respective asymptotic regions. How do we know that these are the
physically appropriate quantities for calculating the potential
energy? Furthermore, there exist initial data sets (e.g., Misner
data~\cite{misner1, misner2}) in which each black hole does not have
separate asymptotic regions; there are only two common asymptotic
regions connected by two wormholes. For these cases, the use of
ADM quantities is clearly inadequate. The same limitations apply to
the assignment of angular momentum.

A natural way to resolve these conceptual issues is to let the
horizons, rather than the punctures, represent black holes.
Thus, in the spirit of the IH and DH frameworks, it is more
appropriate to calculate the mass and angular momentum using
expressions~(\ref{eq:ihmass2}, \ref{eq:generalJ}) which involve
the geometry of the two apparent horizons. (This requires the
apparent horizons to be axi-symmetric, but this limitation could be
overcome following the procedure suggested in Section~\ref{s8}.)
Similarly, the physical distance between the black holes should be
the smallest proper distance between the two apparent horizons. To
test the viability of this approach, one can repeat the original
Brill--Lindquist calculation in the limit when the black holes are
far apart~\cite{bkthesis}. One first approximately locates the
apparent horizon, finds the proper distance $d$ between them, and
then calculates the horizon masses (and thereby $E_\mathrm{b}$) as a power
series in $1/d$. The leading-order term does turn out to be the
usual Newtonian gravitational potential energy but the higher
order terms are now different from~\cite{bl}. Similarly, it would
be interesting to repeat this for the case of spinning black holes
and recover the leading order term of~\cite{sd} within this more
physical paradigm using, say, the Bowen--York initial data. This
result would re-enforce the physical ideas and the approach can
then be used as a well defined method for calculating binding
energy in more general situations.

\begin{description}
\item[Remark]~\\
  In the examples mentioned above, we focussed on
mass and angular momentum; linear momentum was not considered. A
meaningful definition for the linear momentum of a black hole
would be useful for several problems: It would enable one to
define the orbital angular momentum, locate quasi-circular orbits
and enable a comparison between different initial data sets.
However, just as angular momentum is associated with rotational
symmetry, linear momentum ought to be associated with a space
translational symmetry. In a general curved space-time, one does
not have even an approximate notion of translations except at
infinity. Thus, whenever one speaks of linear momentum of initial
data (e.g., in the Bowen--York construction~\cite{by}) the
quantity is taken to be the appropriate component of the ADM
momentum \emph{at infinity}. For reasons discussed above, it is
more appropriate to associate black hole parameters with the
respective horizons. So, a question naturally arises: Can one
define linear momentum as an integral over the apparent horizon?
If $z^a$ is a vector field on $M$ which, for some reason, can be
regarded as an approximate translational symmetry, and $S$ the
apparent horizon, then one might naively define the linear
momentum associated with $z^a$ as
\begin{equation}
  \label{mom}
  P_S^{(z)} = \frac{1}{8\pi}\oint_{S} \!
  \left({\bar{K}}_{ab}-\bar{K}\bar{q}_{ab}\right) z^a R^b \, d^2S.
\end{equation}
Unfortunately, this definition is not satisfactory. For example,
in the boosted Kerr--Schild metric, if one lets $z^a$ to be the
translational vector field of the background Minkowski space,
Equation~(\ref{mom}) does not yield the physically expected linear momentum
of the boosted black hole. See~\cite{bkthesis} for further
discussion and for other possible ways to define linear momentum.
Further work is needed to address this question satisfactorily.
\end{description}


\subsection{Black hole multipole moments}
\label{s5.3}

Let us begin by considering the notion of source and field
multipole moments in Newtonian gravity and in flat space
electrodynamics. Field multipoles appear in the asymptotic
expansions of the fields at infinity while the source multipoles
are defined in terms of the mass or charge distribution of the
source. These two sets of multipole moments are related to each
other via field equations. The same is true in \emph{linearized}
general relativity~\cite{sb}. Also, the well known quadrupole
formula relates the rate of change of the quadrupole moment to the
energy flux at infinity due to gravitational waves.

The situation in exact, non-linear general relativity is not so
simple. Using the geometric structure of the gravitational field
near spatial infinity, the field multipoles for stationary
space-times were studied by Geroch, Hansen, Beig, Simon, and
others~\cite{rg, rh, bs1, bs2, rb1, bs3}. They found that, just as in
electrodynamics, the gravitational field has two sets of
multipoles: The mass multipoles $M_n$ and the angular momentum
multipoles $J_n$. The knowledge of these multipole moments
suffices to determine the space-time geometry in a neighborhood of
spatial infinity~\cite{bs2, rb1, bs3}. Thus, at least in the
context of stationary space-times, the field multipole moments are
well understood. However, in problems involving equations of motion,
it is the \emph{source multipoles} that are of more direct
interest. It is natural to ask if these can be defined for black
holes.

The answer is affirmative for black holes in equilibrium, which
can be represented by isolated horizons. For simplicity, we will
consider only type II (i.e., axisymmetric), non-extremal isolated
horizons in vacuum. The source multipoles are two sets $M_n$ and
$J_n$ of numbers which provide a diffeomorphism invariant
characterization of the horizon geometry.

As before, let $S$ be a cross-section of $\Delta$. We denote the
intrinsic Riemannian metric on it by $\tilde{q}_{ab}$, the
corresponding area 2-form by $\tilde{\epsilon}_{ab}$, and the
derivative operator by $\tilde{D}_a$. Since the horizon is of
type II, there exists a vector field $\varphi^a$ on $S$ such that
$\mathcal{L}_\varphi\tilde{q}_{ab}=0$. The two points where $\varphi^a$
vanishes are called the \emph{poles} of $S$. The integral curves
of $\varphi^a$ are natural candidates for the `lines of latitude'
on $S$, and the lines of longitude are the curves which connect the
two poles and are orthogonal to $\varphi^a$. This leads to an
invariantly defined coordinate $\zeta \in [-1, 1]$ -- the analog of the
function $\cos\theta$ in usual spherical coordinates -- defined by
\begin{equation}
  \label{eq:zetadef}
  \tilde{D}_a\zeta = \frac{1}{R^2}\tilde{\epsilon}_{ba}\varphi^b,
\end{equation}
where $R$ is the area radius of $S$. The freedom of adding a constant
to $\zeta$ is removed by requiring $\oint_S\zeta\tilde{\epsilon} = 0$.
With $\phi\in [0, 2\pi)$ being an affine parameter
along $\varphi^a$, the 2-metric $\tilde{q}_{ab}$ takes the
canonical form
\begin{equation}
  \label{eq:canonical2metric}
  \tilde{q}_{ab} = R^2(f^{-1}\tilde{D}_a\zeta\tilde{D}_b\zeta +
  f\tilde{D}_a\phi\tilde{D}_b\phi),
\end{equation}
where $f=\varphi_a\varphi^a/R^2$. The only remaining freedom in the
choice of coordinates is a rigid shift in $\phi$. Thus, in any
axi-symmetric geometry, there is an invariantly defined coordinate
$\zeta$, and multipole moments are defined using the Legendre
polynomials in $\zeta$ as weight functions.

Recall from Section~\ref{s2.1.3} that the invariant content in the
geometry of an isolated horizon is coded in (the value of its area
and) $\Psi_2$. The real part of $\Psi_2$ is proportional to the
scalar curvature $\tilde{R}$ of $\tilde{q}_{ab}$ and captures
distortions~\cite{dis3, dis4}, while the imaginary part of $\Psi_2$
yields the curl of $\omega_a$ and captures the angular momentum
information. (The free function $f$ in Equation~(\ref{eq:canonical2metric})
determines and is completely determined by the scalar curvature
$\tilde{R}$.) Multipoles are constructed directly from $\Psi_2$.
The angular momentum multipoles are defined as
\begin{equation}
  J_n = -\sqrt{\frac{4\pi}{2n+1}}\frac{R_\Delta^{n+1}}{4\pi G}
  \oint_S Y_n^0(\zeta) \imaginary[\Psi_2] \,d^2V,
\end{equation}
where $Y_n^0(\zeta)$ are the spherical harmonics. (Axi-symmetry
ensures that $Y_n^m$ vanish if $m \not=0$.) The normalization
factors have been chosen to ensure that the dimensions of multipoles
are the physically expected ones and $J_1$ is the angular momentum
of the isolated horizon. The mass multipoles are defined similarly
as the moments of $\tilde{R}$:
\begin{equation}
  M_n = - \sqrt{\frac{4\pi}{2n+1}}
  \frac{M_\Delta R_\Delta^n}{2\pi}
  \oint_S Y_n^0(\zeta) \real[\Psi_2] \,d^2V,
\end{equation}
where the normalization factor has been chosen in order to ensure
the correct physical dimensions and $M_0= M_\Delta$.

These multipoles have a number of physically desired properties:

\begin{itemize}
\item The angular momentum monopole vanishes, $J_0=0$. (This is
  because we are considering only smooth fields. When $\omega_a$ has a
  wire singularity similar to the Dirac monopole in electrodynamics,
  i.e., when the horizon has NUT charge, then $J_0$ would be
  non-zero.)
\item The mass dipole vanishes, $M_1=0$. This tells us that we are in
  the rest frame of the horizon.
\item For a type I (i.e., spherically symmetric) isolated horizon,
  $\imaginary[\Psi_2]=0$ and $\real[\Psi_2]$ is constant. This implies
  that $M_0$ is the only non-zero multipole moment.
\item If the horizon is symmetric under reflections as in the Kerr
  solution (i.e., $\Psi_2 \mapsto \Psi_2^\star$ when
  $\zeta\mapsto-\zeta$), then $M_n$ vanishes for odd $n$ while $J_n$
  vanishes for even $n$.
\end{itemize}

There is a one-one correspondence between the multipole moments
$\{J_n, M_n\}$ and the geometry of the horizon: Given the horizon
area $a_\Delta$ and multipoles $\{J_n, M_n\}$, assuming the
multipoles satisfy a convergence condition for large $n$, we can
reconstruct a non-extremal isolated horizon geometry
$(\Delta, q_{ab},\mathcal{D}_a)$, uniquely up to diffeomorphisms, such that
the area of $\Delta$ is $a_\Delta$ and its multipole moments are
the given $\{J_n, M_n\}$. In vacuum, stationary space-times, the
multipole moments also suffice to determine the space-time
geometry in the vicinity of the horizon. Thus, we see that the
horizon multipole moments have the expected properties. In the
extremal case, because of a surprising uniqueness result~\cite{lp3},
the $M_n, J_n$ are universal -- the same as those on
the extremal Kerr IH and the `true multipoles' which can
distinguish one extremal IH from another are constructed using
different fields in place of $\Psi_2$~\cite{aepv}. Finally, note
that there is no a-priori reason for these source multipoles to
agree with the field multipoles at infinity; there could be matter
fields or radiation outside the horizon which contribute to the
field multipoles at infinity. The two sets of quantities need not
agree even for stationary, vacuum space-times because of
contributions from the gravitational field in the exterior region.
For the Kerr space-time, the source and field moments are indeed
different for $n\ge 2$. However, the difference is small for low
$n$~\cite{aepv}.

See~\cite{aepv} for further discussion and for the inclusion of
electromagnetic fields, and~\cite{bks} for the numerical
implementation of these results.


\subsection{Waveform extraction}
\label{s5.4}

The prospect of receiving gravitational waves offers exciting
opportunities for astrophysics and for testing the
\emph{dynamical, strong field regime} of general relativity (see,
e.g., \cite{sourcesreview}). Black holes are among the most
promising sources both for terrestrial and space-based
observatories. Extracting gravitational waveform resulting from,
say, gravitational collapse~\cite{collapsereview} or black hole
mergers~\cite{lhreview} is one of the important goals of numerical
relativity. The eventual aim is to provide waveforms which, in
conjunction with gravitational wave detectors, can be used to
study the astrophysics of these gravitational wave sources. In
this section, we summarize a method of waveform extraction based
on the isolated horizon formalism.

The theory of gravitational radiation in exact general relativity is
based on structures defined at future null infinity $\mathcal{I}^+$.
In particular, associated with every cross-section of $\mathcal{I}^+$
-- which represents a retarded `instant of time' -- there is a well
defined notion of mass, introduced by Bondi~\cite{bondiflux}, which
decreases as gravitational radiation flows across $\mathcal{I}^+$. On
the other hand, except for those based on conformal methods, most
simulations only deal with a finite portion of space-time and thus
have no direct access to $\mathcal{I}^+$. Instead, one usually uses
scalars such as the Weyl tensor component $\Psi_4$ to define the
radiation waveform. However, $\Psi_4$ depends on the choice of a null
tetrad as well as coordinates. While a natural tetrad is available for
the perturbation theory on Kerr back ground, this is not true in
general. In this section we will sketch an approach to solve both
these problems using the isolated horizon framework: One can construct
an approximate analog of $\mathcal{I}^+$ for a suitable, \emph{finite}
portion of space-time, and introduce a geometrically defined null
tetrad and coordinates to extract gauge invariant, radiative
information from simulations.

\epubtkImage{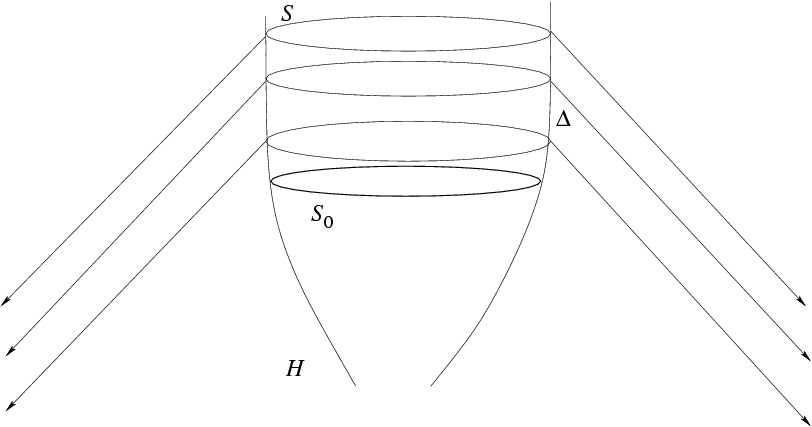}
{\begin{figure}[hptb]
   \def\epsfsize#1#2{0.5#1}
   \centerline{\epsfbox{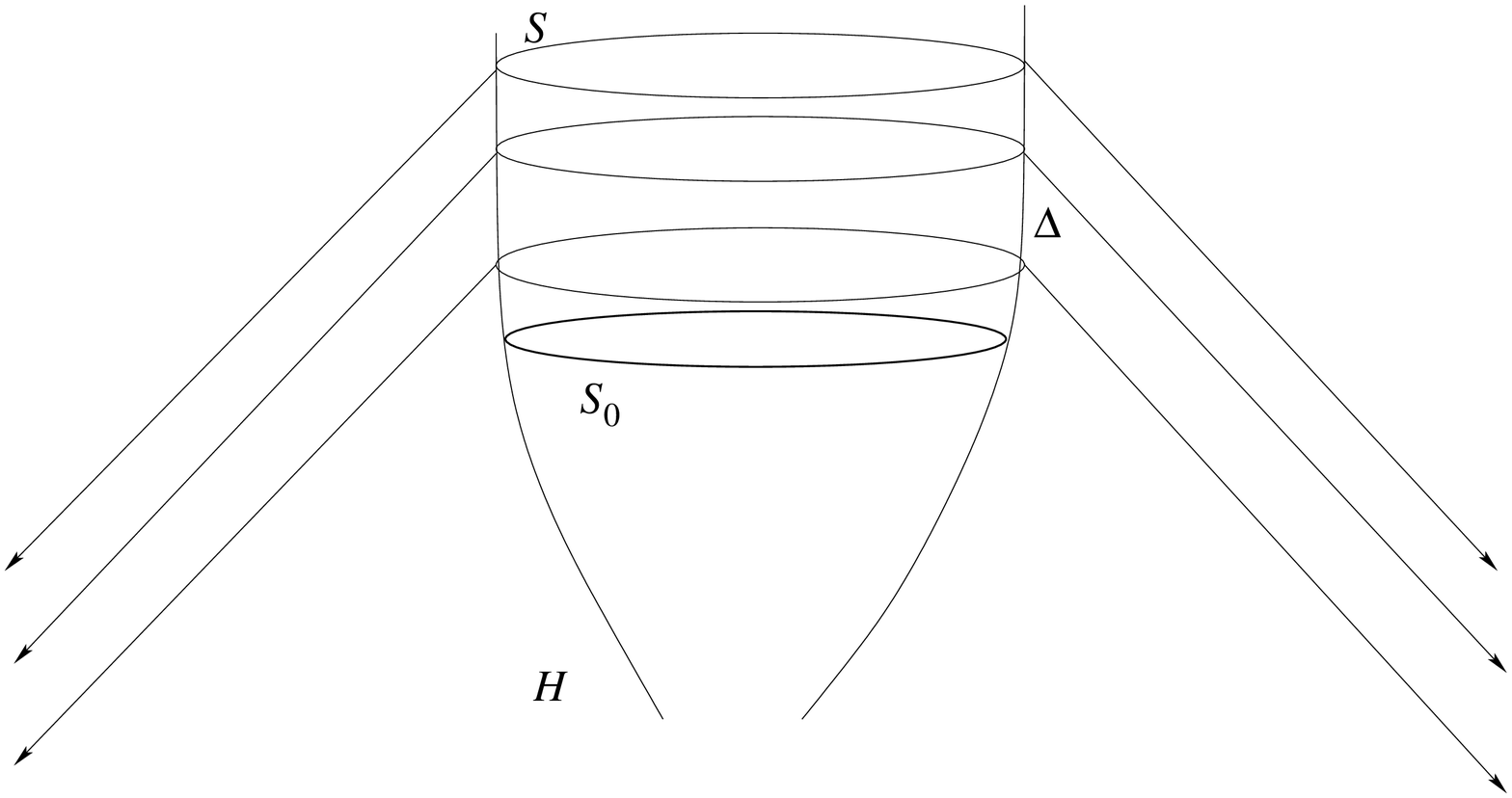}}
   \caption{\it Bondi-like coordinates in a neighborhood of $\Delta$.}
   \label{fig:bondi}
 \end{figure}
}

Our procedure is analogous to the one used at null infinity to
construct the Bondi-type coordinates starting from
$\mathcal{I}^+$. Let us assume that $(\Delta, [\ell])$ is an IH and
consider its preferred foliation (i.e., the ``rest frame'' of the
horizon) defined in Section~\ref{s2.1.3}. Using terminology from
null infinity, we will refer to the leaves of the foliation as the
\emph{good cuts} of $\Delta$. Pick a null normal $\ell^a$ from the
$[\ell^a]$; this can be done, e.g., by choosing a specific value for
surface gravity. Let $n_a$ be the unique 1-form satisfying
$\ell^an_a=-1$ and which is orthogonal to the good cuts. Let
$(v,\theta,\phi)$ be coordinates on $\Delta$ such that $v$ is an
affine parameter along $\ell^a$, and the good cuts are given by
surfaces of constant $v$. Here $\theta$ and $\phi$ are coordinates
on the good cut satisfying $\mathcal{L}_\ell\theta = 0$ and
$\mathcal{L}_\ell\phi=0$. Next, consider past null geodesics emanating
from the good cuts with $-n^a$ as their initial tangent vector.
Let the null geodesics be affinely parameterized and let the
affine parameter be called $r$ and set $r=r_0$ on $\Delta$. Lie
drag the coordinates $(v,\theta,\phi)$ along null geodesics. This
leads to a set of coordinates $(r, v,\theta,\phi)$ in a
\emph{neighborhood} of the horizon. The only arbitrariness in
this coordinate system is in the choice of $(\theta,\phi)$ on one
good cut and the choice of a number $r_0$. Using vacuum Einstein's
equations, one can obtain a systematic expansion of the metric in
inverse powers of $(r-r_0)$~\cite{ihprl}.

One can also define a null tetrad in the neighborhood in a similar
fashion. Let $m^a$ be an arbitrary complex vector tangent to the
good cuts such that $(\ell, n,m,\bar{m})$ is a null tetrad on
$\Delta$. Using parallel transport along the null geodesics, we
can define a null tetrad in the neighborhood. This tetrad is
unique up to the spin rotations of $m$: $m\rightarrow
e^{2i\theta}m$ on the fiducial good cut. This construction is
shown in Figure~\ref{fig:bondi}. The domain of validity of the
coordinate system and tetrads is the space-time region in which
the null geodesics emanating from good cuts do not cross. 

Numerically, it might be possible to adapt 
existing event horizon finders to locate the outgoing past null cone
of a cross-section of the horizon. This is because event horizon
trackers also track null surfaces backward in time~\cite{pd1, pd2}; the
event horizon 
is the ingoing past null surface starting from sufficiently close to
future timelike infinity while here we are interested in the
\emph{outgoing} past directed null surface starting from an apparent
horizon. 

In the Schwarzschild solution, it can be shown analytically that this
coordinate system covers an entire asymptotic region. In the 
Kerr space-time, the domain of validity is not explicitly known,
but in a numerical implementation, this procedure does not
encounter geodesic crossing in the region of interest to the
simulation~\cite{pd1}. In general space-times, the extraction of wave forms 
requires the construction to go through only along the past light
cones of good cuts which lie in the distant future, whence problems
with geodesic-crossing are unlikely to prevent one from covering a
sufficiently large region with these coordinates. This
invariantly defined structure provides a new approach to extract
waveforms. First, the null tetrad presented above can be used to
calculate $\Psi_4$. There is only a phase factor ambiguity (which
is a function independent of $v$ and $r$) inherited from the
ambiguity in the choice of $m^a$ on the fiducial good cut. Second,
the past null cone of a good cut at a sufficiently late time can
be used as an approximate null infinity. This should enable one
to calculate dynamical quantities such as the analogs of the Bondi
mass and the rate of energy loss from the black hole, now on the
`approximate' $\mathcal{I}^+$. However, a detailed framework to
extract the approximate expressions for fluxes of energy and Bondi
mass, with sufficient control on the errors, is yet to be
developed. This is a very interesting analytical problem since its
solution would provide numerical relativists with an algorithm to
extract waveforms and fluxes of energy in gravitational waves in
an invariant and physically reliable manner. Finally, the
invariant coordinates and tetrads also enable one to compare late
time results of distinct numerical simulations.

\newpage


\section{Applications in Mathematical Physics}
\label{s6}

For simplicity, in Section~\ref{s4} we restricted our review of
the laws of black hole mechanics to the Einstein--Maxwell theory.
However, there is a large body of literature on black holes in
more general theories with dilatonic, Yang--Mills, Higgs, Proca, and
Skyrme fields. These fields are not expected to be physically
significant in the macroscopic, astrophysical world. However, they
are of considerable interest from a mathematical physics
perspective because their inclusion brings about qualitative,
structural changes in the theory. The most dramatic of these is
that the uniqueness theorems that play a central role in the
Einstein--Maxwell theory are no longer valid. Consequently, the
structure of these `colored' or `hairy' black-holes is much more
complicated than those in the Einstein--Maxwell theory, and most of
the work in this area has been carried out through a combination
of analytic and numerical methods. By now a very large body of
facts about stationary black holes with hair has accumulated. A
major challenge is to unify this knowledge through a few, general
principles.

The isolated horizon framework has provided surprising insights
into  the properties of hairy black holes in
equilibrium~\cite{acdilaton, cs, cns, afk, acs, cns2}. While the
zeroth and first laws go through in a straightforward manner, the
notion of the horizon mass now becomes much more subtle and its
properties have interesting consequences. The framework also suggests
a new phenomenological model of colored black holes as \emph{bound
states} of ordinary, uncolored black holes and solitons. This
model successfully explains the qualitative behavior of these
black holes, including their stability and instability, and
provides unexpected \emph{quantitative} relations between colored
black holes and their solitonic analogs.

In these theories, matter fields are minimally coupled to gravity.
If one allows non-minimal couplings, the first law itself is
modified in a striking fashion: Entropy is no longer given by the
horizon area but depends also on the matter fields. For globally
stationary space-times admitting bifurcate Killing horizons, this
result was first established by Jacobson, Kang, and Myers~\cite{jkm},
and by Iyer and Wald~\cite{iw1, iw2} for a general class of
theories. For scalar fields non-minimally coupled to gravity, it has
been generalized in the setting of Type II WIHs~\cite{acs2}. While the
procedure does involve certain technical subtleties, the overall
strategy is identical to that summarized in
Section~\ref{s4.1}. Therefore we will not review this issue in detail.

This section is divided into two parts. In the first, we discuss
the mechanics of weakly isolated horizons in presence of dilatons
and Yang--Mills fields. In the second, we discuss three
applications. This entire discussion is
in the framework of isolated horizons because the effects of these
fields on black hole dynamics remain largely unexplored.


\subsection{Beyond Einstein--Maxwell theory}
\label{s6.1}

Our primary purpose in this section is to illustrate the
differences from the Einstein--Maxwell theory. These stem from
`internal charges' and other `quantum numbers' that are unrelated
to angular momentum. Therefore, for simplicity, we will restrict
ourselves to non-rotating weakly isolated horizons. Extension to
include angular momentum is rather straightforward.


\subsubsection{Dilatonic couplings}
\label{s6.1.1}

In dilaton gravity, the Einstein--Maxwell theory is supplemented
with a scalar field -- called the dilaton -- and (in the Einstein
frame) the Maxwell part of the action is replaced by
\begin{equation}
  S_\mathrm{dil} (\phi, A) = -\frac{1}{16\pi}
  \int_\mathcal{M} \!\! \left[ 2(\nabla\phi)^2 +
  e^{-2\alpha\phi}\mathbf{F}_{ab}\mathbf{F}^{ab} \right], d^4v,
  \label{dil:act}
\end{equation}
where $\alpha\ge0 $ is a free parameter which governs the strength
of the coupling of $\phi$ to the Maxwell field $\mathbf{F}_{ab}$. If
$\alpha=0$, one recovers the standard
Einstein--Maxwell--Klein--Gordon system, while $\alpha =1$ occurs in a
low energy limit of string theory. For our  illustrative purposes,
it will suffice to consider the $\alpha =1$ case.

At spatial infinity, one now has three charges: the ADM mass
$M_\mathrm{ADM}$, the usual electric charge $Q_\infty$, and another
charge $\tilde{Q}_\infty$:
\begin{equation}
  Q_\infty = \frac{1}{4\pi} \oint_{S_\infty} \!\!\!\!\! {}^\star\mathbf{F},
  \qquad
  \tilde{Q}_\infty = \frac{1}{4\pi} \oint_{S_\infty} \!\!\!\!\!
  e^{-2\phi}\, {}^\star\mathbf{F}.
\end{equation}
$\tilde{Q}_\infty$ is conserved in space-time (i.e., its value does
not change if the 2-sphere of integration is deformed) while
$Q_\infty$ is not. From the perspective of weakly isolated
horizons, it is more useful to use $a_\Delta$, $Q_\Delta$, and
$\tilde{Q}_\Delta$ as the basic charges:
\begin{equation}
  Q_\Delta = \frac{1}{4\pi} \oint_{S} \! {}^\star\mathbf{F},
  \qquad
  \tilde{Q}_\Delta = \frac{1}{4\pi} \oint_{S} e^{-2\phi}
  {}^\star\mathbf{F},
\end{equation}
where $S$ is any cross-section of $\Delta$. Although the standard
electric charge is not conserved in space-time, it \emph{is}
conserved along $\Delta$ whence $Q_\Delta$ is well-defined.

It is straightforward to extend the Hamiltonian framework of
Section~\ref{s4.1} to include the dilaton. To define energy, one
can again seek \emph{live} time-translation vector fields $t^a$,
evolution along which is Hamiltonian. The necessary and
sufficient condition now becomes the following: There should exist
a phase space function $E^t_\Delta$, constructed from horizon fields,
such that
\begin{equation}
  \label{dilaton1}
  \delta E^t_\Delta = \frac{\kappa_{(t)}}{8\pi G} \delta a_\Delta +
  \Phi_{(t)}\, \delta \tilde{Q}_\Delta.
\end{equation}
Thus, the only difference from the Einstein--Maxwell case is that
$Q_\Delta$ is now replaced by $\tilde{Q}_\Delta$. Again, there exists an
infinite number of such live vector fields and one can construct
them systematically starting with any (suitably regular) function
$\kappa_0(a_\Delta, \tilde{Q}_\Delta)$ of the horizon area and charge
and requiring $\kappa_{(t)}$ should equal $\kappa_0$.

The major difference arises in the next step, when one attempts to
construct a preferred $t_0^a$. With the dilatonic coupling, the
theory has a unique \emph{three} parameter family of static
solutions which can be labelled by $(a_\Delta, Q_\Delta,
\tilde{Q}_\Delta )$~\cite{gm, ghs1, ghs2, masood}. As in the Reissner
Nordstr\"om family, these solutions are spherically symmetric. In
terms of these parameters, the surface gravity $\kappa_{(\xi)}$ of
the static Killing field $\xi^a$, which is unit at infinity, is given
by
\begin{equation}
  \kappa_{(\xi)} = \frac{1}{2R_\Delta}
  \left( 1 + 2G \frac{Q_\Delta \tilde Q_\Delta}{R_\Delta^2} \right)
  \left( 1 - 2G \frac{Q_\Delta \tilde Q_\Delta}{R_\Delta^2}
  \right)^{-\frac{1}{2}}\!\!\!\!\!\!\!\!.
\end{equation}
The problem in the construction of the preferred $t_0^a$ is that
we need a function $\kappa_0$ which depends only on $a_\Delta$ and
$\tilde{Q}_\Delta$ and, since $\kappa_{(\xi)}$ depends on all three
horizon parameters, one can no longer set $\kappa_0 =
\kappa_{(\xi)}$ on the entire phase space. Thus, there is no live
vector field $t^a_0$ which can generate a Hamiltonian evolution
\emph{and} agree with the time-translation Killing field in all
static solutions. It was the availability of such live vector
fields that provided a canonical notion of the horizon mass in the
Einstein--Maxwell theory in Section~\ref{s4.1.3}.

One can weaken the requirements by working on sectors of phase
space with fixed values of $Q_\Delta$. On each sector,
$\kappa_{(\xi)}$ trivially depends only on $a_\Delta$ and
$\tilde{Q}_\Delta$. So one can set $\kappa_0 = \kappa_{(\xi)}$,
select a canonical $t_0$, and obtain a mass function $M_\Delta =
E^{t_0}_\Delta$. However, now the first law~(\ref{dilaton1}) is
satisfied only if the variation is restricted such that $\delta
Q_\Delta =0$. For general variation, one has the modified
law~\cite{afk}
\begin{equation}
  \label{dilaton2}
  \delta{M}_\Delta = \frac{\kappa}{8\pi G} \delta{a_\Delta} +
  \hat{\Phi} \, \delta \hat{Q}_\Delta,
\end{equation}
where $\kappa = \kappa_{(t_0)}$, $\hat\Phi^2 = (Q_\Delta
\tilde{Q}_\Delta/R_\Delta^2)$ and $\hat{Q}_{\Delta}^2 = Q_\Delta
\tilde{Q}_\Delta$. Thus, although there is still a first law in
terms of $t^a_0$ and $M_\Delta$, it does not have the canonical
form~(\ref{dilaton1}) because $t_0^a$ does not generate
Hamiltonian evolution on the entire phase space. More generally,
in theories with multiple scalar fields, if one focuses only on
static sectors, one obtains similar `non-standard' forms of the
first law with new `work terms' involving scalar
fields~\cite{gkk}. From the restricted perspective of static sector,
this is just a fact. The isolated horizon framework provides a
deeper underlying reason: In these theories, there is no evolution
vector field $t^{a}$ defined for all points of the phase space,
which coincides with the properly normalized Killing field on all
static solutions, \emph{and} evolution along which is Hamiltonian
on the full phase space.


\subsubsection{Yang--Mills fields}
\label{s6.1.2}

In the Einstein--Maxwell theory, with and without the dilaton, one
can not construct a quantity with the dimensions of mass from the
fundamental constants in the theory. The situation is different
for Einstein--Yang--Mills theory because the coupling constant $g$
has dimensions $(LM)^{-1/2}$. The existence of such a
dimensionful quantity has interesting consequences.

\epubtkImage{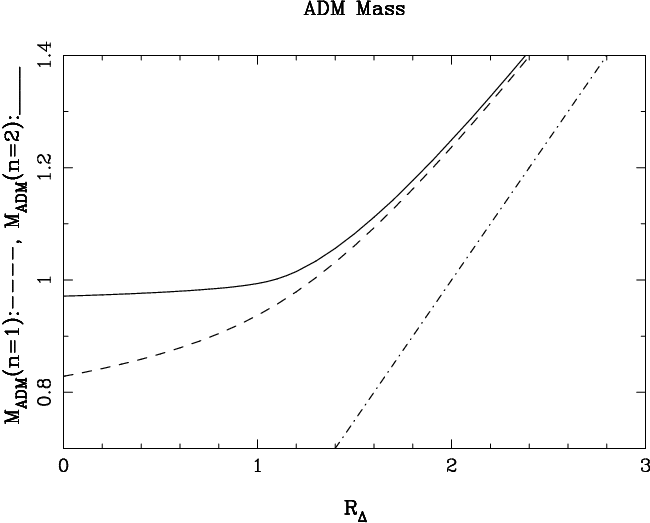}
{\begin{figure}[hptb]
   \def\epsfsize#1#2{0.5#1}
   \centerline{\epsfbox{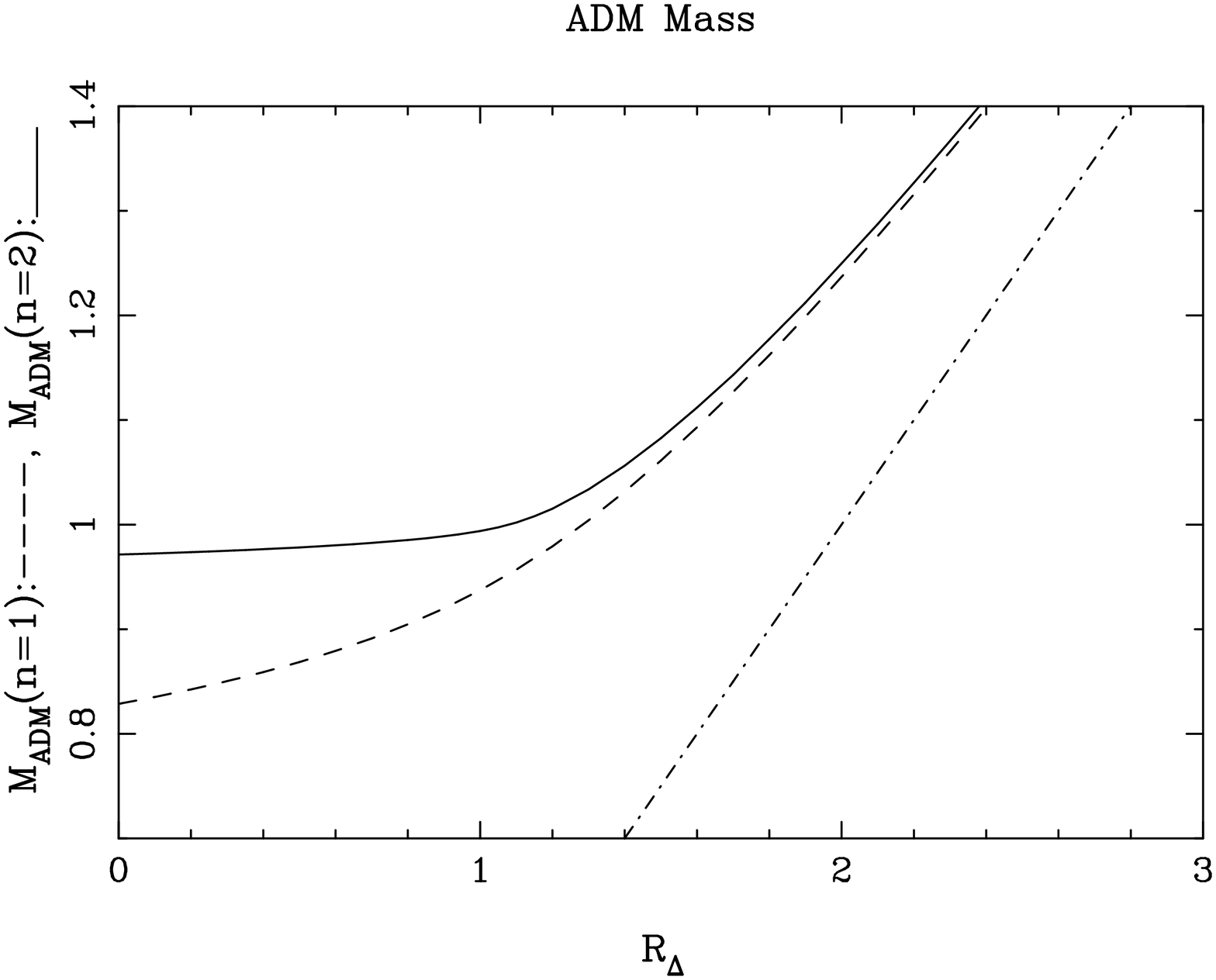}}
   \caption{\it The ADM mass as a function of the horizon radius
     $R_{\Delta}$ of static spherically symmetric solutions to the
     Einstein--Yang--Mills system (in units provided by the Yang--Mills
     coupling constant). Numerical plots for the colorless ($n=0$) and
     families of colored black holes ($n=1, 2$) are shown. (Note that
     the $y$-axis begins at $M = 0.7$ rather than at $M=0$.)}
   \label{hairyfig1}
 \end{figure}
}

For simplicity, we will restrict ourselves to  $\mathrm{SU}(2)$
Yang--Mills fields, but results based on the isolated horizon
framework go through for general compact groups~\cite{afk}. Let us
begin with a summary of the known static solutions. First, the
Reissner--Nordstr\"om family constitutes a continuous 2-parameter
set of static solutions of the Einstein--Yang--Mills theory,
labelled by $(a_\Delta, Q^\mathrm{YM}_\Delta)$. In addition, there is a
1-parameter family of `embedded Abelian solutions' with (a fixed)
magnetic charge $P_\Delta^0$, labelled by $(a_\Delta, P_\Delta^0)$.
Finally, there are families of `genuinely non-Abelian solutions'.
For these, the analog of the Israel theorem for Einstein--Maxwell
theory fails to hold~\cite{kk1, kk2, kk3}; the theory admits static
solutions which need not be spherically symmetric. In particular,
an infinite family of solutions labelled by two integers $(n_1,
n_2)$ is known to exist. All static, spherically symmetric
solutions are known and they correspond to the infinite sub-family
$(n_1, n_2 =0)$, labelled by a single integer. However, the two
parameter family is obtained using a specific ansatz, and other
static solutions also exist. Although the available information on the
static sector is quite rich, in contrast to the
Einstein--Maxwell-dilaton system, one is still rather far from
having complete control.

However, the existing results are already sufficient to show that,
in contrast to the situation in the Einstein--Maxwell theory, the
ADM mass is not a good measure of the black hole (or horizon) mass
\emph{even in the static case.} Let us consider the simplest case,
the spherically symmetric static solutions labelled by a single
integer $n$ (see Figure~\ref{hairyfig1}). Let us decrease the
horizon area along any branch $n\not= 0$. In the zero area limit,
the solution is known to converge point-wise to a regular, static,
spherical solution, representing an Einstein--Yang--Mills
\emph{soliton}~\cite{bm, solitons1, solitons2}. This solution
has, of course, a non-zero ADM mass $M_{\mathrm{ADM}}^\mathrm{sol}$, which
equals the limiting value of $M_{\mathrm{ADM}}^\mathrm{BH}$. However, in
this limit, there is no black hole \emph{at all}!  Hence, this
limiting value of the ADM mass can not be meaningfully identified
with any horizon mass. By continuity, then, $M_{\mathrm{ADM}}^\mathrm{BH}$
can not be taken as an accurate measure of the horizon mass for
any black hole along an $n\not= 0$ branch. Using the isolated
horizon framework, it \emph{is} possible to introduce a meaningful
definition of the horizon mass on any given static branch.

To establish laws of black hole mechanics, one begins with
appropriate boundary conditions. In the Maxwell case, the gauge
freedom in the vector potential is restricted on the horizon by
requiring ${\mathcal L}_\ell A_a =0$ on $\Delta$. The analogous
condition ensuring that the Yang--Mills potential $\mathbf{A}$ is
in an `adapted gauge' on $\Delta$ is more subtle~\cite{afk}. However,
it does exist and again ensures that (i) the action principle is
well defined, and (ii) the Yang--Mills electric potential
$\mathbf\Phi_{(\ell)} := - |\ell\cdot \mathbf{A}|$ is constant on the
horizon, where the absolute sign stands for the norm in the
internal space. The rest of the boundary conditions are the same
as in Section~\ref{s2.1.1}. The proof of the zeroth law and the
construction of the phase space is now straightforward. There is a
well-defined notion of conserved horizon charges
\begin{equation}
  \label{ymcharge}
  Q^\mathrm{YM}_{\Delta} := -\frac{1}{4\pi}
  \oint_{S}|{}^\star{\mathbf{F}}|\, d^2V,
  \qquad
  P^\mathrm{YM}_{\Delta} := - \frac{1}{4\pi}
  \oint_{S} \! |\mathbf{F}| \,d^2V,
\end{equation}
where $| \mathbf{F}| := [(\epsilon^{ab} \mathbf{F}_{ab}^i)
(\epsilon^{ab} \mathbf{F}_{ab}^j)\, K_{ij}]^{\frac{1}{2}}$, with
$K_{ij}$ being the Cartan--Killing metric on $\mathrm{SU}(2)$,
$\epsilon^{ab}$ the alternating tensor on the cross-section $S$ of
$\Delta$, and where $|{}^\star\mathbf{F}|$ is defined by replacing $\mathbf{F}$ with
${}^\star\mathbf{F}$. Finally, one can again introduce live vector fields
$t^a$, time evolution along which generates a Hamiltonian flow on
the phase space, and establish a first law for each of these
$t^a$:
\begin{equation}
  \label{ym1law}
  \delta E_{\Delta}^{t} = \frac{1}{8\pi G}\kappa_{(t)}
  \delta a_{\Delta} + \mathbf\Phi_{(t)} \delta Q^\mathrm{YM}_{\Delta}.
\end{equation}
Note that, even though the magnetic charge $P^\mathrm{YM}_\Delta$ is in
general non-zero, it does \emph{not} enter the statement of the
first law. In the Abelian case, a non-zero magnetic charge
requires non-trivial $\mathrm{U}(1)$ bundles, and Chern numbers
characterizing these bundles are discrete. Hence the magnetic
charge is quantized and, if the phase space is constructed from
connections, $\delta P^\mathrm{EM}$ vanishes identically for any
variation $\delta$. In the non-Abelian case, one can work with a
trivial bundle and have non-zero $P^\mathrm{YM}_\Delta$. Therefore, $\delta
P^{YM}_\Delta$ does not automatically vanish and absence of this term
in the first law is somewhat surprising.

A more significant difference from the Abelian case is that,
because the uniqueness theorem fails, one can not use the static
solutions to introduce a canonical function $\kappa_0$ on the
entire phase space, whence as in the dilatonic case, there is no
longer a canonical horizon mass $M_\Delta$ function on the entire
phase space. In the next Section~\ref{s6.2} we will see that it is
nonetheless possible to introduce an extremely useful notion of
the horizon mass for each static sequence.


\subsection{Structure of colored, static black holes}
\label{s6.2}

We will briefly summarize research in three areas in which the
isolated horizon framework has been used to illuminate the
structure of static, colored black holes and associated
solitons\epubtkFootnote{We are grateful to Alejandro Corichi for
  correspondence on the recent results in this area.}.


\subsubsection{Horizon mass}
\label{s6.2.1}

Let us begin with Einstein--Yang--Mills theory considered in the
last Section~\ref{s6.1}. As we saw, the ADM mass fails to be a good
measure of the horizon mass for colored black holes. The failure
of black hole uniqueness theorems also prevents the isolated
horizon framework from providing a canonical notion of horizon
mass on the full phase space. However, one can repeat the strategy
used for dilatonic black holes to define horizon mass
unambiguously for the static solutions~\cite{cs, afk}.

Consider a connected component of the known static solutions,
labelled by $\vec{n}_0 \equiv (n^0_1, n^0_2)$. Using for $\kappa_0$ the
surface gravity of the properly normalized static Killing vector,
in this sector one can construct a live vector field $t_0^a$ and
obtain a first law. The energy $E^{t_0}_\Delta$ is well-defined on
the full phase space and can be naturally interpreted as the
horizon mass $M_\Delta^{(\vec{n}_0)}$ for colored black holes with
`quantum number' $\vec{n}_0$. The explicit expression is given by
\begin{equation}
  \label{hormass}
  M_\Delta^{(\vec{n}_0)} = \frac{1}{2G}
  \int_0^{R_\Delta} \!\!\!\!\! \beta_{\vec{n}_0}(R) \, dR,
\end{equation}
where, following a convention in the literature on hairy black
holes~\cite{cs}, we have used
\begin{equation}
  \beta_{\vec{n}_0}(R) = 2\,[\kappa_{\vec{n}_0}(R)]\,R
\end{equation}
rather than the surface gravity $\kappa_{\vec{n}_0}(R)$ of static
solutions. Now, as in the Einstein--Maxwell case, the Hamiltonian
generating evolution along $t_0^a$ contains only surface terms:
$H_{t_0} = E^{t_0}_\mathrm{ADM} - E^{t_0}_\mathrm{\Delta}$ and is
constant on each connected, static sector if $t_0^a$ coincides
with the static Killing field on that sector. By construction, our
$t_0^a$ has this property for the $\vec{n}_0$-sector under
consideration. Now, in the Einstein--Maxwell case, since there is
no constant with the dimension of energy, it follows that the
restriction of $H^{t_0}_\mathrm{ADM}$ to the static sector must
vanish. The situation is quite different in Einstein--Yang--Mills
theory where the Yang--Mills coupling constant $g$ provides a
scale. In  $c = 1$ units, $(g\sqrt{G})^{-1} \sim \mathrm{mass}$.
Therefore, we can only conclude that
\begin{equation}
  M_\mathrm{ADM}^{(\vec{n})} = M_{\Delta}^{(\vec{n}_0)} +
  (g\sqrt{G})^{-1} C^{(\vec{n}_0)}
\end{equation}
for some $\vec{n}_0$-dependent constant $C^{(\vec{n}_0)}$. As the horizon
radius shrinks to zero, the static solution~\cite{kk1, vgreview}
under consideration tends to the solitonic solution with the same
`quantum numbers' $\vec{n}_0$. Hence, by taking this limit, we conclude
$(g\sqrt{G})^{-1} C^{(\vec{n}_0)} = M_\mathrm{ADM}^{\mathrm{sol},\,(\vec{n}_0)}$. Therefore, on any $\vec{n}$-static
sector, we have the following interesting relation between the
black hole and solitonic solutions:
\begin{equation}
  \label{hairyM}
  M_\mathrm{ADM}^{\mathrm{BH},\,(\vec{n})} = \frac{1}{2G}
  \int_0^{R_\Delta} \!\!\!\!\! \beta_{\vec{n}}(R) \, dR +
  M_\mathrm{ADM}^{\mathrm{sol},\,(\vec{n})} =
  M_{\Delta}^{\mathrm{BH}, (\vec{n})} +
  M_\mathrm{ADM}^{\mathrm{sol},\,(\vec{n})}\!\!\!.
\end{equation}
Thus, although the main motivation behind the isolated horizon
framework was to go beyond globally time-independent situations,
it has led to an interesting new relation between the ADM masses
of black holes and their solitonic analogs already in the static
sector.

The relation~(\ref{hairyM}) was first proposed for spherical
horizons in~\cite{cs}, verified in~\cite{cns}, and extended to
distorted horizons in~\cite{afk}. It provided impetus for new work
by mathematical physicists working on colored black holes. The
relation has been confirmed in three more general and non-trivial
cases:

\begin{itemize}
\item Non-spherical, non-rotating black holes parameterized by two
  quantum numbers~\cite{kksw}.
\item Non-spherical solutions to the more general
  Einstein--Yang--Mills--Higgs theory~\cite{hkk}, where distortions
  are caused by `magnetic dipole hair'~\cite{kk-dipole}.
\item Static solutions in the Born--Infeld theory~\cite{Breton}.
\end{itemize}


\subsubsection{Phenomenological model of colored black holes}
\label{s6.2.2}

Isolated horizon considerations suggested the following simple
heuristic model of colored black holes~\cite{acs}: \emph{A
colored black hole with quantum numbers $\vec{n}$ should be thought of
as `bound states' of a ordinary (colorless) black hole and a
soliton with color quantum numbers $\vec{n}$}, where $\vec{n}$ can be more
general than considered so far. Thus the idea is that an uncolored
black hole is `bare' and becomes `colored' when `dressed' by the
soliton.

The mass formula~(\ref{hairyM}) now suggests that the total ADM
mass $M^{(\vec{n})}_{\mathrm{ADM}}$ has three components: the mass
$M_{\Delta}^{(0)}$ of the bare horizon, the mass
$M_{\mathrm{ADM}}^{\mathrm{sol},\,\vec{n}}$ of the colored soliton, and
a binding energy given by
\begin{equation}
  \label{binding}
  E_\mathrm{binding} = M^{(\vec{n})}_\Delta (R_\Delta) -
  M^{(0)}_\Delta (R_\Delta).
\end{equation}
If this picture is correct, being gravitational binding energy,
$E_\mathrm{binding}$ would have to be negative. This expectation is
borne out in explicit examples. The model has several predictions
on the qualitative behavior of the horizon mass~(\ref{hormass}),
the surface gravity, and the relation between properties of black
holes and solitons. We will illustrate these with just three
examples (for a complete list with technical caveats, see~\cite{acs}):

\begin{itemize}
\item For any fixed value of $R_\Delta$ and of all quantum numbers
  except $n$, the horizon mass and surface gravity decrease
  monotonically with $n$.
\item For fixed values of all quantum numbers $\vec{n}$, the horizon
  mass $M^{(\vec{n})}_{\Delta}(R_\Delta) $ is non-negative, vanishing
  if and only if $R_\Delta$ vanishes, and increases monotonically with
  $R_\Delta$. $\beta_{(\vec{n})}$ is positive and bounded above by
  $1$.
\item For fixed $\vec{n}$, the binding energy decreases as the horizon
  area increases.
\end{itemize}

The predictions for fixed $\vec{n}$ have recently been verified beyond
spherical symmetry: for the distorted, axially symmetric
Einstein--Yang--Mills solutions in~\cite{kksw} and for the distorted
`dipole' solutions in Einstein--Yang--Mills--Higgs solutions
in~\cite{hkk}. Taken together, the predictions of this model can
account for all the qualitative features of the plots of the
horizon mass and surface gravity as functions of the horizon
radius and quantum numbers. More importantly, they have
interesting implications on the stability properties of colored
black holes.

\epubtkImage{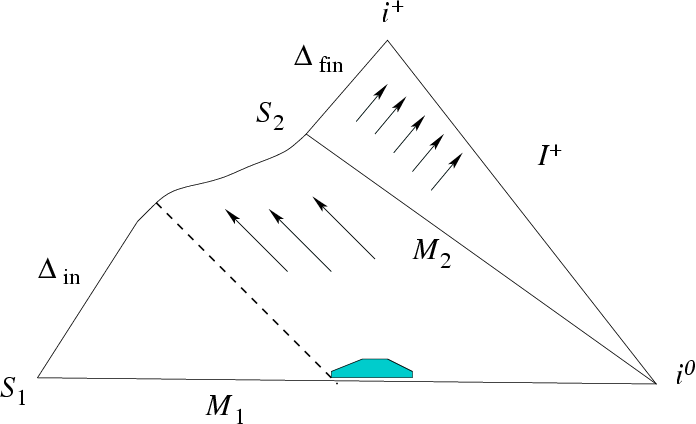}
{\begin{figure}[hptb]
   \def\epsfsize#1#2{0.55#1}
   \centerline{\epsfbox{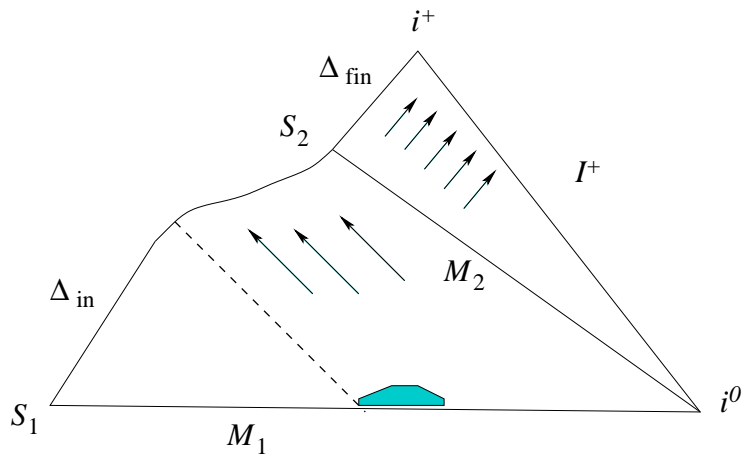}}
   \caption{\it An initially static colored black hole with horizon
     $\Delta_\mathrm{in}$ is slightly perturbed and decays to a
     Schwarzschild-like isolated horizon $\Delta_\mathrm{fin}$, with
     radiation going out to future null infinity $\mathcal{I}^+$.}
   \label{hairyfig2}
 \end{figure}
}

One begins with an observation about solitons and deduces
properties of black holes. Einstein--Yang--Mills solitons are known
to be unstable~\cite{unstable1}; under small
perturbations, the energy stored in the `bound state' represented
by the soliton is radiated away to future null infinity
$\mathcal{I}^+$. The phenomenological model suggests that colored
black holes should also be unstable and they should decay into
ordinary black holes, the excess energy being radiated away to
infinity. In general, however, even if one component of a bound
system is unstable, the total system may still be stable if the
binding energy is sufficiently large. An example is provided by
the deuteron. However, an examination of energetics reveals that
this is not the case for colored black holes, so instability should
prevail. Furthermore, one can make a few predictions about the
nature of instability. We summarize these for the simplest case of
spherically symmetric, static black holes for which there is a
single quantum number $n$:

\begin{enumerate}
\item All the colored black holes on the $n$th branch have the same
  number (namely, $2n$) of unstable modes as the $n$th soliton. (The
  detailed features of these unstable modes can differ especially
  because they are subject to different boundary conditions in the two
  cases.)
  \label{item_1}
\item For a given $n$, colored black holes with larger horizon area
  are less unstable. For a given horizon area, colored black holes
  with higher value of $n$ are more unstable.
  \label{item_2}
\item The `available energy' for the process is given by
  \begin{equation}
    \label{avail}
    E_\mathrm{available}^{(n)}= M_\mathrm{sol}^{(n)} -
    |E_\mathrm{binding}^\mathrm{initial}|.
  \end{equation}
  Part of it is absorbed by the black hole so that its horizon area
  increases and the rest is radiated away to infinity. Note that
  $E_\mathrm{available}^{(n)}$ can be computed \emph{knowing just the
  initial configuration}.
  \label{item_3}
\item In the process the horizon area necessarily
  increases. Therefore, the energy radiated to infinity is strictly
  less than $E_\mathrm{available}^{(n)}$.
  \label{item_4}
\end{enumerate}

Expectation~\ref{item_1} of the model is known to be
correct~\cite{unstable2}. Prediction~\ref{item_2} has been shown to be
correct in the $n=1$, colored black holes in the sense that the
frequency of all unstable modes is a decreasing function of the area,
whence the characteristic decay time grows with area~\cite{vgreview,
  unstable4}. To our knowledge a detailed analysis of instability,
needed to test Predictions~\ref{item_3} and~\ref{item_4} are yet to be
made.

Finally, the notion of horizon mass and the associated stability
analysis has also provided an `explanation' of the following fact
which, at first sight, seems puzzling. Consider the `embedded
Abelian black holes' which are solutions to Einstein--Yang--Mills
equations with a specific magnetic charge $P^0_\Delta$. They are
isometric to a family of magnetically charged Reissner--Nordstr\"om
solutions and the isometry maps the Maxwell field strength to the
Yang--Mills field strength. The only difference is in the form of
the connection; while the Yang--Mills potential is supported on a
trivial $\mathrm{SU}(2)$ bundle, the Maxwell potential requires a
non-trivial $\mathrm{U}(1)$ bundle. Therefore, it comes as an initial
surprise that the solution is stable in the Einstein--Maxwell
theory but unstable in the Einstein--Yang--Mills theory~\cite{bw,
  bfm}. It turns out that this difference is naturally
explained by the WIH framework. Since the solutions are isometric,
their ADM mass is the same. However, since the horizon mass arises
from Hamiltonian considerations, it is theory dependent: It is
\emph{lower} in the Einstein--Yang--Mills theory than in the
Einstein--Maxwell theory! Thus, from the Einstein--Yang--Mills
perspective, part of the ADM mass is carried by the soliton and
there is positive $E^{YM}_\mathrm{available}$ which can be radiated
away to infinity. In the Einstein--Maxwell theory,
$E^\mathrm{EM}_\mathrm{available}$ is zero. The stability analysis
sketched above therefore implies that the solution should be unstable
in the Einstein--Yang--Mills theory but stable in the
Einstein--Maxwell theory. This is another striking example of the
usefulness of the notion of the horizon mass.


\subsubsection{More general theories}
\label{s6.2.3}

We will now briefly summarize the most interesting result obtained
from this framework in more general theories. When one allows
Higgs or Proca fields in addition to Yang--Mills, or considers
Einstein--Skyrme theories, one acquires additional dimensionful
constants which trigger new phenomena~\cite{pb:tc, tt:km, vgreview}.
One of the most interesting is the `crossing phenomena' of
Figure~\ref{hairyfig3} where curves in the `phase diagram' (i.e., a plot
of the ADM mass versus horizon radius) corresponding to the two
distinct static families cross. This typically occurs in theories
in which there is a length scale even in absence of gravity, i.e.,
even when Newton's constant is set equal to zero~\cite{nqs, acs}.

\epubtkImage{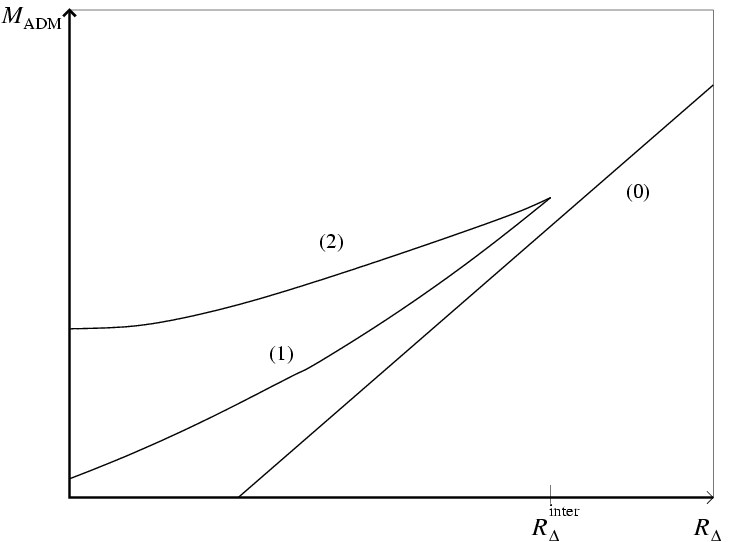}
{\begin{figure}[hptb]
   \def\epsfsize#1#2{0.55#1}
   \centerline{\epsfbox{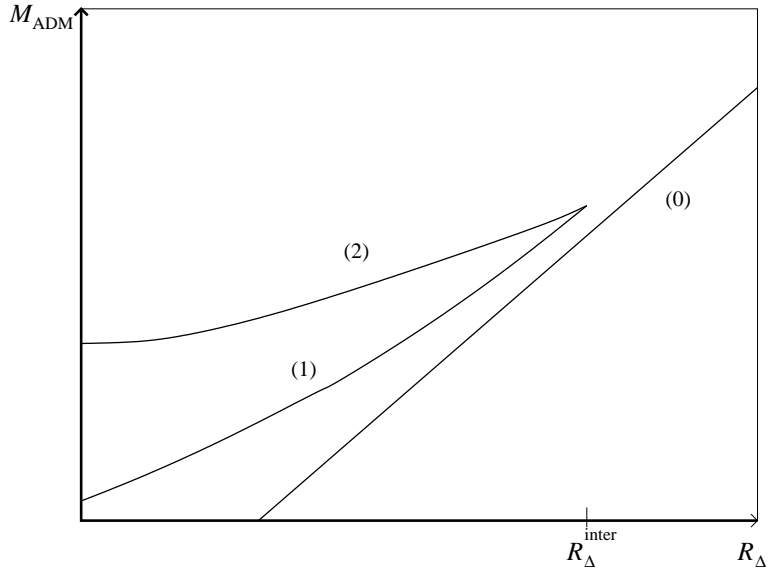}}
   \caption{\it The ADM mass as a function of the horizon radius
     $R_\Delta$ in theories with a built-in non-gravitational length
     scale. The schematic plot shows crossing of families labelled by
     $n=1$ and $n=2$ at $R_\Delta = R^\mathrm{inter}_{\Delta}$.}
   \label{hairyfig3}
 \end{figure}
}

In this case, the notion of the horizon mass acquires further
subtleties. If, as in the Einstein--Yang--Mills theory considered
earlier, families of static solutions carrying distinct quantum
numbers do not cross, there is a well-defined notion of horizon
mass for each static solution, although, as the example of
`embedded Abelian solutions' shows, in general its value is theory
dependent. When families cross, one can repeat the previous
strategy and use Equation~(\ref{hormass}) to define a mass $M_\Delta^{(n)}$
along each branch. However, at the intersection point
$R^\mathrm{inter}_\Delta$ of the $n$th and $(n+1)$th branches, the mass is
discontinuous. This discontinuity has an interesting implication.
Consider the closed curve $C$ in the phase diagram, starting at
the intersection point and moving along the $(n+1)$th branch in
the direction of decreasing area until the area becomes zero, then
moving along $a_\Delta = 0$ to the $n$th branch and moving up to the
intersection point along the $n$th branch (see
Figure~\ref{hairyfig3}). Discontinuity in the horizon mass implies
that the integral of $\beta(R_\Delta)$ along this closed curve is
non-zero. Furthermore, the relation between the horizon and the
soliton mass along each branch implies that the value of this
integral has a direct physical interpretation:
\begin{equation}
  M^{(n+1)}_\mathrm{sol}-M^{(n)}_\mathrm{sol} =
  \frac{1}{2G} \oint_C \! \beta (r) \, dr.
\end{equation}
This is a striking prediction because it relates differences in
masses of \emph{solitons} to the knowledge of horizon properties of
the corresponding \emph{black holes}! Because of certain
continuity properties which hold as one approaches the static
Einstein--Yang--Mills sector in the space to static solutions to
Einstein--Yang--Mills--Higgs equations~\cite{cns2}, one can
also obtain a new formula for the ADM masses of
Einstein--Yang--Mills solitons. If $(1- \beta_{(n)}(r))$ for the
black hole solutions of this theory is integrable over the entire
positive half line, one has~\cite{acs}:
\begin{equation}
  M^{(n)}_\mathrm{sol} = \frac{1}{2G} \int_{0}^{\infty} \!\!\!
  (1 - \beta_{(n)} (r))\, dr.
\end{equation}
Both these predictions of the phenomenological model~\cite{acs}
have been verified numerically in the spherically symmetric
case~\cite{cns2}, but the axi-symmetric case is still open.

\newpage


\section{Applications in Quantum Gravity}
\label{s7}

As discussed in the introduction, laws of black hole mechanics,
discovered in the early seventies, provided a concrete challenge
to candidate quantum theories of gravity: Account for the
thermodynamic, black hole entropy through a detailed, statistical
mechanical counting of appropriate micro-states. Indeed, this is
essentially the only {concrete} quantitative hint we have had
about the nature of quantum space-time geometry. The isolated
horizon framework has been used to address this issue
systematically and has led to the only available detailed
calculations within a full-fledged approach to quantum gravity
that encompass realistic black holes (which carry no or negligible
gauge charges and may be distorted). As we will discuss in
Section~\ref{s8}, what we know about dynamical horizons does suggest
that there should be interesting generalizations of these results also
to non-equilibrium situations. But so far there has been no work
along these lines.


\subsection{Preliminaries}
\label{s7.1}

The first and the second laws of black hole mechanics discussed in
Section~\ref{s4},
\begin{equation}
  \delta E = \frac{\kappa}{8\pi G} \delta a + \mathrm{work},
  \qquad
  d \delta a \ge 0,
\end{equation}
have a close similarity with the first and second laws of
thermodynamics
\begin{equation}
  \delta E = T \, \delta S + \mathrm{work},
  \qquad
  \delta S \ge 0.
\end{equation}
This suggest that one should assign to black holes an entropy
proportional to their area. Indeed, already in the early
seventies, using imaginative thought experiments,
Bekenstein~\cite{jdb1, jdb2} argued that such an assignment is forced
upon us if, in presence of black holes, we are not going to simply
abandon the second law of thermodynamics. Specifically, consider an
experiment in which a box containing radiation is slowly lowered
into a black hole. At the end of the process, the entropy of the
exterior region decreases in apparent violation of the second law.
Bekenstein argued that the area of the horizon increases in the
process so that the total entropy would not decrease if the black
hole is assigned entropy equal to an appropriate multiple of the
area. Hawking's celebrated calculation of black hole
evaporation~\cite{h} provided the precise numerical coefficient. For,
it suggested that we should assign the black hole a temperature
$T = \hbar \kappa/2\pi$. By comparing the forms of the first laws of
black hole mechanics and thermodynamics, one is then led to set
\begin{equation}
  \label{entropy1}
  S_\mathrm{bh} = \frac{a}{4\ell_\mathrm{Pl}^2}
  \qquad \mathrm{with\ }
  \ell_\mathrm{Pl}^2 = G\hbar.
\end{equation}
Note that the factor of $\hbar$ is absolutely essential: In the
limit $\hbar \mapsto 0$, the black hole temperature goes to zero
and its entropy tends to infinity just as one would expect
classically.

The relation~(\ref{entropy1}) is striking and deep because it
brings together the three pillars of fundamental physics
-- general relativity, quantum theory, and statistical mechanics.
However, the argument itself is a rather hodge-podge mixture of
classical and semi-classical ideas, reminiscent of the Bohr theory
of atom. A natural question then is: What is the analog of the
more fundamental, Pauli--Schr\"odinger theory of the hydrogen atom?
More precisely, what is the statistical mechanical origin of black
hole entropy? To answer this question, one has to isolate the
microscopic degrees of freedom responsible for this entropy. For a
classical ideal gas, these are given by the positions and momenta
of all molecules; for a magnet, by the states of each individual
spin at lattice sites. How do we represent the analogous
microscopic degrees of freedom for a black hole? They can not be
described in terms of quantum states of physical gravitons because
we are dealing with black holes in equilibrium. In the approach
based on weakly isolated horizons, they are captured in \emph{the
quantum states of the horizon geometry.} Just as one would expect
from Bekenstein's thought experiments, these degrees of freedom
can interact with the exterior curved geometry, and the resulting
notion of black hole entropy is tied to observers in the exterior
regions.

A heuristic framework for the calculation of entropy was suggested
by John Wheeler in the nineties, which he christened `\emph{It}
from \emph{bit}'. Divide the black hole horizon into elementary
cells, each with one Planck unit of area $\ell_\mathrm{Pl}^2$ and assign to
each cell two microstates. Then the total number of states ${\cal
N}$ is given by $\mathcal{N} = 2^n$, where $n = ({a/ \ell_\mathrm{Pl}^2})$ is the
number of elementary cells, whence entropy is given by $S = \ln
\mathcal{N} \sim a$. Thus, apart from a numerical coefficient, the
entropy (`\emph{It}') is accounted for by assigning two states
(`\emph{bit}') to each elementary cell. This qualitative picture
is simple and attractive. But it raises at least three central
questions:

\begin{itemize}
\item What is the origin of the `elementary cells' and why is each
  endowed with an area $\ell_\mathrm{Pl}^2$?
\item What is the origin of the microstates carried by each elementary
  cell and why are there \emph{precisely} two microstates?
\item What does all this have to do with a black hole? Why doesn't it
  apply to \emph{any} 2-surface, including a 2-sphere in Minkowski
  space-time?
\end{itemize}

An understanding of geometry of \emph{quantum} WIHs provides a
detailed framework which, in particular, answers these questions.
The precise picture, as usual, is much more involved than that
envisaged by Wheeler. Eigenvalues of area operator turn out to be
discrete in quantum geometry and one can justify dividing the
horizon 2-sphere into elementary cells. However, there are many
permissible area eigenvalues and cells need not all carry the same
area. To identify horizon surface states that are responsible for
entropy, one has to \emph{crucially} use the WIH boundary
conditions. However, the number of surface states assigned to each
cell is not restricted to two. Nonetheless, Wheeler's picture
turns out to be qualitatively correct.


\subsection{Quantum horizon geometry}
\label{s7.2}

For simplicity of presentation, in this section we will
restrict ourselves to type I WIHs, i.e., the ones for which the
only non-zero multipole moment is the mass monopole. The extension
to include type II horizons with rotations and distortion will be
briefly summarized in the next Section~\ref{s7.3}. Details can be found
in~\cite{abck, abk, ac, aa4, aepv}.

The point of departure is the classical Hamiltonian formulation
for space-times $\mathcal{M}$ with a type I WIH $\Delta$ as an internal
boundary, with fixed area $a_0$ and charges $Q^{\alpha}_0$, where
$\alpha$ runs over the number of distinct charges (Maxwell,
Yang--Mills, dilaton, \ldots) allowed in the theory. As we noted in
Section~\ref{s4.1}, the phase space ${\mathbf\Gamma}$ can be
constructed in a number of ways, which lead to equivalent
Hamiltonian frameworks and first laws. However, so far, the only
known way to carry out a background independent, non-perturbative
quantization is through connection variables~\cite{alreview}.

As in Figure~\ref{fig:phasespace} let us begin with a partial
Cauchy surface $M$ whose internal boundary in $\mathcal{M}$ is a 2-sphere
cross-section $S$ of $\Delta$ and whose asymptotic boundary is a
2-sphere $S_\infty$ at spatial infinity. The configuration
variable is an $\mathrm{SU}(2)$ connection $A_a^i$ on $M$, where $i$ takes
values in the 3-dimensional Lie-algebra $\mathrm{su}(2)$ of $\mathrm{SU}(2)$. Just
as the standard derivative operator acts on tensor fields and
enables one to parallel transport vectors, the derivative operator
constructed from $A_a^i$ acts on fields with internal indices and
enables one to parallel transport spinors. The conjugate momentum
is represented by a vector field $P^a_i$ with density weight 1
which also takes values in $\mathrm{su}(2)$; it is the analog of the
Yang--Mills electric field. (In absence of a background metric,
momenta always carry a density weight 1.) $P^a_i$ can be regarded
as a (density weighted) triad or a `square-root' of the intrinsic
metric $\tilde{q}_{ab}$ on $S$: $(8\pi G\gamma)^2 P^a_i P^b_j
\delta_{ij} = \tilde{q}\, \tilde{q}^{ab}$, where $\delta_{ij}$ is
the Cartan Killing metric on $\mathrm{su}(2)$, $\tilde{q}$ is the
determinant of $\tilde{q}_{ab}$ and $\gamma$ is a positive real
number, called the Barbero--Immirzi parameter. This parameter
arises because there is a freedom in adding to Palatini action a
multiple of the term which is `dual' to the standard one, which
does not affect the equations of motion but changes the definition
of momenta. This multiple is $\gamma$. The presence of $\gamma$
represents an ambiguity in quantization of geometry, analogous to
the $\theta$-ambiguity in QCD. Just as the classical Yang--Mills
theory is insensitive to the value of $\theta$ but the quantum
Yang--Mills theory has inequivalent $\theta$-sectors, classical
relativity is insensitive to the value of $\gamma$ but the quantum
geometries based on different values of $\gamma$ are (unitarily)
inequivalent~\cite{gp} (for details, see, e.g., \cite{alreview}).

Thus, the gravitational part of the phase space ${\mathbf \Gamma}$
consists of pairs $(A_a^i, P^a_i)$ of fields on $M$ satisfying the
boundary conditions discussed above. Had there been no internal
boundary, the gravitational part of the symplectic structure would
have had just the expected volume term:
\begin{equation}
  \label{sym1}
  \mathbf{\Omega}_V(\delta_1, \delta_2) =
  \int_M \! \left( \delta_2 A^i \wedge \delta_1 \Sigma_i -
  \delta_1 A^i \wedge \delta_2 \Sigma_i \right),
\end{equation}
where $\Sigma_{ab}{}_i := \eta_{abc} P^c_i$ is the 2-form dual to
the momentum $P^c_i$ and $\delta_1$, and $\delta_2$ denote any two
tangent vectors to the phase space. However, the presence of the
internal boundary changes the phase space-structure. The type I
WIH conditions imply that the non-trivial information in the
pull-back $\pullback{A}_a^i$ of $A_a^i$ to $S$ is contained in a
$\mathrm{U}(1)$ connection $W_a:= \pullback{A}_a^ir_i$ on $S$, where $r^i$ is
the unit, internal, radial vector field on $S$. Its curvature
2-form $F_{ab}$ is related to the pull-back
$\pullback{\Sigma}^i_{ab}$ of the 2-form $\Sigma_{ab}^i$ to $S$ via
\begin{equation}
  \label{hbc}
  F \equiv dW = - 8\pi G\gamma \frac{2\pi}{a_0}\,\pullback{\Sigma}^i r_i.
\end{equation}
The restriction is called the \emph{horizon boundary condition}.
In the Schwarzschild space-time, all the 2-spheres on which it is
satisfied lie on the horizon. Finally, the presence of the
internal boundary modifies the symplectic structure: It now
acquires an additional boundary term
\begin{equation}
  \label{sym2}
  \mathbf{\Omega}(\delta_1, \delta_2) =
  \mathbf{\Omega}_V(\delta_1, \delta_2) +
  \mathbf{\Omega}_S(\delta_1, \delta_2),
\end{equation}
with
\begin{equation}
  \mathbf{\Omega}_S (\delta_1, \delta_2) =
  \frac{1}{2\pi}\frac{a_0}{4\pi G\gamma }
  \oint_S \! \delta_1 W \wedge \delta_2 W,
\end{equation}
The new surface term is precisely the symplectic structure of a
well-known topological field theory -- the $\mathrm{U}(1)$-Chern--Simons
theory. \emph{The symplectic structures of the Maxwell,
Yang--Mills, scalar, and dilatonic fields do not acquire surface
terms.} Conceptually, this is an important point: This, in
essence, is the reason why (for minimally coupled matter) the
black hole entropy depends just on the area and not, in addition,
on the matter charges.

In absence of internal boundaries, the quantum theory has been
well-understood since the mid-nineties (for recent reviews, see,
\cite{crreview, ttreview, alreview}). The fundamental quantum
excitations are represented by Wilson lines (i.e., holonomies)
defined by the connection and are thus 1-dimensional, whence the
resulting quantum geometry is \emph{polymer-like}. These
excitations can be regarded as \emph{flux lines of area} for the
following reason. Given any 2-surface $\mathbb{S}$ on $M$, there
is a self-adjoint operator $\hat{A}_{\mathbb{S}}$ \emph{all} of
whose eigenvalues are known to be discrete. The simplest
eigenvectors are represented by a single flux line, carrying a
half-integer $j$ as a label, which intersects the surface
$\mathbb{S}$ exactly once, and the corresponding eigenvalue
$a_{\mathbb{S}}$ of $\hat{A}_{\mathbb{S}}$ is given by
\begin{equation}
  a_{\mathbb{S}} = 8 \pi \gamma \ell_\mathrm{Pl}^2 \sqrt{j(j+1)}.
\end{equation}
Thus, while the general form of the eigenvalues is the same in all
$\gamma$-sectors of the quantum theory, their numerical values
\emph{do depend on} $\gamma$. Since the eigenvalues are distinct
in different $\gamma$-sectors, it immediately follows that these
sectors provide unitarily inequivalent representations of the
algebra of geometric operators; there is `super-selection'. Put
differently, there is a quantization ambiguity, and which
$\gamma$-sector is actually realized in Nature is an experimental
question. One appropriate experiment,  for example a measurement
of the smallest non-zero area eigenvalue, would fix the value of
$\gamma$
and hence the quantum theory. Every further experiment -- e.g.,
the measurement of higher eigenvalues or eigenvalues of other
operators such as those corresponding to the volume of a region -- 
would provide tests of the theory. While such direct measurements
are not feasible today we will see that, somewhat surprisingly,
the Hawking--Bekenstein formula~(\ref{entropy1}) for the entropy of
large black holes provides a thought experiment to fix the value
of $\gamma$.

Recall next that, because of the horizon internal boundary, the
symplectic structure now has an additional surface term. In the
classical theory, since all fields are smooth, values of fields on
the horizon are completely determined by their values in the bulk.
However, a key point about field theories is that their quantum
states depend on fields which are arbitrarily discontinuous.
Therefore, in quantum theory, a decoupling occurs between fields
in the surface and those in the bulk, and independent surface
degrees of freedom emerge. \emph{These describe the geometry of
the quantum horizon and are responsible for entropy}.

In quantum theory, then, it is natural to begin with a total
Hilbert space
$\mathcal{H} = \mathcal{H}_{V}\otimes \mathcal{H}_{S}$ where
$\mathcal{H}_V$ is the
well-understood bulk or volume Hilbert space with `polymer-like
excitations', and $\mathcal{H}_S$ is the surface Hilbert space of the
$\mathrm{U}(1)$-Chern--Simons theory. As depicted in
Figure~\ref{entropyfig1}, the polymer excitations puncture the
horizon. An excitation carrying a quantum number $j$ `deposits' on $S$ an
area equal to $8\pi \ell_\mathrm{Pl}^2\, \sqrt{j(j+1)}$. These contributions
add up to endow $S$ a total area $a_0$. The surface Chern--Simons
theory is therefore defined on the punctured 2-sphere $S$. To
incorporate the fact that the internal boundary $S$ is not
arbitrary but comes from a WIH, we still need to incorporate the
residual boundary condition~(\ref{hbc}). This key condition is
taken over as an operator equation. Thus, in the quantum theory,
neither the triad nor the curvature of $W$ are frozen at the
horizon; neither is a classical field. Each is allowed to undergo
quantum fluctuations, but the quantum horizon boundary condition
requires that they have to fluctuate in tandem.

\epubtkImage{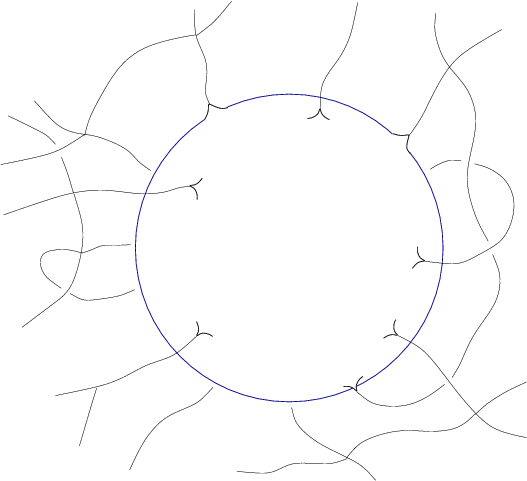}
{\begin{figure}[hptb]
   \def\epsfsize#1#2{0.4#1}
   \centerline{\epsfbox{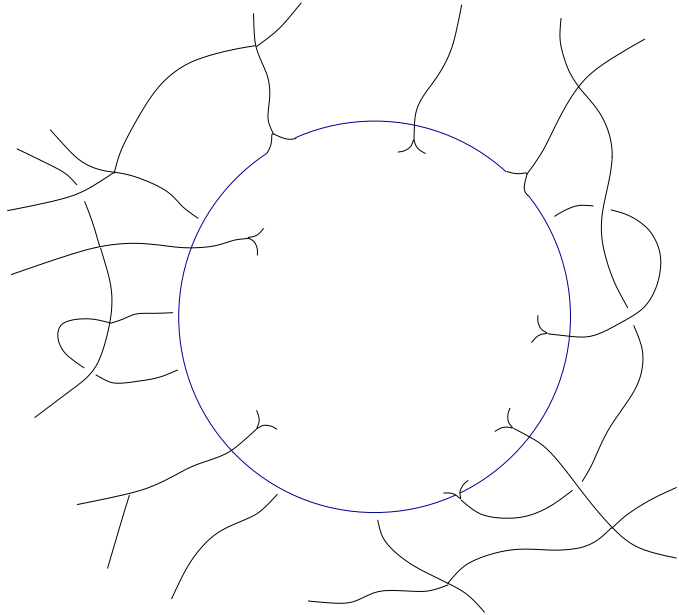}}
   \caption{\it Quantum horizon. Polymer excitations in the bulk
     puncture the horizon, endowing it with quantized
     area. Intrinsically, the horizon is flat except at punctures
     where it acquires a quantized deficit angle. These angles add up
     to endow the horizon with a 2-sphere topology.}
   \label{entropyfig1}
 \end{figure}
}

An important subtlety arises because the operators corresponding
to the two sides of Equation~(\ref{hbc}) act on different Hilbert spaces:
While $\hat{F}$ is defined on $\mathcal{H}_S$, $\hat{\pullback{\Sigma}}\cdot r$
is defined on $\mathcal{H}_V$. Therefore, the quantum horizon boundary
condition introduces a precise intertwining between the bulk and
the surface states: \emph{Only those states  $\Psi = \sum_i
\Psi_V^i\otimes \Psi_S^i$ in $\mathcal{H}$ which satisfy}
\begin{equation}
  \label{qbc}
  (1 \otimes \hat{F}) \Psi = \left( -\frac{2 \pi \gamma}{a_0}
  \hat{\pullback{\Sigma}} \cdot r \otimes 1 \right) \Psi
\end{equation}
\emph{can describe quantum geometries with WIH as inner
boundaries.} This is a stringent restriction: Since the operator
on the left side acts only on surface states and the one on the
right side acts only on bulk states, the equation can have
solutions \emph{only if} the two operators have the same
eigenvalues (in which case we can take $\Psi$ to be the tensor
product of the two eigenstates). Thus, for solutions to
Equation~(\ref{qbc}) to exist, there has to be a very delicate
matching between eigenvalues of the triad operators
$\hat{\pullback{\Sigma}}\cdot r$ calculated from bulk quantum
geometry, and eigenvalues of $\hat{F}$, calculated from
Chern--Simons theory. The precise numerical coefficients in the
surface calculation depend on the numerical factor in front of the
surface term in the symplectic structure, which is itself
determined by our type I WIH boundary conditions. Thus, the
existence of a coherent quantum theory of WIHs requires that the
three cornerstones
-- classical isolated horizon framework, quantum mechanics of bulk
geometry, and quantum Chern--Simons theory -- be united
harmoniously. Not only should the three frameworks be mutually
compatible at a conceptual level, but certain \emph{numerical
coefficients}, calculated independently within each framework,
have to match delicately. Remarkably, {these delicate constraints
are met}, whence the quantum boundary conditions do admit a
sufficient number of solutions.

We will conclude by summarizing the nature of geometry of the
quantum horizon that results. Given any state satisfying
Equation~(\ref{qbc}), the curvature $F$ of $W$ vanishes everywhere
\emph{except} at the points at which the polymer excitations in
the bulk puncture $S$. The holonomy around each puncture is
non-trivial. Consequently, the \emph{intrinsic} geometry of the
quantum horizon is flat except at the punctures. At each puncture,
there is a deficit angle, whose value is determined by the
holonomy of $W$ around that puncture. Each deficit angle is
\emph{quantized} and these angles add up to $4\pi$ as in a
discretized model of a 2-sphere geometry. Thus, the quantum
geometry of a WIH is quite different from its smooth classical
geometry.


\subsection{Entropy}
\label{s7.3}

Let us now summarize the ideas behind counting of surface
microstates that leads to the expression of entropy. To
incorporate dynamics in this canonical approach, we have to first
construct physical states by imposing quantum Einstein equations
(i.e., quantum constraints). While the procedure is technically
quite involved, the result is simple to state: What matters is
only the number of punctures and not their locations. To calculate
entropy, then, one constructs a micro-canonical ensemble as
follows. Fix the number $n$ of punctures and allow only those
(non-zero) spin-labels $j_I$ and charge labels $q^\alpha_I$ on the
polymer excitations which endow the horizon with a total area in
an interval $(a_0-\epsilon, a_0 + \epsilon)$ and charges in an
interval $(Q^\alpha_0 -\epsilon^\alpha, Q^\alpha_0 +
\epsilon^\alpha)$. (Here $I=1, 2, \ldots, n$ and $\epsilon$ and
$\epsilon^\alpha$ are suitably small. Their precise values will
not affect the leading contribution to entropy.) We denote by
$\mathcal{H}_n^\mathrm{BH}$ the sub-space of
$\mathcal{H}= \mathcal{H}_V \otimes \mathcal{H}_S$ in which
the volume states $\Psi_V$ are chosen with the above restrictions
on $j_I$ and $q^\alpha_I$, and the total state $\Psi$ satisfies the
quantum horizon boundary condition as well as quantum Einstein
equations. Then the desired micro-canonical ensemble consists of
states in $\mathcal{H}^\mathrm{BH}= \oplus_n \mathcal{H}^\mathrm{BH}_n$. Note that,
because there is no contribution to the symplectic structure from
matter terms, surface states in $\mathcal{H}^\mathrm{BH}$ refer only
to the gravitational sector.

The next step is to calculate the entropy of this quantum,
micro-canonical ensemble. Note first that what matters are only the
surface states. For, the `bulk-part' describes, e.g., states of
gravitational radiation and matter fields far away from $\Delta$ and
are irrelevant for the entropy $S_\Delta$ of the WIH. Heuristically,
the idea then is to `trace over' the bulk states, construct a
density matrix $\rho_\mathrm{BH}$ describing a maximum-entropy
mixture of surface states and calculate $\tr \rho_\mathrm{BH} \,
\ln \rho_\mathrm{BH}$. As is usual in entropy calculations, this
translates to the evaluation of the dimension $\mathcal{N}$ of a
well-defined sub-space $\mathcal{H}_S^\mathrm{BH}$ of the surface Hilbert
space, namely the linear span of those surface states which occur
in $\mathcal{H}^\mathrm{BH}$. Entropy $S_{\Delta}$ is given by $\ln \mathcal{N}$.

A detailed calculation~\cite{dl, km3} leads to the following
expression of entropy:
\begin{equation}
  \label{entropy2}
  S_\Delta = \frac{\gamma_0}{\gamma} \frac{a_0}{4\ell_\mathrm{Pl}^2} -
  \frac{1}{2} \ln \frac{a_0}{\ell_\mathrm{Pl}^2} +
  \mathcal{O} \left( \ln \frac{a_0}{\ell_\mathrm{Pl}^2} \right),
  \qquad \mathrm{with\ }
  \gamma_0 = \approx 0.2375,
\end{equation}
where $\mathcal{O}(\ln(a_0/\ell_\mathrm{Pl}^2))$ is a term which, when
divided by $a_0/\ell_\mathrm{Pl}^2$, tends to zero as
$a_0/\ell_\mathrm{p}^2$ tends to infinity. Thus, for large
black holes, the leading term in the expression of entropy is
proportional to area. This is a non-trivial result. For example
if, as in the early treatments~\cite{ee1, ee2, ee3, ee4, ee5, ee6} one
ignores the horizon boundary conditions and the resulting
Chern--Simons term in the symplectic structure, one would find that
$S_\Delta$ is proportional to $\sqrt{a_0}$. However, the theory does
not have a unique prediction because the numerical coefficient
depends on the value of the Barbero--Immirzi parameter $\gamma$.
The appearance of $\gamma$ can be traced back directly to the fact
that, in the $\gamma$-sector of the theory, the area eigenvalues
are proportional to $\gamma$.

One adopts a `phenomenological' viewpoint to fix this ambiguity.
In the infinite dimensional space of geometries admitting $\Delta$ as
their inner boundary, one can fix one space-time, say the
Schwarzschild space-time with mass $M_0 \gg M_\mathrm{Pl}$, (or, the
de Sitter space-time with the cosmological constant $\Lambda_0 \ll
1/\ell_\mathrm{Pl}^2$, or, \ldots ). For agreement with semi-classical
considerations in these cases, the leading contribution to entropy
should be given by the Hawking--Bekenstein
formula~(\ref{entropy1}). This can happen only in the sector $\gamma =
\gamma_0$. The quantum theory is now completely determined through
this single constraint. We can go ahead and calculate the entropy
of any other type I WIH in \emph{this} theory. The result is
again $S_\Delta = S_\mathrm{BH}$. Furthermore, in this $\gamma$-sector,
the \emph{statistical mechanical} temperature of any type I WIH is
given by Hawking's semi-classical value $\kappa\hbar/(2\pi)$~\cite{ak,
  ee6}. Thus, we can do one thought experiment -- observe
the temperature of one large black hole from far away -- to
eliminate the Barbero--Immirzi ambiguity and fix the theory. This
theory then predicts the correct entropy and temperature for
\emph{all} WIHs with $a_0 \gg \ell^2$, \emph{irrespective of
other parameters such as the values of the electric or dilatonic
charges or the cosmological constant}. An added bonus comes from
the fact that the isolated horizon framework naturally
incorporates not only black hole horizons but also the
cosmological ones for which thermodynamical considerations are
also known to apply~\cite{gh}. The quantum entropy calculation is
able to handle both these horizons in a single stroke, again for
the same value $\gamma=\gamma_0$ of the Barbero--Immirzi parameter.
In this sense, the prediction is robust.

Finally, these results have been subjected to further robustness
tests. The first comes from non-minimal couplings. Recall from
Section~\ref{s6} that in presence of a scalar field which is
\emph{non-minimally} coupled to gravity, the first law is
modified~\cite{jkm, iw1, iw2}. The modification suggests that the
Hawking--Bekenstein formula $S_\mathrm{BH} = a_0/(4\ell_\mathrm{Pl}^2)$
is no longer valid. If the non-minimal coupling is dictated by the
action
\begin{equation}
  \widetilde{S}[g_{ab}, \phi] =
  \int \left( \frac{1}{16\pi G} f(\phi) R -
  \frac{1}{2}  g^{ab} \nabla_a \phi \nabla_b \phi - V(\phi)\right)
  \sqrt{- g}\, d^4x,
  \label{Action1}
\end{equation}
where $R$ is the scalar curvature of the metric $g_{ab}$ and $V$
is a potential for the scalar field, then the entropy should be
given by~\cite{acs2}
\begin{equation}
  \label{entropy3}
  S_\Delta = \frac{1}{4G\hbar} \oint f(\phi)\, d^2V.
\end{equation}
An immediate question arises: Can the calculation be extended to
this qualitatively new situation? At first, this seems very
difficult within an approach based on quantum geometry, such as
the one described above, because the relation is non-geometric.
However, the answer was shown to be in the affirmative~\cite{ac}
for type I horizons. It turns out that the metric $\bar{q}_{ab}$
on $M$  is no longer coded in the gravitational momentum $P^a_i$
alone but depends also on the scalar field. This changes the
surface term $\mathbf{\Omega}_S$ in the symplectic
structure~(\ref{sym2}) as well as the horizon boundary
condition~(\ref{hbc}) just in the right way for the analysis to go through. Furthermore,
state counting now leads to the desired expression~(\ref{entropy3})
\emph{precisely for the same value $\gamma=
\gamma_0$ of the Barbero--Immirzi parameter}~\cite{ac}.

Next, one can consider type II horizons which can be distorted and
rotating. In this case, all the (gravitational, electro-magnetic,
and scalar field) multipoles are required as \emph{macroscopic}
parameters to fix the system of interest. Therefore, now the
appropriate ensemble is determined by fixing all these multipoles
to lie in a small range around given values. This ensemble can be
constructed by first introducing multipole moment operators and
then restricting the quantum states to lie in the subspace of the
Hilbert space spanned by their eigenvectors with eigenvalues in
the given intervals. Again recent work shows that the state
counting yields the Hawking--Bekenstein formula~(\ref{entropy1})
for minimally coupled matter and its modification~(\ref{entropy3})
for non-minimally coupled scalar field, \emph{for the same value
$\gamma= \gamma_0$ of the Barbero--Immirzi parameter}~\cite{aa4, aev}.

To summarize, the isolated horizon framework serves as a natural
point of departure for a statistical mechanical calculation of
black hole entropy based on quantum geometry. How does this
detailed analysis compare with the `It from Bit'
scenario~\cite{wheeler} with which we began? First, the quantum
horizon boundary conditions play a key role in the construction of a
consistent quantum theory of the horizon geometry. Thus, unlike in
the `It from Bit' scenario, the calculation pertains only to those
2-spheres $S$ which are cross-sections of a WIH. One can indeed
divide the horizon into elementary cells as envisaged by Wheeler:
Each cell contains a single puncture. However, the area of these
cells is not fixed but is dictated by the $j$-label at the
puncture. Furthermore, there are not just 2 but rather $2j+1$
states associated with each cell. Thus, the complete theory is
\emph{much} more subtle than that envisaged in the `It from Bit'
scenario.

\newpage


\section{Outlook}
\label{s8}

In the last six sections, we summarized the isolated and dynamical
horizon frameworks and their applications. These provide a
quasi-local and more physical paradigm for describing black holes
both in the equilibrium and dynamical regimes. One of the most
pleasing aspects of the paradigm is that it provides a unified
approach to a variety of problems involving black holes, ranging
from entropy calculations in quantum gravity, to analytical issues
related to numerical simulations, to properties of hairy black
holes, to laws of black hole mechanics. More importantly, as
summarized in Section~\ref{s1}, these frameworks enable one to
significantly extend the known results based on Killing and event
horizons, and provide brand new laws in the dynamical regime.

In this section, we will discuss some of the open issues whose
resolution would lead to significant progress in various areas.

\begin{description}
\item[Isolated horizons]~\\
  This is the best understood part of the new paradigm. Nonetheless,
  several important issues still remain. We will illustrate these with
  a few examples:
  \begin{description}
  \item[\newline \it Black hole mechanics]~\\
    Throughout, we assumed that the space-time metric is $C^k$ (with
    $k \ge 3$) and the topology of $\Delta$ is $S^2 \times R$. These
    assumptions rule out, by fiat, the presence of a NUT charge. To
    incorporate a non-zero NUT charge in black hole mechanics, one
    must either allow $\Delta$ to be topologically $S^3$, or allow
    for `wire singularities' in the rotation 1-form $\omega_a$ on
    $\Delta$. The zeroth law goes through in these more general
    situations. What about the first law? Arguments based on Euclidean
    methods~\cite{shch, mann1, mann2} show that entropy is no longer
    given by the horizon area but there is also contribution due to
    `Misner strings'. However, to our knowledge, a systematic
    derivation in the Lorentzian regime is not yet available. Such a
    derivation would provide a better understanding of the
    \emph{physical} origin of the extra terms. The covariant phase
    space methods used in the isolated horizon framework should be
    applicable in this case.
  \item[\newline \it Application to numerical relativity]~\\
    We saw in Section~\ref{s5.4} that, in the IH framework, one can
    introduce an approximate analog of future null infinity
    ${\mathcal{I}}^+$ and invariant coordinate systems and tetrads in
    its neighborhood. With this structure, it is feasible to extract
    waveforms and energy fluxes in a reliable manner within the
    standard 3\,+\,1 Cauchy evolution of numerical relativity, without
    having to do a Cauchy characteristic matching or use conformal
    field equations. The challenge here is to develop, on the
    approximate ${\mathcal{I}^+}$, the analog of the basics of the
    Bondi framework~\cite{bondiflux, as, wz} at null infinity.
  \item[\newline \it Colored black holes]~\\
    As discussed in Section~\ref{s6}, black hole uniqueness theorems of
    the Einstein--Maxwell theory fail once non-Abelian fields are
    included. For example, there are black hole solutions to the
    Einstein--Yang--Mills equations with non-trivial Yang--Mills fields
    whose only non-zero charge at infinity is the ADM mass. Thus, from
    the perspective of infinity, they are indistinguishable from the
    Schwarzschild solution. However, their horizon properties are quite
    different from those of the Schwarzschild horizon. This example
    suggests that perhaps the uniqueness theorems fail because of the
    insistence on evaluating charges at infinity. Corichi, Nucamendi,
    and Sudarsky~\cite{cns} have conjectured that the uniqueness
    theorems could be restored if they are formulated in terms of all
    the relevant horizon charges. This is a fascinating idea and it has
    been pursued numerically. However, care is needed because the list
    of all relevant charges may not be obvious a priori. For example, in
    the case of static but not necessarily spherical Yang--Mills black
    holes, the conjecture seemed to fail~\cite{kksw} until one realized
    that, in addition to the standard Yang--Mills charges at the
    horizon, one must also include a topological, `winding charge' in
    the list~\cite{aacommun}. Once this charge is included, uniqueness
    is restored not only in the static sector of the
    Einstein--Yang--Mills theory, but also when Higgs fields~\cite{hkk}
    and dilatons~\cite{Kleihaus:2003sh} are included. The existence of
    these semi-numerical proofs suggests that it should be possible to
    establish uniqueness completely analytically. A second set of
    problems involves the surprising relations, e.g., between the
    soliton masses and horizon properties of colored black holes,
    obtained using isolated horizons. Extensions of these results to
    situations with non-zero angular momentum should be possible and may
    well provide yet new insights.
  \item[\newline \it Quantum black holes]~\\
    In the approach based on isolated horizons, the microscopic
    degrees of freedom responsible for the statistical mechanical
    entropy of black holes are directly related to the quantum
    geometry of horizons. Therefore, their relation to the curved
    space-time geometry is clearer than in, say, the string theory
    calculations based on D-branes. Therefore, one can now attempt to
    calculate the Hawking radiation directly in the physical
    space-time as a process in which quanta of area are converted to
    quanta of matter. However, such a calculation is yet to be
    undertaken. A direct approach requires quantum field theory (of
    matter fields) on a quantum geometry state which is approximated
    by the classical black hole space-time. Elements of this theory
    are now in place. The concrete open problem is the calculation of
    the absorption cross-section for quantum matter fields propagating
    on this `background state'. If this can be shown to equal the
    classical absorption cross-section to the leading order, it would
    follow~\cite{ak} that the spectrum of the outgoing particles would
    be thermal at the Hawking temperature. Another, perhaps more
    fruitful, avenue is to introduce an effective model whose
    Hamiltonian captures the process by which quanta of horizon area
    are converted to quanta of matter. Both these approaches are
    geared to large black holes which can be regarded as being in
    equilibrium for the process under consideration, i.e., when
    $a_\Delta \gg \ell_\mathrm{Pl}^2$. However, this approximation
    would fail in the Planck regime whence the approaches can not
    address issues related to `information loss'. Using ideas first
    developed in the context of quantum cosmology, effects of the
    quantum nature of geometry on the black hole singularity have
    recently been analyzed~\cite{ab}. As in the earlier analysis of
    the big-bang singularity, it is found that the black hole
    singularity is resolved, but the classical space-time dissolves in
    the Planck regime. Therefore, the familiar Penrose diagrams are no
    longer faithful representations of the physical situation. Suppose
    that, evolving through what was singularity in the classical
    theory, the quantum state becomes semi-classical again on the
    `other side'. Then, the indications are that information would not
    be lost: It would be recovered on \emph{full} ${\mathcal{I}}^+$,
    although observers restricted to lie in the part of space-time
    which is completely determined by the data on ${\mathcal{I}}^-$
    would see an approximate Hawking radiation. If on the other hand
    the evolved state on the `other side' never becomes
    semi-classical, information would not be recovered on the
    available ${\mathcal{I}}^+$. An outstanding open issue is which of
    this possibility is actually realized.
  \end{description}
\end{description}

\begin{description}
\item[Dynamical horizons]~\\
  The DH framework is less developed and the number of open issues is
  correspondingly higher. At least for the classical applications of
  the framework, these problems are more important.
  \begin{description}
  \item[\newline \it Free data and multipoles]~\\
    Since $H$ is space-like, to find the fields $(q_{ab}, K_{ab})$
    which constitute the DH geometry, one has to solve just the
    initial value equations, subject to the condition that $H$ admits
    a foliation by marginally trapped surfaces. A general solution of
    this problem would provide the `free data' on DHs. In the case
    when the marginally trapped surfaces are round spheres, this
    problem has been analyzed by Bartnik and Isenberg~\cite{bi}. As
    noted in Section~\ref{s2.2}, in this case there are no DHs in
    absence of matter sources. In presence of matter, one can freely
    specify the trace $K$ of the extrinsic curvature and the (radial
    component of the) momentum density $T_{ab}\hat\tau^a \hat{r}^b$,
    and determine the geometry by solving a non-linear ordinary
    differential equation whose solutions always exist locally. It
    would be interesting to analyze the necessary and sufficient
    conditions on the free data which guarantee that global solutions
    exist and the DH approaches the Schwarzschild horizon
    asymptotically. From the point of view of numerical relativity, a
    more pressing challenge is to solve the constraint equations in
    the vacuum case, assuming only that the marginally trapped
    surfaces are axi-symmetric. Using the free data, as in the case of
    isolated horizons~\cite{aepv}, one could define multipoles. Since
    $\Psi_2$ is again defined unambiguously, a natural starting point
    is to use it as the key geometrical object as in the IH
    case. However, just as the Bondi mass aspect acquires shear terms
    in presence of gravitational radiation~\cite{bondiflux}, it is
    likely that, in the transition from isolated to dynamical
    horizons, $\Psi_2$ would have to be supplemented with terms
    involving, e.g., $\sigma_{ab}$ (and perhaps also $\zeta^a$). For
    instance, by adding a suitable combination of such terms, one may
    be able to relate the rate of change of the mass quadrupole moment
    with the flux of energy across $H$.
  \item[\newline \it Geometric analysis]~\\
    The dynamical horizon framework provides new inputs for the proof
    of Penrose inequalities which, when applied to time symmetric data
    (i.e., when the extrinsic curvature vanishes), say that the total
    (ADM) mass of space-time must be greater than half the radius of
    the apparent horizon on any Cauchy slice. This conjecture was
    recently proved by Bray~\cite{hb}, and by Huisken and
    Ilmamen~\cite{hi}. Recently, for the non-time symmetric case,
    Ben--Dov has constructed an example where the apparent horizon
    does not satisfy this inequality~\cite{ibd}. This is not a
    contradiction with the original Penrose inequality which referred
    to the area of cross-sections of the event horizon, however it
    does show that extending the results beyond time-symmetry would be
    quite non-trivial. The `flows' which led to the area law in
    Section~\ref{s3} and balance equations in Section~\ref{s4.2.2} may
    be potentially useful for this purpose. This approach could lead
    to an inequality relating the area of certain marginally trapped
    surfaces (the ones connected to future time like infinity via a
    dynamical horizon) to the future limit of the Bondi mass. The
    framework also suggests a program which could shed much light on
    what John Wheeler called `the issue of the final state': what are
    the final equilibrium states of a dynamical black hole and how, in
    detail, is this equilibrium reached? From a mathematical
    perspective an important step in addressing this issue is to
    analyze the non-linear stability of Kerr black holes. Consider,
    then, a neighborhood of the initial data of the Kerr solution in
    an appropriate Sobolev space. One would expect the space-time
    resulting from evolution of this data to admit a dynamical horizon
    which, in the distant future, tends to an isolated horizon with
    geometry that is isomorphic to that of a Kerr horizon. Can one
    establish that this is what happens? Can one estimate the `rate'
    with which the Kerr geometry is approached in the asymptotic
    future? One avenue is to first establish that the solution admits
    a Kerr--Schild type foliation, each leaf of which admits an
    apparent horizon, then show that the world tube of these apparent
    horizons is a DH, and finally study the decay rates of fields
    along this DH.
  \item[\newline \it Angular momentum]~\\
    As we saw in Section~\ref{s5}, most of the current work on IHs and
    DHs assumes the presence of an axial symmetry $\phi^a$ on the
    horizon. A natural question arises: Can one weaken this requirement
    to incorporate situations in which there is only an approximate --
    rather than an exact -- symmetry of the horizon geometry? The answer
    is in the affirmative in the following sense. Recall first that the
    Newman--Penrose component $\Psi_2$ is gauge invariant on IHs and
    DHs. Let $F = |\Psi_2|^2$. (While any geometric field could be used
    here, $F$ is the most natural candidate because, for IHs, $\Psi_2$
    encodes the horizon geometry.) If the horizon geometry admits a
    symmetry, the orbits of the symmetry field $\phi^a$ are the level
    surfaces of $F$. More generally, let us suppose that the level
    surfaces of $F$ provide a foliation of each good cut $S$ (minus two
    points) of the horizon. Then, using the procedure of section 2.1
    of~\cite{aepv}, one can introduce on $S$ a vector field $\phi_0^a$,
    tangential to the foliation, which has the property that it agrees
    with the symmetry vector field $\phi^a$ whenever the horizon
    geometry admits a symmetry. The procedure fails if the metric on $S$
    is at of a round sphere but should work generically. One can then
    use $\phi^a_0$ to define angular momentum. On IHs this angular
    momentum is conserved; on DHs it satisfies the balance law of
    Section~\ref{s4.2.2}; and in terms of the initial data, angular
    momentum $J^{(\phi_0)}$ has the familiar
    form~(\ref{eq:generalJ}). IS this proposal viable for numerical
    simulations? In cases ready analyzed, it would be interesting to
    construct $\phi_0^a$ d compare it with the symmetry vector field
    obtained via Killing transport. A better test would be provided by
    non-axisymmetric Brill waves. A second issue associated with angular
    momentum is ether the Kerr inequality $J \le GM^2$ can be violated
    in the early stages of black hole formation or merger, particularly
    in a non-axisymmetric context. Equations that must hold on DHs
    provide no obvious obstruction~\cite{ak2}. Note that such an
    occurrence is \emph{not} incompatible with the DH finally settling
    down to a Kerr horizon. For, there is likely to be radiation trapped
    between the DH and the `peak' of the effective gravitational
    potential that could fall into the DH as time elapses, reducing its
    angular momentum and increasing its mass. The issue of whether a
    black hole can violate the Kerr inequality when it is first formed
    is of considerable interest to astrophysics~\cite{sf}. Again,
    numerical simulations involving, say, Brill waves would shed
    considerable light on this possibility.
  \item[\newline \it Black hole thermodynamics]~\\
    The fact that an integral version of the first law is valid even for
    non-equilibrium processes, during which the horizon makes a
    transition from a given state to one which is far removed, has
    interesting thermodynamic ramifications. In \emph{non}-equilibrium
    thermodynamical processes, in general the system does not have time
    to come to equilibrium, whence there is no canonical notion of its
    temperature. Therefore, while one can still interpret the difference
    $E_2 - E_1 - (\mathrm{work})$ as the heat $\Delta Q$ absorbed by the
    system, in general there is no longer a clean split $\Delta Q =
    T\Delta S$ of this term into a temperature part and a change in
    entropy part. If the process is such that the system remains close
    to equilibrium throughout the process, i.e., can be thought of as
    making continuous transitions between a series of equilibrium
    states, then the difference can be expressed as $\int T dS$, where
    the temperature $T$ varies slowly during the transition. The
    situation on dynamical horizons is analogous. It is only when the
    horizon geometry is changing slowly that the effective surface
    gravity $\bar\kappa$ of Section~\ref{s4.2.2} would be a good measure
    of temperature, and the horizon area a good measure of entropy (see
    Section~5.3 of~\cite{ak2}). These restricted situations are
    nonetheless very interesting. Can the black hole entropy derivations
    based on counting of micro-states, such as those of~\cite{abck}, be
    extended to such DHs? In the case of event horizons one would not
    expect such a procedure to be meaningful because, as we saw in
    Equation~\ref{s2.2.2}, an event horizon can be formed and grow in a
    flat space region in anticipation of a future gravitational
    collapse. It is difficult to imagine how a quasi-local counting of
    micro-states can account for this phenomenon.
  \end{description}
\end{description}

Perhaps the most surprising aspect of the current status of the
theory of black holes is that so little is known about their
properties in the fully dynamical and non-linear regime of general
relativity. Indeed, we do not even have a fully satisfactory
\emph{definition} of a dynamical black hole. Traditionally, one
uses event horizons. But as we discussed in detail, they have
several undesirable features. First, they are defined only in
space-times where one can unambiguously identify infinity. Even in
these restricted contexts, event horizons are teleological, can
form in a flat region of space-time and grow even though there is
no flux of energy of any kind across them. When astronomers tell
us that there is a black hole in the center of Milky Way, they are
certainly not referring to event horizons.

Numerical simulations~\cite{bks, gv2} suggest that the outermost
marginally trapped world-tubes become dynamical horizons soon
after they are formed. As we saw, dynamical horizons have a number
of attractive properties that overcome the limitations of event
horizons: they are defined quasi-locally, can not be formed in
flat space-time, and their growth is dictated by balance laws with
direct physical interpretation. Physically, then, dynamical
horizons satisfying the additional physical condition $\mathcal{L}_n
\Theta_{(\ell)} <0 $ (i.e., SFOTHs) appear to be good
candidates to represent the surface of a black hole. But so far,
our understanding of their uniqueness is rather limited. If a
canonical dynamical horizon could be singled out by imposing
physically reasonable conditions, one could use it as the
physical representation of an evolving black hole.

A plausibility argument for the existence of a canonical dynamical
horizon was given by Hayward. Note first that on physical grounds
it seems natural to associate a black hole with a connected,
trapped region ${\mathcal{T}}$ in space-time (see
Section~\ref{s2.2.1}). Hayward~\cite{sh} sketched a proof that, under
seemingly natural but technically strong conditions, the dynamical
portion of its boundary, $\partial {\mathcal{T}}$, would be a
\emph{dynamical horizon} $H$. This $H$ could serve as the
canonical representation of the surface of an evolving black hole.
However, it is not clear whether Hayward's assumptions are not too
strong. To illustrate the concern, let us consider a single black
hole. Then, Hayward's argument implies that there are no trapped
surfaces outside $H$. On the other hand there has been a general
expectation in the community that, given any point in the interior
of the event horizon, there passes a (marginally) trapped surface
through it (see, e.g., \cite{de}). This would imply that the
boundary of the trapped region is the \emph{event horizon} which,
being null, can not qualify as a dynamical horizon. However, to
our knowledge, this result has not been firmly established. But it
is clear that this expectation contradicts the conclusion based on
Hayward's arguments. Which of these two expectations is correct?
It is surprising that such a basic issue is still unresolved. The
primary reason is that very little is known about trapped and
marginally trapped surfaces which fail to be spherical symmetric.
Because of this, we do not know the boundary of the trapped region
even in the Vaidya solution.

If it should turn out that the second expectation is correct, one
would conclude that Hayward's assumptions on the properties of the
boundary of the trapped region are not met in physically
interesting situations. However, this would not rule out the
possibility of singling out a canonical dynamical horizon through
some other conditions (as, e.g., in the Vaidya solution). But
since this dynamical horizon would not be the boundary of the
trapped region, one would be led to conclude that in the dynamical
and fully non-linear regime, one has to give up the idea that
there is a single 3-manifold that can be interpreted as the black
hole surface without further qualifications. For certain questions
and in certain situations, the dynamical horizon may be the
appropriate concept, while for other questions and in different
situations, the boundary of the trapped region (which may be the
event horizon) may be more appropriate.

\newpage


\section{Acknowledgements}

We are grateful to John Baez, Chris Beetle, Alejandro Corichi, Sergio
Dain, Olaf Dreyer, Jonathan Engle, Stephen Fairhurst, Greg Galloway,
Sean Hayward, Jose Luis Jaramillo, Jerzy Lewandowski, Kirill Krasnov,
Tomasz Pawlowski, Erik Schnetter, Deirdre Shoemaker, Daniel Sudarsky,
Chris van den Broeck, and Jacek Wisniewski for collaboration on
several of the results summarized in this review. We have also
benefited from stimulating discussions with numerous colleagues,
especially among them Marcus Ansorg, Ivan Booth, Bobby Beig, Piotr
Bizo\'{n}, Piotr Chru\'{s}ciel, Peter Diener, Sam Finn, \'{E}anna
Flanagan, Jim Hartle, Ian Hawke, Gary Horowitz, Gerhard Huisken, Jim
Isenberg, Pablo Laguna, Luis Lehner, Richard Matzner, Jorge Pullin,
Oscar Reula, Bernd Schmidt, Walter Simon, Ken Smith, Josh Willis, and
Jeff Winicour. This work was supported in part by the National Science
Foundation grants PHY-0090091, the National Science Foundation
Cooperative Agreement PHY-0114375, the Albert-Einstein-Institut, the
Erwin-Schr\"odinger-Institut, the Kavli Institute of Theoretical
Physics, the Alexander von Humboldt Foundation, and the Eberly research
funds of Penn State.

\newpage


\bibliography{refs}

\end{document}